\documentclass[prb,twocolumn,showpacs]{revtex4}

\usepackage{graphicx}
\usepackage{dcolumn}
\usepackage{amsmath}
\usepackage{amssymb}

\begin{document}
\graphicspath{{/EPSF/}{figures/}{./}} 

\title{Renormalization group for 2D fermions with a flat Fermi surface}

\author{S\'ebastien Dusuel$^{1,2}$, Fern\~{a}o Vistulo de Abreu$^{3}$, Beno\^{\i}t Dou\c{c}ot$^{2,1}$}

\address{
$^1$ Laboratoire de Physique Th\'eorique et Hautes
 Energies, CNRS UMR 7589, Universit\'e Paris VII-Denis Diderot,\\
4, Place Jussieu, 75252 Paris Cedex 05, France.}

\address{$^2$ Laboratoire de Physique de la mati\`ere condens\'ee, CNRS UMR 8551, Ecole Normale Sup\'erieure,\\
24, Rue Lhomond, 75231 Paris Cedex 05, France}

\address{$^3$ Departamento de F\'{\i}sica, Universidade de Aveiro,\\
3810 Aveiro, Portugal.}

\begin{abstract}
We present a renormalization group analysis of two-dimensional interacting fermion systems with a closed and partially flat Fermi surface. Numerical solutions of the one-loop flow equations show that for a bare local repulsion, the system evolves through three different regimes as the typical energy is lowered~: a high-energy transient with a strong competition between particle-particle and particle-hole channels, an intermediate regime with dominant spin density wave correlations, and finally a ``hot spot'' regime exhibiting d-wave superconductivity. We study, mostly by analytical methods, how this flow pattern depends on the number $N$ of Fermi surface patches used in the numerical solution. This clearly indicates that the final regime requires vanishingly small microscopic interactions, for the one-loop approximation to be valid, as $N$ is going to infinity.
\end{abstract}


\maketitle

\section{INTRODUCTION}
\label{sec:intro}
Over the past years, several groups have been using renormalization group (RG) methods to study two-dimensional (2D) interacting electron systems.\cite{Zanchi96,Zanchi00,Halboth00,Furukawa98,Honerkamp01,Gonzalez97,Gonzalez00,Shan01,Irkhin01} One of the main advantage of these approaches is that in principle they are not biased towards any special type of instability. Charge and spin ordering and superconductivity are treated on the same footing, which is highly desirable in the modelling of many materials such as oxide superconductors or organic conductors. The quoted works often deal with rather complex situations, since they are inspired by real systems. In fact, this is another advantage with the RG schemes based on successive elimination of high-energy modes~: basically no assumption (such as renormalizability) has to be made on the underlying microscopic model. It has therefore been possible to study the effect of Van-Hove singularities, in the absence\cite{Gonzalez97,Gonzalez00,Irkhin01} or in the presence\cite{Zanchi96,Zanchi00,Halboth00,Furukawa98,Honerkamp01,Shan01,Binz01} of partial or perfect Fermi surface nesting. Some very encouraging results have been obtained in this way, showing how d-wave superconductivity may occur in a doped Mott insulating antiferromagnet\cite{Zanchi96,Zanchi00,Halboth00} or even how the metallic character of such systems gradually disappears in momentum space, starting from saddle points to extend progressively over the whole Fermi line, as the typical energy scale is reduced.\cite{Furukawa98,Honerkamp01}

The present work is motivated by the desire to analyze in detail a slightly simpler situation, which still captures most of the complexity of some two-dimensional electron systems. We shall consider the case of a closed Fermi line which exhibits some flat segments. In this situation, we have perfect Fermi surface nesting, but no singularity in the single-particle density of states. Therefore, quantum corrections to the effective interaction produce the same logarithmic divergence in the particle-particle (or Cooper) channel as in the particle-hole (or Peierls) channel. As recognized already long ago by the Soviet School, this leads to a rather intricate competition between various possible instabilities. The corresponding ``parquet'' diagrams have been investigated in various situations~: a semi-metal in a magnetic field,\cite{Brazovskii72_1,Brazovskii72_2} the itinerant antiferromagnetic state of chromium,\cite{Dzyaloshinskii72} or the phase diagram of quasi one-dimensional conductors.\cite{Gorkov75} These important works have discussed the existence of two types of solutions, referred to as ``fixed poles'' and ``moving poles'' according to whether all the effective couplings diverge at a single energy scale or instead decouple into various classes which each have their own instability scale. The moving poles occur mostly in situations where a single channel emerges in the low-energy regime, after a complicated transient at higher energies. This corresponds to the simple picture of particle-hole or particle-particle bound state formation. Since the total momentum of these bound states is a good quantum number in a pure system, we do get decoupled poles, one for each value of this momentum. For a nearly square Fermi surface, and with repulsive local interactions, this scenario has already been supported,\cite{Dzyaloshinskii72} and confirmed more recently thanks to more extensive numerical calculations.\cite{Zheleznyak97} The mobile poles then occur in the particle-hole channel, with the formation of a spin density wave.

We have recently analyzed numerically the same problem, with special focus on the relative evolution of various couplings in the immediate vicinity of the singularity.\cite{VdA01} This was done using differential equations for the normalized couplings (i.e. the othogonal projection of the coupling vector onto the unit sphere in coupling space). To our surprise, we found that the spin density wave regime is not the final stage of the RG flow (according to the one-loop approximation). Seen in terms of normalized couplings, this regime corresponds to an intermediate fixed point which is ultimately unstable towards a d-wave superconductor~! It was then very tempting to relate this remarkable behavior of the RG flow to the temperature evolution of physical properties of superconducting cuprates in the underdoped regime,\cite{Timusk99} in which strong but local antiferromagnetic correlations built up in the pseudo gap state above the superconducting transition. The present article aims at providing a detailed explanation for the occurence of this unforeseen RG flow. Unfortunately, we shall establish that its experimental relevance is severely limited by the fact it requires vanishingly small couplings if the one-loop approximation we are using remains under good control. Numerical solutions of RG equations naturally introduce a discrete set of $N$ patches instead of the continuous Fermi line. At finite $N$, we may observe the final fixed point provided the initial couplings are chosen in a finite interval whose width shrinks to zero as $N$ goes to infinity.

We should note that our work has some common features with a series of studies on coupled one-dimensional chains,\cite{Lin97,Lin98,Ledermann00,Ledermann01} and we shall borrow some of the tools developed in these articles, specially in the works by Lin et al.\cite{Lin97,Lin98} However these investigations do not make the assumption of flat Fermi surfaces, so we get quite different results, simply because the allowed low-energy couplings are different.

This paper is organized as follows. Sec.~\ref{sec:RG2Dnested} describes the two-dimensional model, the one-loop RG equations and our numerical results, in an expanded version of our previous work.\cite{VdA01} In Sec.~\ref{sec:1loopRGforNcoupch}, we analyze a simplified model of coupled chains (the same as in the work by Zheleznyak et al\cite{Zheleznyak97}), focusing on generalized Luttinger liquid types of fixed points which were first mentioned in one of our works.\cite{Vda97} Sec.~\ref{sec:RPA} discusses the fate of ``moving poles'' solutions as $N$ is large but finite, so that perturbations from the Cooper channel are still able to affect the leading Peierls channel. In Sec.~\ref{sec:high_sym_FD} we introduce special intermediate fixed directions with large SU$(N)$ symmetry, and establish their physical properties, via the computation of some response functions. Sec.~\ref{sec:stability} discusses the local stability of these special directions in coupling space, and confirms the existence of unstable perturbations at finite $N$, as observed numerically and expected on general grounds in Sec.~\ref{sec:RPA}. The ideas used in this work are then applied in Sec.~\ref{sec:symmetry_restoration} to some other aspects of symmetry restoration under RG flows. An Appendix also contains some considerations regarding the interplay between RG and continuous global symmetries.

\section{ONE-LOOP RENORMALIZATION GROUP ANALYSIS OF TWO-DIMENSIONAL MODELS WITH A NESTED FERMI SURFACE}
\label{sec:RG2Dnested}
\subsection{CHOICE OF THE MODEL}
\begin{figure}[h]
\includegraphics[width=8cm]{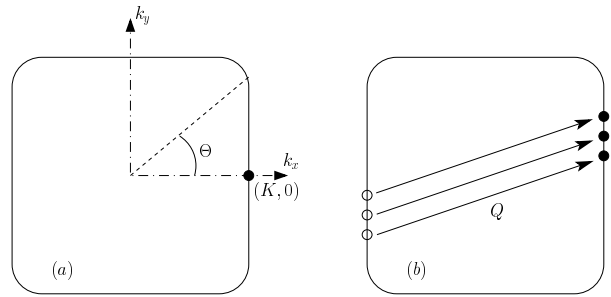}
\caption{$(a)$ : The square with rounded corners shaped Fermi surface considered in this work. The distance between the origin and any flat region is denoted by $K$, and $2\Theta$ is the angular size of the segments seen from the origin. We have $0\leq\Theta\leq\pi/4$. $(b)$ : Particle-hole pairs with vanishing excitation energy and a common total momentum $Q$.}
\label{fig:FS_AND_Q}
\end{figure}
Let us consider a two-dimensional model of non-interacting electrons for which the Fermi surface exhibits the shape shown on Fig.~\ref{fig:FS_AND_Q}$(a)$, namely a square with rounded corners. The flat regions will be referred to as nested regions, since it is possible to excite a large family of particle-hole pairs with vanishing excitation energy, provided the particle and the hole are created in the vicinity of the Fermi surface. A remarquable feature of these Fermi surfaces is that they yield continuous one-parameter families of wave vectors (see Fig.~\ref{fig:FS_AND_Q}$(b)$) for which the static charge or spin density response functions diverge in the zero temperature limit. With the notations of Fig.~\ref{fig:FS_AND_Q}$(a)$, these special wave vectors $Q$ are given by $Q=(\pm 2K,k_y)$, or $Q=(k_x,\pm 2K)$ with $|k_x|$ and $|k_y|$ freely chosen in the interval $[0,2K \tan \Theta]$. We couldn't find simple tight binding Hamiltonians which would lead to precisely this Fermi surface. But the tight binding Hamiltonian with nearest neighbor hopping on a square lattice is well known to generate a perfect square Fermi surface at half-filling. For a slightly less than half filled band, the Fermi surface is closed and remains highly anisotropic, with a diamond-like shape, although its curvature never vanishes. Experimentally, some very flat Fermi surfaces are seen on photoemission studies of $Bi_2 Sr_2 Ca Cu_2 0_8$ compounds.\cite{Fretwell00} Note that however some controversies do not seem to have been completely settled in the experimental community. Alternative views are discussed for instance by Bogdanov et al\cite{Bogdanov00} and Feng et al.\cite{Feng99} 

Our simple model is expected to be valid as long as the temperature is not too small, so that the difference between the small curvature regions and the idealized flat segments lies in the thermally excited layer around the Fermi surface. 

\begin{figure}[h]
\includegraphics[width=8cm]{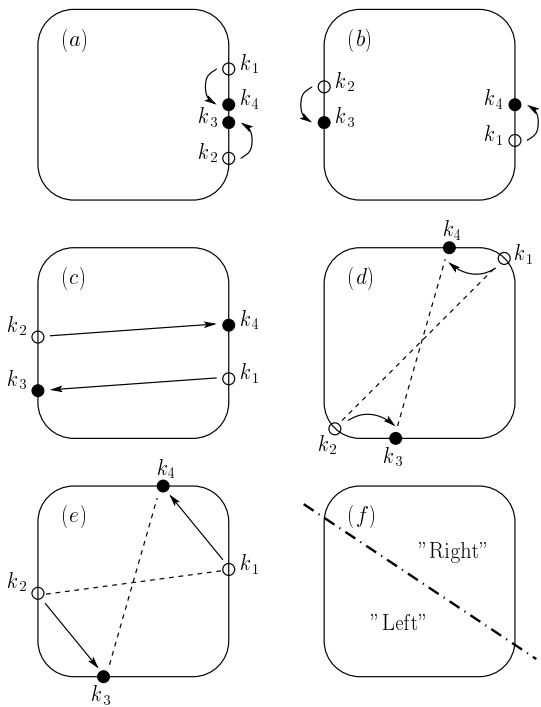}
\caption{Examples of low-energy interaction processes. $(a)$~: $g_4$ process, $(b)$~: $g_2$ process, $(c)$~: $g_1$ process, $(d)$ and $(e)$~: BCS interaction. $(e)$ is peculiar because it involves the two pairs of flat FS segments. $(f)$ shows one of the many ways to define ``Right'' and ``Left'' parts of the Fermi surface.}
\label{fig:g124BCS}
\end{figure}

Once the Fermi surface is given, we have to specify which interactions will be kept in an effective model, in order to capture the dominant processes, involving excitation energies already small with respect to the Fermi energy. Following the picture developed by Shankar,\cite{Shankar94} we consider all scattering processes with incoming momenta denoted by $k_1$ and $k_2$, and outgoing momenta by $k_3$ and $k_4$. These are constrained by momentum conservation~: $k_1+k_2=k_3+k_4$, and all the $k_i$'s have to lie on the Fermi surface. The allowed processes can be classified in four families. Three of these involve three continuous parameters each. By analogy to the standard notation for quasi one-dimensional systems,\cite{Solyom79} a first family which may be denoted by $g_4(k_1,k_2,k_3,k_4)$, involves all the $k_i$'s on the same flat segment of the Fermi surface (see Fig.~\ref{fig:g124BCS}$(a)$). 
Experience with one-dimensional systems suggest they won't play a leading role at low energies, since their effect is mostly to renormalize the single-particle Fermi velocity. We shall therefore discard them here, since we wish to focus on the possible static instabilities of our system. A second set of processes, denoted by $g_2(k_1,k_2,k_3,k_4)$, involves incoming particles on opposite parallel segments, and the transferred momentum is parallel to the Fermi surface (see Fig.~\ref{fig:g124BCS}$(b)$). Note that for generic systems, namely anisotropic Fermi surfaces with non-vanishing curvature, the effective parameters depend continuously on two variables. Indeed, in the low-energy limit, momentum conservation usually requires $(k_3,k_4)=(k_2,k_1) \mbox{ or } (k_1,k_2)$. The presence of flat segments greatly enlarges the space of the available processes, and a third variable emerges, since the transferred momentum may be varied at will along the flat directions. A third three parameter family generalizes the back-scattering processes of quasi one-dimensional system and will be denoted by $g_1(k_1,k_2,k_3,k_4)$. They are depicted on Fig.~\ref{fig:g124BCS}$(c)$. Finally, we have the two parameter continuous family of BCS couplings for which the total incoming momentum vanishes (see Fig.~\ref{fig:g124BCS}$(d)$ and $(e)$). By contrast to the previous scattering configurations, the BCS couplings are able to connect one pair of flat regions to the other pair (Fig.~\ref{fig:g124BCS}$(e)$) or to the curved regions (Fig.~\ref{fig:g124BCS}$(d)$). So they constitute at this stage the most important difference between the present 2D model and quasi 1D systems of weakly coupled chains. We haven't considered Umklapp terms in this work. On physical grounds, they are expected to play a very important role, in the appearance of Mott insulating phases. The main reason for discarding them was to keep the space of effective Hamiltonians as small as possible, since even the simplest situation manifests already a fair amount of complexity, as we shall illustrate below. Some methodological questions are better addressed in a simple setting, which will help in further investigations of more realistic models.

Given the resemblance of our model with quasi 1D systems, it is natural to recast these effective couplings in terms of two species of fermions, which shall be referred to as ``right'' and ``left'' fermions, as shown on Fig.~\ref{fig:g124BCS}$(f)$. This partitioning is to a large extent arbitrary. It is motivated by the desire to consider the two parallel Fermi surface segments of a given orientation as different species of electrons. Let us consider a situation where $k_1$ and $k_4$ are on the ``right'' vertical segment and $k_2$ and $k_3$ are on the ``left'' one. According to the previous discussion, they are connected by the two processes $g_2(k_1,k_2,k_3,k_4) c^\dagger_{\tau}(k_4) c^\dagger_{\tau'}(k_3) c_{\tau'}(k_2) c_{\tau}(k_1)$ and $g_1(k_1,k_2,k_4,k_3) c^\dagger_{\tau}(k_3) c^\dagger_{\tau'}(k_4) c_{\tau'}(k_2) c_{\tau}(k_1)$ (see Fig.~\ref{fig:g124BCS}). In the previous expressions, and in all this article, except otherwise stated, repeated indices are to be summed over. $c^\dagger_{\tau}(k)$ is of course a fermion creation operator, with momentum $k$ and spin projection $\tau$. The Pauli principle enables us to write the second term as~: $-g_1(k_1,k_2,k_4,k_3) c^\dagger_{\tau'}(k_4) c^\dagger_{\tau}(k_3) c_{\tau'}(k_2) c_{\tau}(k_1)$. So the $g_1$ interaction can be cast in a current-current type (right goes into right and left goes into left) which reads~:
\begin{eqnarray}
-\frac{1}{2} g_1(k_1,k_2,k_4,k_3) \Bigl[ c^\dagger_{\tau}(k_4) c^\dagger_{\tau'}(k_3) c_{\tau'}(k_2) c_{\tau}(k_1) \nonumber\\
+ \sum_{a=1}^3 \sigma^a_{\lambda \mu} \sigma^a_{\lambda' \mu'} c^\dagger_{\lambda}(k_4) c^\dagger_{\lambda'}(k_3) c_{\mu'}(k_2) c_{\mu}(k_1) \Bigl]\;.
\end{eqnarray}
Here $\sigma^1$, $\sigma^2$ and $\sigma^3$ are the usual Pauli matrices. Note that the BCS couplings can also be treated in the same way. In the limit where all excited states have been integrated out, we therefore have the following form of the effective Hamiltonian~: 
\begin{widetext}
\begin{eqnarray}
H_{\rm int}^{\rm eff}=\frac{1}{\Omega} \sum_{k_1,k_2,k_3,k_4} \delta_{k_1+k_2,k_3+k_4} \times \Bigl\lbrace g^{\rm c}(k_1,k_2,k_3,k_4) c^\dagger_{{\rm R},\tau}(k_4) c^\dagger_{{\rm L},\tau'}(k_3) c_{{\rm L},\tau'}(k_2) c_{{\rm R},\tau}(k_1) \hspace{3cm}\\
+g^{\rm s}(k_1,k_2,k_3,k_4) \sigma^a_{\lambda \mu} \sigma^a_{\lambda' \mu'} c^\dagger_{{\rm R},\lambda}(k_4) c^\dagger_{{\rm L},\lambda'}(k_3) c_{{\rm L},\mu'}(k_2) c_{{\rm R},\mu}(k_1) \Bigl\rbrace\nonumber\;.
\end{eqnarray}
\end{widetext}
In this expression, the wave vectors are chosen in a very narrow shell around the Fermi surface, and the subscripts ${\rm R}$ and ${\rm L}$ denote the side of the partition. $g^{\rm c}$ and $g^{\rm s}$ will be referred to as charge and spin couplings respectively. We quantize the system in a finite box of volume $\Omega$, so the $k$ vectors are chosen on a discrete set, and therefore $\delta_{k_1+k_2,k_3+k_4}$ is a Kronecker symbol. With this normalization~: $\lbrace c_{\tau}(k),c^\dagger_{\tau'}(k')\rbrace=\delta_{k,k'}\delta_{\tau,\tau'}$. In a renormalization group approach, it is common to introduce some cut-offs. Since we shall be working at the one-loop approximation, we have decided to use a single high energy cut-off which will be denoted by $\Lambda$. This means we consider only the one-particle states lying at a distance smaller than $\Lambda$ to the Fermi surface. Following Anderson's ``poor man scaling'' procedure, we will compute the effect of a small change in $\Lambda$ on the interaction vertex, and show this change can be compensated by a small modification of the effective couplings. So the effective Hamiltonian written above (where states were restricted to the Fermi surface) motivates the following form for the finite cut-off interacting Hamiltonian~:
\begin{widetext}
\begin{eqnarray}
H_{\rm int}^{\rm eff}(\Lambda)=\frac{1}{\Omega} \sum_{k_1,k_2,k_3,k_4}^{(\Lambda)} \delta_{k_1+k_2,k_3+k_4} \Bigl\lbrace g^{\rm c}_\Lambda ({\mathcal P} k_1,{\mathcal P} k_2,{\mathcal P} k_3,{\mathcal P} k_4) c^\dagger_{{\rm R},\tau}(k_4) c^\dagger_{{\rm L},\tau'}(k_3) c_{{\rm L},\tau'}(k_2) c_{{\rm R},\tau}(k_1)\hspace{3cm}\\
+g^{\rm s}_\Lambda ({\mathcal P} k_1,{\mathcal P} k_2,{\mathcal P} k_3,{\mathcal P} k_4) \sigma^a_{\lambda \mu} \sigma^a_{\lambda' \mu'} c^\dagger_{{\rm R},\lambda}(k_4) c^\dagger_{{\rm L},\lambda'}(k_3) c_{{\rm L},\mu'}(k_2) c_{{\rm R},\mu}(k_1) \Bigl\rbrace\nonumber\;.
\end{eqnarray}
\end{widetext}
In this expression, the wave vectors lie in the strip of width $2\Lambda$ around the Fermi surface. The charge and spin couplings acquire a $\Lambda$ dependence, as physical observables (Green's functions) are required not to depend on $\Lambda$. Following arguments exposed for instance in Shankar's review, we shall neglect the dependence of $g^{\rm c}_\Lambda$ and $g^{\rm s}_\Lambda$ with respect to the distance of the $k_i$'s from the Fermi surface, since such a dependence is irrelevant in the low-energy limit. We have therefore introduced the projection ${\mathcal P}$ onto the Fermi surface, along normal directions. In the same spirit, we shall neglect any frequency dependence of these functions. Note that the BCS couplings are naturally taken into account in this description, if we choose $k_1+k_2=0=k_3+k_4$. 

In a RG treatment, the effective couplings are obtained as solutions of a set of differential equations involving derivatives with respect to $\Lambda$. The initial conditions of these effective couplings' flow have to be specified on the basis of a microscopic model. The initial value of $\Lambda$ will be denoted by $\Lambda_0$, which is then of the order of the inverse lattice spacing. As the previous discussion has shown, the space of effective couplings is extremely large, since they correspond to continuous functions of two or three variables. Instead of trying to explore the whole space, we shall restrict ourselves to a very simple two-dimensional set of initial conditions, namely $g^{\rm c}_{\Lambda_0} (k_1,k_2,k_3,k_4)=G^{\rm c}$ and $g^{\rm s}_{\Lambda_0} (k_1,k_2,k_3,k_4)=G^{\rm s}$. For a local initial interaction, the microscopic couplings are indeed independent of momenta. The standard Hubbard model corresponds to $G^{\rm c}=U/2$ and $G^{\rm s}=-U/2$, where $U$ is positive (resp. negative) in the case of local repulsion (resp. attraction). If we were to start with an initial pair potential whose Fourier transform will be denoted by $\tilde{U}(q)$, a reasonable choice for the initial couplings would be given by~:
\begin{equation}
\left\lbrace \begin{array}{l}
g^{\rm c}_{\Lambda_0} (k_1,k_2,k_3,k_4)=\tilde{U}(k_4-k_1)-\frac{1}{2}\tilde{U}(k_3-k_1)\\
g^{\rm s}_{\Lambda_0} (k_1,k_2,k_3,k_4)=-\frac{1}{2}\tilde{U}(k_3-k_1)
\end{array}\right..
\end{equation}

\subsection{ONE-LOOP RENORMALIZATION GROUP EQUATIONS}
\label{sec:oneloopRGeq}
\begin{figure}[h]
\includegraphics[width=8cm]{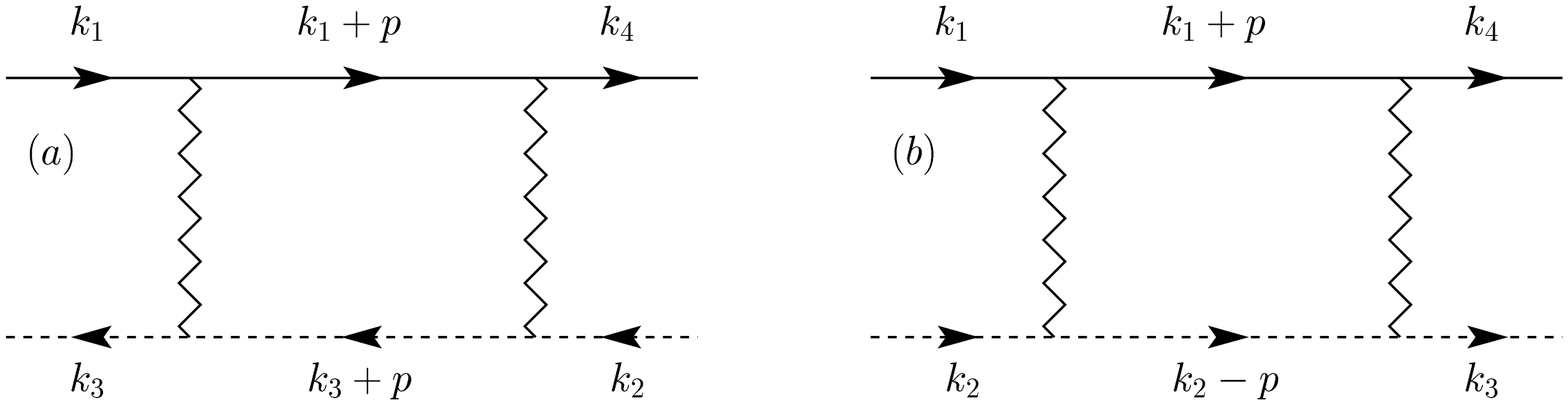}
\caption{The two graphs which contribute to the flow of effective couplings at the one-loop approximation~: $(a)$ corresponds to the Peierls channel and $(b)$ to the Cooper channel. As often, we shall use full lines for ``right'' electrons and dashed lines for ``left'' electrons.}
\label{fig:feynman2D}
\end{figure}

\begin{figure}[h]
\includegraphics[width=8cm]{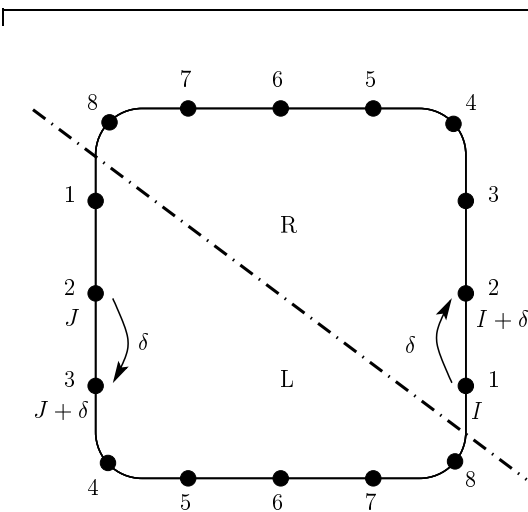}
\caption{A ``tomographic'' labelling of Fermi surface patches in the $N=3$ and $M=1$ case. In this description, two opposite points on the Fermi surface are assigned the same positive integer, but they are distinguished by their ``right''(R) or ``left''(L) nature. Arrows illustrate the $C_\delta(I,J)=C_1(1,2)$ coupling (see text).}
\label{fig:param_tomographique}
\end{figure}

Because of the very simple form of the effective Hamiltonian, the one-loop corrections to the irreducible two-particle vertex functions are given by the sum of the two contributions attached to the graphs shown on Fig.~\ref{fig:feynman2D}. 
Examining the particle-particle and particle-hole density of states, we find that these bubbles are diverging logarithmically as a function of the high energy cut-off $\Lambda$, for some special configurations of external leg momenta. In the Cooper channel, this occurs if the total incoming momentum is vanishing or if it is parallel to one of the flat Fermi surface segments. In the particle-hole channel, the divergence is obtained if the transferred momentum $k_1-k_3$ is of the form $(\pm 2 K, k_y)$ or $(k_x,\pm 2 K)$, with $|k_x|$ and $|k_y|$ smaller than $2K\tan\Theta$. With this type of external configurations, we have the simple result~:
\begin{widetext}
\begin{eqnarray}
\Delta\Gamma(k_1,\omega_1,\ldots,k_4,\omega_4)=-\frac{1}{2\pi v_{\rm F}} \ln\left| \frac{v_{\rm F}\Lambda}{\omega_1+\omega_2}\right| \int_{\rm F.S.} \frac{{\rm d}\mu(k_1+p)}{2\pi} g_\Lambda(k_1,k_2,k_2-p,k_1+p) g_\Lambda(k_1+p,k_2-p,k_3,k_4)\nonumber\\
+\frac{1}{2\pi v_{\rm F}} \ln\left| \frac{v_{\rm F}\Lambda}{\omega_1-\omega_3}\right| \int_{\rm F.S.} \frac{{\rm d}\mu(k_1+p)}{2\pi} g_\Lambda(k_1,k_3+p,k_3,k_1+p) g_\Lambda(k_1+p,k_2,k_3+p,k_4)\;.
\end{eqnarray}
\end{widetext}
In these expressions, the $k_i$'s are assumed to be on the Fermi surface, with $k_1$ and $k_4$ on the ``right'' side of the partition and $k_2$ and $k_3$ on the ``left'' side. In the first integral (Cooper channel), $p$ is chosen so that $k_1+p$ lies on the ``right'' part of the Fermi surface, and $k_2-p$ on the ``left'' part. In the second (Peierls channel), $k_1+p$ again lies on the ``right'' side, but now $k_3+p$ lies on the ``left'' one. The one-dimensional measure ${\rm d}\mu(k_1+p)$ is just the infinitesimal length element on the Fermi surface. We have not explicited the charge and spin aspects of the couplings at this stage to keep the equations simple. Note that if the external momenta are not on a single pair of flat segments, $k_1-k_3$ is not a nesting vector, so only the first term contributes. In fact, for such processes, we have $k_1+k_2=0$, which means they are of BCS type, and they renormalize only in the Cooper channel. From these expressions, we see the irreducible two-particle vertex function remains unchanged (to this one-loop approximation), provided the effective couplings depend on $\Lambda$ according to~:
\begin{widetext}
\begin{eqnarray}
\frac{\partial g_\Lambda(k_1,k_2,k_3,k_4)}{\partial \ln\Lambda}=\frac{1}{2\pi v_{\rm F}} \int_{\rm F.S.} \frac{{\rm d}\mu(k_1+p)}{2\pi} g_\Lambda(k_1,k_2,k_2-p,k_1+p) g_\Lambda(k_1+p,k_2-p,k_3,k_4)\nonumber\\
\label{eq:eq_RG_2D}
-\frac{1}{2\pi v_{\rm F}} \int_{\rm F.S.} \frac{{\rm d}\mu(k_1+p)}{2\pi} g_\Lambda(k_1,k_3+p,k_3,k_1+p) g_\Lambda(k_1+p,k_2,k_3+p,k_4)\;.
\end{eqnarray}
\end{widetext}
As previously, all the wave vectors on the RHS which appear as arguments in effective couplings are on the Fermi surface. It is now important to keep track of the charge and spin channels of the interaction. We obtain two equations with the same structure as Eq.~(\ref{eq:eq_RG_2D}), but the quadratic form on the RHS should now be written according to Table \ref{tab:flotc_et_s}.
\begin{table}
\caption{Quadratic form on the RHS of Eq.~(\ref{eq:eq_RG_2D}) when one takes account of the spin.}
\label{tab:flotc_et_s}
\begin{tabular}{|c|c|c|}
\hline
& First term & Second Term\\
& (Cooper Channel) & (Peierls Channel)\\ \hline
Flow of & $g_\Lambda^{\rm c} g_\Lambda^{\rm c} + 3 g_\Lambda^{\rm s} g_\Lambda^{\rm s}$ & $g_\Lambda^{\rm c} g_\Lambda^{\rm c} + 3 g_\Lambda^{\rm s} g_\Lambda^{\rm s}$ \\
Charge couplings & &\\ \hline
Flow of & $g_\Lambda^{\rm c} g_\Lambda^{\rm s} + g_\Lambda^{\rm s} g_\Lambda^{\rm c} $ & $g_\Lambda^{\rm c} g_\Lambda^{\rm s} + g_\Lambda^{\rm s} g_\Lambda^{\rm c}$ \\
Spin couplings & $-2g_\Lambda^{\rm s} g_\Lambda^{\rm s}$ & $+2g_\Lambda^{\rm s} g_\Lambda^{\rm s}$ \\
\hline
\end{tabular}
\end{table}
Finally, these equations are still difficult to solve, even on a computer. So the integrals on the RHS are replaced by finite sums, which amounts to discretizing the Fermi surface with a finite set of patches. Let us denote by $N$ (resp. $M$) the number of these patches on each straight segment (resp. curved arc) of the Fermi surface. We therefore have $2(M+N)$ patches of each type (right and left). If we assume patches of equal length, the shape of the Fermi surface, i.e. the angle $\Theta$, is related to $M/N$ by~: $M/N=(\pi-4\Theta)/(4\sin\Theta)$. Using the ``tomographic'' parametrization of the couplings illustrated on Fig.~\ref{fig:param_tomographique}, we may replace the continuous functions $g_\Lambda(k_1,k_2,k_3,k_4)$ by ``vectors'' in coupling space $C^{(\Lambda)}_\delta(I,J)$. 
This notation means that a pair of particles on patches $I$ on the right and $J$ on the left scatters into another pair of patches $I+\delta$ on the right and $J+\delta$ on the left. This labelling scheme is consistent with the constraint of momentum conservation. We therefore get the discretized version of Eq.~(\ref{eq:eq_RG_2D})~:
\begin{widetext}
\begin{eqnarray}
\frac{\partial C^{(\Lambda)}_\delta(I,J)}{\partial \ln\Lambda}=\frac{P}{16 \pi^2 v_{\rm F}(M+N)} \left\lbrace \sum_\alpha f_{\rm C}(I,J,\delta,\alpha) C^{(\Lambda)}_\alpha(I,J)C^{(\Lambda)}_{\delta-\alpha}(I+\alpha,J+\alpha)\right.\nonumber\\
\label{eq:eq_RG_2D_discr}
\left. -\sum_\alpha f_{\rm P}(I,J,\delta,\alpha) C^{(\Lambda)}_\alpha(I,J+\delta-\alpha)C^{(\Lambda)}_{\delta-\alpha}(I+\alpha,J)\right\rbrace\;.
\end{eqnarray}
\end{widetext}
Here, $P$ is the perimeter of the Fermi surface, so $P=[(2\pi-8\Theta)/\cos\Theta+8\tan\Theta]K$. $f_{\rm C}$ and $f_{\rm P}$ enforce phase space restrictions for the Cooper and Peierls channels. In the Peierls channel, our previous discussion has shown that patches $I$, $I+\alpha$ and $I+\delta$ have to lie on a given flat segment on the right side, with patches $J$, $J+\delta-\alpha$ and $J+\delta$ on the opposite parallel segment. If this condition is met, $f_{\rm P}$ is set equal to one, and vanishes otherwise. In the Cooper channel, $f_{\rm C}$ is equal to one if a similar condition is satisfied. But this channel allows for processes which connect one set of parallel flat segments to the remainder of the Fermi surface. In this case, $I=J$ and $f_{\rm C}$ is also equal to one. In any other situation, $f_{\rm P}$ vanishes. By using Table \ref{tab:flotc_et_s} and with a minor change in the couplings' normalization, we obtain the flow equations (2) and (3) of our previous work\cite{VdA01}, which we won't reproduce here to save space.

We have performed numerical evaluations of the discretized equations for various initial conditions of the form~:
\begin{equation}
{C^{\rm c}_\delta}^{(\Lambda_0)}(I,J)=C^{\rm c}, \mbox{ and } {C^{\rm s}_\delta}^{(\Lambda_0)}(I,J)=C^{\rm s}\;.
\end{equation}
Again the simple Hubbard model corresponds to $C^{\rm c}+C^{\rm s}=0$ with $C^{\rm c}$ positive in the case of repulsion. In most cases we find that the effective couplings diverge after reaching a low-energy scale $\Lambda_{\rm c}<\Lambda_0$. Although we are working in a zero temperature formalism, it is natural (and often done in the literature) to interpret $\Lambda_{\rm c}$ as a typical temperature scale for the onset of some kind of long range order, either in the particle-particle channel (superconductivity) or in the particle-hole channel (charge or spin density waves). As will be described in detail in Sec.~\ref{sec:reponse} below, the response functions for the corresponding order parameters can be evaluated in the one-loop approximation, and are found to diverge in general as power laws of $\ln(\Lambda/\Lambda_{\rm c})$, where $\Lambda$ is a typical frequency scale at which the system is probed.

In order to get a better understanding of these one-loop RG flows, we have found useful to decompose them in two different (although not unrelated) aspects~: the evolution of the magnitude of the couplings, and the evolution of the direction in the space of all possible couplings. This is possible because one-loop flows have a nice invariance under dilations of the effective couplings. To see this, we note that our RG equations take the simple form~: $\partial_t g_i=A_{ijk}g_j g_k$. Here, the flow's time $t$ is related to the current cut-off $\Lambda$ by $t=\ln(\Lambda_0/\Lambda)$. The $g_i$'s are for instance the set of $C_\delta(I,J)$'s defined above and used in numerical computations. The initial conditions (microscopic couplings) correspond to $t=0$, and $t$ increases as the high energy cut-off is gradually reduced. Since the RHS of the one-loop RG equations is homogeneous and quadratic in the couplings, we have the simple property that if $g_i(t)$ is a solution so is $g_i^{(\lambda)}(t) \equiv \lambda g_i(\lambda t)$ for any $\lambda$. This implies that $g_i^{(\lambda)}(t)/g_j^{(\lambda)}(t)=g_i(\lambda t)/g_j(\lambda t)$, namely that the direction of the coupling vector follows a trajectory which is independent of the magnitude of the initial couplings. It is therefore natural to introduce normalized couplings $h_i(t)$ defined by
\begin{equation}
h_i(t)=\frac{g_i(t)}{{\mathcal N}(t)},\mbox{ with } {\mathcal N} (t) = {\left( \sum_i {g_i (t)}^2 \right)}^{\frac{1}{2}}\;.
\end{equation}
We then have for any positive $\lambda$, and with obvious notation~: ${\mathcal N}^{(\lambda)} (t)=\lambda {\mathcal N}(\lambda t)$ and ${h}^{(\lambda)}_i(t)=h_i(\lambda t)$. Next we wish to write down differential equations for ${\mathcal N}(t)$ and for $h_i(t)$ instead of $g_i(t)$. It is straightforward to show that one has~:
\begin{equation}
\label{eq:normedfloweq}
\left\{
\begin{array}{l}
\partial_s h_i = A_{ijk} h_j h_k - (A_{jkl} h_j h_k h_l)h_i\\
\\
\partial_s {\mathcal N} = (A_{jkl} h_j h_k h_l) {\mathcal N}  \;
\end{array}
\right.,
\end{equation}
with $s$ being the ``time'' adapted to the flow of $h_i$, defined by ${\rm d} s={\mathcal N}(t) {\rm d} t$ and $s(t=0)=0$. Changing from $t$ to $s$ amounts to zooming near the singularity at $t_{\rm c}=\ln(\Lambda_0/\Lambda_{\rm c})$, and $s$ can therefore be taken as large as possible without actually reaching the divergence of ${\mathcal N}$. However, we should keep in mind that our RG equations are valid only if ${\mathcal N}(t)$ is not too large. How does this translate on the flows regarded as functions of $s$ ? From Eq.~(\ref{eq:normedfloweq}), we have~: 
\begin{eqnarray}
{\mathcal N}(s)={\mathcal N}(0) e^{F(s)}, \mbox{ with }\hspace{1cm} \nonumber\\
F(s)=\int_0^{\rm s} {\rm d}s' A_{jkl} h_j(s') h_k(s') h_l(s')\;.
\end{eqnarray}
In most cases, $F(s)$ becomes infinite as $s$ goes to infinity, so there will be a typical value $s_{\rm max}$ of $s$ beyond which the one-loop equations are no longer valid. But $s_{\rm max}$ depends on the magnitude of the initial couplings, and can be made as large as desired, if ${\mathcal N}(0)$ is replaced by $\lambda{\mathcal N}(0)$ with $\lambda$ small enough, while keeping the initial direction $h_i(0)$ fixed. Part of this work deals with the large $s$ limit, which may be equivalently viewed as a very weak coupling limit.

Since the flow of the normalized couplings defined above is constrained to stay on a compact space (the unit sphere), it is expected to reach either limit cycles or fixed points in the large $s$ limit. We have never observed cycles, but fixed points (in the space of directions) are always found for large $s$. These correspond to solutions of the type $g_i(t)=u_i/(t_{\rm c}-t)$, with $u_i$ satisfying $u_i=A_{ijk}u_j u_k$. Fixed directions are interesting since they are characterized by a well defined set of exponents for the growth of various response functions in both Cooper and Peierls channels (see Sec.~\ref{sec:reponse}). This point of view has already been emphasized by several authors.\cite{Dzyaloshinskii72,Gorkov75,Lin98} The new feature of the present work lies in our attempt to reach conclusions for large values of the number $N$ of patches, as required for the description of a macroscopic 2D system. Note that when the system is flowing along a fixed direction, $F(s)$ defined above is equal to $s/c$, with $c={( \sum_i {u_i}^2 )}^{1/2}$. So ${\mathcal N}(s)={\mathcal N}(0)\exp(s/c)$. We also have $t(s)=c[1-\exp(-s/c)]/{\mathcal N}(0)$, so $t_{\rm c}=c /{\mathcal N}(0)$.

\subsection{NUMERICAL RESULTS FOR THE 2D MODEL WITH NESTED FERMI SURFACE}
\label{subsec:numres2D}
\begin{figure}[h]
\includegraphics[width=8cm]{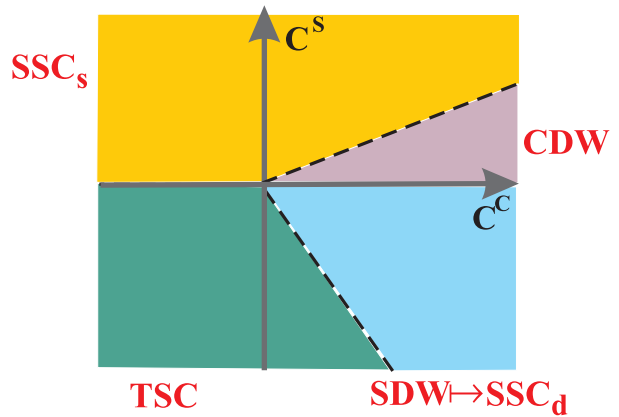}
\caption{Phase diagram for a nested system~. The flows of the couplings for the charge density wave phase and the d-wave superconducting phase are shown in Figs.~\ref{fig:flotcdw} and \ref{fig:flotsdw} respectively.}
\label{fig:diagramme_phase}
\end{figure}

\begin{figure}[h]
\includegraphics[width=8cm]{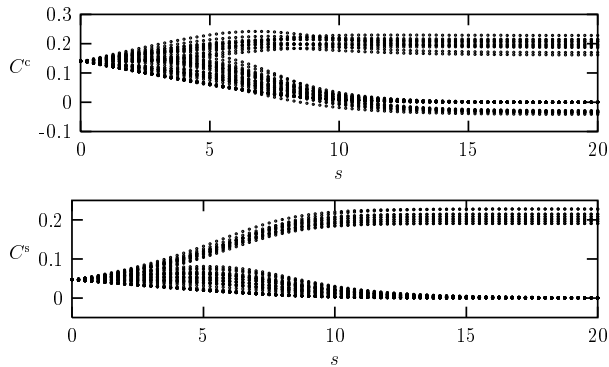}
\caption{Flow of all the charge (top) and spin (bottom) couplings, in the case of the charge density wave phase of Fig.~\ref{fig:diagramme_phase}.}
\label{fig:flotcdw}
\end{figure}

\begin{figure}[h]
\includegraphics[width=8cm]{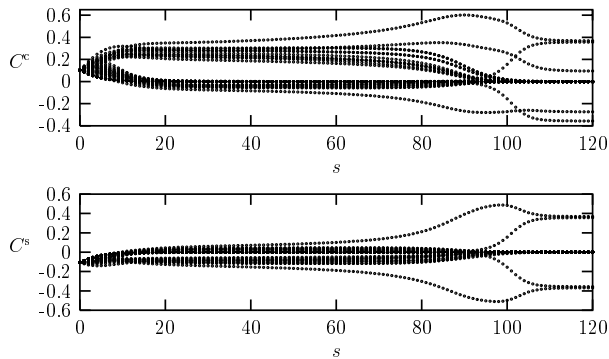}
\caption{Flow of all the charge (top) and spin (bottom) couplings, in the case of the d-wave superconducting phase diagram of Fig.~\ref{fig:diagramme_phase}.}
\label{fig:flotsdw}
\end{figure}

\begin{figure}[h]
\includegraphics[width=8cm]{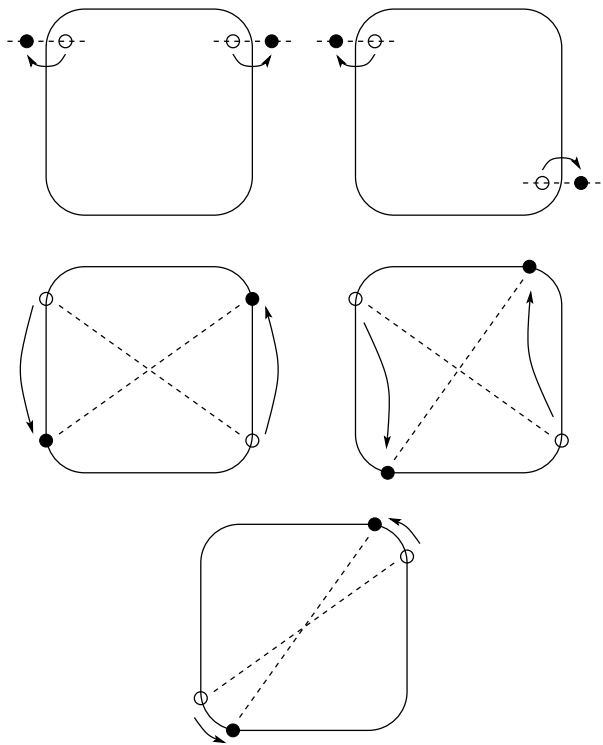}
\caption{Some of the few processes which dominate in the final stage of the flow (d-wave superconductivity fixed direction) for $C^{\rm s}<0$, $C^{\rm c}>0$ and $|C^{\rm s}/C^{\rm c}|$ not too large.}
\label{fig:inthotspots}
\end{figure}

We have performed numerical evaluations of the RG flows for a large set of ratios $C^{\rm s}/C^{\rm c}$ of the initial couplings and for various values of $M$ and $N$, where $M/N$ is the length of the curved regions divided by the length of the flat regions on the Fermi surface. We have used the set of Eqs.~\ref{eq:normedfloweq} for the normalized couplings $h_i$. Besides the theoretical motivations just discussed for introducing them, they have the practical advantage of remaining bounded during the RG flow, by contrast to the $g_i$'s which diverge at the finite ``time'' $t_{\rm c}$. In Fig.~\ref{fig:diagramme_phase}, we show the corresponding phase diagram for $N=6$ and $M=0$, together with typical flow patterns (using the $s$ variable and normalized couplings) for the charge density wave (Fig.~\ref{fig:flotcdw}) and for the d-wave superconducting phase (Fig.~\ref{fig:flotsdw}). These results have already been discussed.\cite{VdA01}
In this work we shall focus on the two regions which occur for positive charge couplings, and when the absolute value of $C^{\rm s}/C^{\rm c}$ is not too large. In these conditions, the BCS type couplings (i.e. with a vanishing total incoming momentum) are strongly reduced in the early stages of the flow. The couplings which are building up are of the Peierls-type, meaning they can be interpreted as processes with a transferred momentum $Q=(\pm 2K,0)$ or $Q=(0,\pm 2K)$ connecting one flat region to another. Note that the final behavior of the flow (for very large $s$) looks very different according to whether the initial spin coupling $C^{\rm s}$ is positive or not. If $C^{\rm s}>0$, and $C^{\rm s}/C^{\rm c}$ not too large, the system flows towards a charge density wave instability. As will be discussed below, the approach to the transition is qualitatively well described by a RPA like description which singles out the Peierls channel. If $C^{\rm s}<0$, and $|C^{\rm s}/C^{\rm c}|$ not too large, we observe three stages in the RG flow, instead of two for $C^{\rm s}>0$. After a transient regime which eliminates most of the BCS couplings, we find a rather long intermediate regime in which the dominant correlations are spin density wave like. This regime is very similar to the one obtained by Zheleznyak et al.\cite{Zheleznyak97} In agreement with them, we find a tendency for the dominant couplings to be larger for incoming and outgoing electrons close to the end points of the flat regions. But running the flow closer to the singularity shows that this regime is eventually unstable towards the formation of a very different type of correlations. Analysis of the corresponding couplings shows that events involving all the incoming and outgoing particles on the end points of the flat regions dominate all others. Some of these processes are illustrated on Fig.~\ref{fig:inthotspots}. 
So the stable fixed direction reached by the flow in this regime corresponds to an eight-patch model, independently of the actual number of patches involved in the numerical computation of the flow~! In physical terms, these eight patches are reminiscent of the ``hot spots'' observed in angle-resolved photoemission spectra of high temperature superconductors, for electron wave vectors in the vicinity of the Fermi surface intersections with Brillouin zone boudaries.\cite{Norman98} This final regime is also remarkable since the leading instability is in the Cooper channel, with the formation of singlet particle pairs of d-wave like orbital symmetry. So we interpreted the flow in this part of the phase diagram (in which the usual Hubbard model with local repulsion falls) as manifesting a cross-over phenomenon~: between a metallic high temperature regime and a d-wave superconducting phase lies an intermediate state with dominant itinerant antiferromagnetic correlations. Note that superconductivity occurs here in a perfectly nested system. This is to be contrasted with other studies,\cite{Zanchi96,Zanchi00,Halboth00,Furukawa98,Honerkamp01} in which superconductivity is possible only when the nesting is not perfect. With an imperfect nesting, contributions from the Peierls channel are suppressed at low energies, where only the Cooper channel survives. 

One of the motivations for this work was the desire to understand to which extent such a picture has a chance to apply to real macroscopic 2D systems. Three types of objections have been considered. The first one is inspired by the analogy just mentionned with some aspects of high temperature superconductors. One of the most striking features of this high-$T_{\rm c}$ cuprates is the presence of the strange pseudo-gap phase in the underdoped compounds. At least two types of scenarios have been proposed to describe this regime. In the most popular one, local Cooper pairs are formed around a temperature $T^*$ above $T_{\rm c}$, as suggested by a clear reduction of angle resolved photoemission intensity at the chemical potential energy, near the Brillouin zone boundary, where the single-particle spectral weight is pushed below the Fermi level. $T_{\rm c}$ is then interpreted as the onset of phase coherence of these preformed pairs.\cite{Timusk99} In another scenario\cite{Tallon01} superconductivity is argued to compete with another instability, whose typical onset energy scale vanishes as one goes from the underdoped to the overdoped regime. Our intermediate regime would fall in this latter type of scenario. However the simple model considered here doesn't seem to generate a spin gap in the intermediate regime. Our numerical results indicate that the static spin susceptibility at the nesting vector $Q$ steadily increases as the typical energy scale is reduced, showing no evidence of a spin gap. The cross-over to a d-wave superconductor at lower energy is simply deduced from the fact that the superconducting response function grows faster than the N\'eel one as $s$ goes to infinity. Clearly, since we are dealing with a low-dimensional system, we have to investigate the effect of long wavelength fluctuations of the N\'eel order parameter. This question is briefly addressed in Sec.~\ref{sec:phys_pict} below, and we find, within a simple version of the intermediate fixed direction exhibiting SU$(N)$ symmetry, that this ``Goldstone mode'' fluctuation induced gap vanishes like $\exp(-N)$ as $N$ goes to infinity. This is an interesting (though not very surprising) conclusion~: other physical ingredients are required to explain the pseudo-gap regime of high-$T_{\rm c}$ cuprates. A natural candidate for this are Umklapp processes as discussed within a RG approach by Honerkamp et al.\cite{Honerkamp01} On intuitive grounds, the spin gap corresponds to short range singlet correlations. Since there is no frustration on a square lattice, the strongest source of fluctuations comes from the motion of holes which do a lot of damage in an ordered magnetic state. Clearly, this picture makes sense for a moderately doped Mott insulator, and Umklapps are the natural way to describe the vicinity of the Mott insulator in a weak coupling RG approach. Umklapps have already been considered by Zheleznyak et al, for a very similar situation, and were shown not to affect much the spin density wave like regime, in the case of a half-filled Hubbard model. These authors haven't considered doped situations where the Fermi surface would remain perfectly nested, but with twice the nesting vectors $Q$ not exactly on the reciprocal lattice. It would be very interesting to study whether a pseudo-gap behavior appears in spin or uniform charge responses for such a case. We leave this for future investigation. 

Leaving aside the question of whether we may obtain an intermediate regime with a pseudo-gap, we return to the question of assessing the possibility of observing cross-over phenomena involving an intermediate direction as described above. As will be shown below, there does not seem to be any problem with the numerics on which our previous work\cite{VdA01} is based. This would be our second objection. Indeed, the example of an intermediate direction with SU$(N)$ symmetry is treated in great detail below. In the large $N$ limit, we have established analytically that this system admits $(N-1)$ relevant perturbations for the flow of the normalized couplings. So the existence of these instabilities and the tendency for the system to keep only a small finite set of ``hot spots'' in the infinite $s$ limit seem firmly established for any finite value of $N$.

The most serious objection, which hasn't been addressed before, deals with the infinite $N$ limit. Recall that patches were introduced to replace integrals of functions of momenta chosen on a continuous Fermi surface by finite sums. The validity of such an approximation depends on the way effective couplings vary with external momenta. But the large $s$ regime exhibits a very singular variation, since a hot spot picture emerges. In this situation, it is highly desirable to keep $N$ as large as possible. As will be shown, the relevant perturbations around SU$(N)$ fixed direction of spin density wave type become marginal in the infinite $N$ limit. This has the consequence that the RG time required to leave the vicinity of the intermediate fixed direction diverges as $N$ goes to infinity. According to the arguments detailed above, this implies we need to restrict ourselves to vanishingly small initial couplings as $N$ becomes infinite, in order to observe the final stage of the flow in the validity range of one-loop RG equations. At the present stage, we don't have any strong argument to counteract this objection. So the present work reinforces the validity of the more traditional interpretation already presented by Zheleznyak et al~: in a real system, the instability of the couplings occurs while they are close to an intermediate fixed direction.
\begin{figure}[h]
\includegraphics[width=8cm]{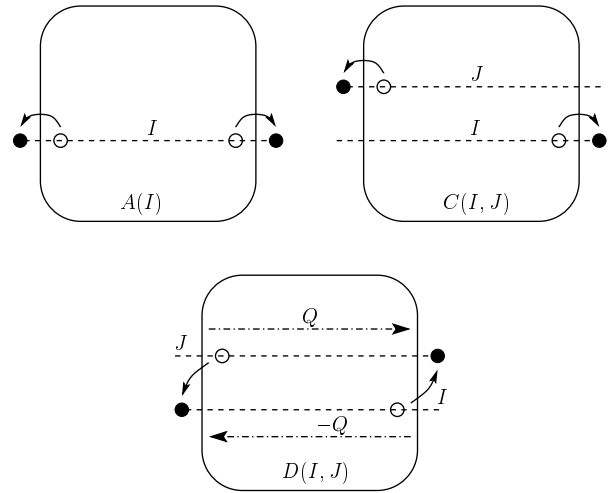}
\caption{The dominant couplings $A(I)$, $C(I,J)$ and $D(I,J)$ involving the vertical Fermi segments in the regimes of density wave like instabilities. The patch indices $I$ and $J$ refer now to the projection of the momentum along the vertical axis. Two Fermi surface points linked by a dashed line thus have the same patch index.}
\label{fig:ACD}
\end{figure}
The remaining parts of this article are mostly dedicated to the physical properties of these intermediate fixed directions. 

Let us now motivate the strategy we have followed. A detailed analysis of the effective couplings in the intermediate regime shows that the BCS couplings which connect a pair of parallel segments of the Fermi surface to its complementary parts are extremely small compared to the dominant couplings. For a system with $N=7$ and $M=0$, we found a typical ratio of $10^{-4}$ between these two classes of couplings. So, we shall adopt the approximation already made by Zheleznyak et al, i.e. we shall focus our attention on a given pair of parallel segments, thus restricting ourselves to a 1D setting.\cite{Zheleznyak97} Next, we find that the dominant couplings in this regime are of the three types depicted on Fig.~\ref{fig:ACD}. The $D(I,J)$ couplings are reinterpreted (using the Pauli principle) as scattering events with a momentum transfer $Q=(\pm 2K,0)$. 
\begin{figure}[h]
\includegraphics[width=8cm]{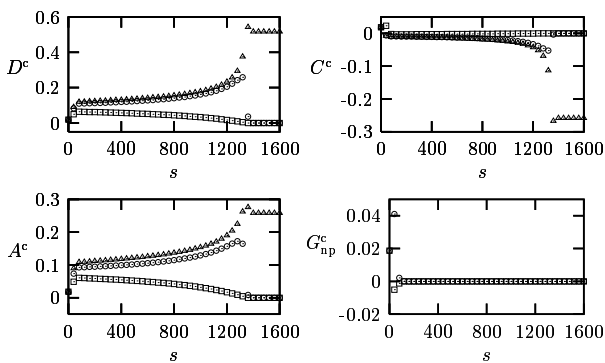}
\caption{Example of a typical 1D RG flow pattern for $N=16$, and initial condition $G^s=-G^c$. Only charge couplings are represented here (see Fig.~\ref{fig:flot_glob_s} for spin couplings). The four figures show the behavior of $D$, $C$, $A$, and non-principal couplings respectively. We have represented the couplings that survive in the end of the flow (that is involving patch numbers 1 and 16) with triangles. With circles and squares, we have shown the behavior of two charge couplings that are a sort of ``envelope'' of the couplings. By this we mean that lots of couplings are found between the values of these two couplings, as was also the case in Figs.~\ref{fig:flotcdw} and \ref{fig:flotsdw} where all couplings were shown (see also Fig.~\ref{fig:histo} for a detailed view of the distribution of the values of the couplings).}
\label{fig:flot_glob_c}
\end{figure}
\begin{figure}[h]
\includegraphics[width=8cm]{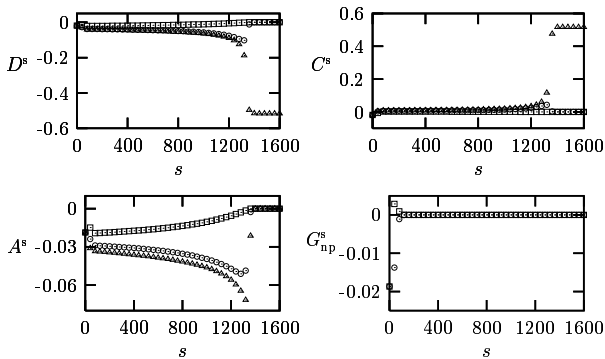}
\caption{Same as Fig.~\ref{fig:flot_glob_c} for the spin couplings.}
\label{fig:flot_glob_s}
\end{figure}
Note that a symmetrical pattern is obtained for the horizontal segments. For the sake of simplicity, it is not shown on the figure. This classification of the couplings according to their magnitude in the intermediate regime of the flow is well illustrated on Figs.~\ref{fig:flot_glob_c} to \ref{fig:flot_debut_s}. 
\begin{figure}[h]
\includegraphics[width=8cm]{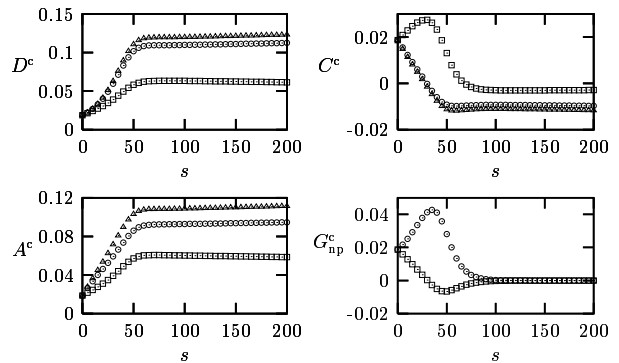}
\caption{Zoom of Fig.~\ref{fig:flot_glob_c} for times $s$ between 0 and 200.}
\label{fig:flot_debut_c}
\end{figure}
\begin{figure}[h]
\includegraphics[width=8cm]{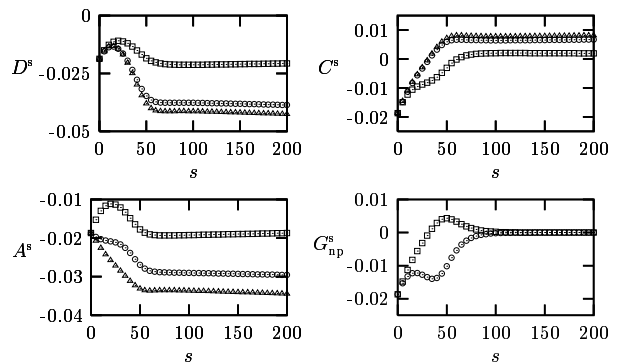}
\caption{Zoom of Fig.~\ref{fig:flot_glob_s} for times $s$ between 0 and 200.}
\label{fig:flot_debut_s}
\end{figure}
These curves have been obtained in a purely 1D setting, since we haven't found much difference between the full 2D case and this simplified version which involves $N$ coupled 1D chains. The corresponding formalism is outlined in section \ref{sec:1loopRGforNcoupch} below. Roughly speaking, the initial transient of the flow is observed for $s\lesssim 50$, and the intermediate regime for $50\lesssim 1000$. The selection of $A$, $C$ and $D$ couplings with respect to the remaining ones (which will be referred to as non-principal couplings) occurs clearly for $s$ between 50 and 100 (see Figs.~\ref{fig:flot_debut_c} and \ref{fig:flot_debut_s}). 

\begin{figure}[h]
\includegraphics[width=8cm]{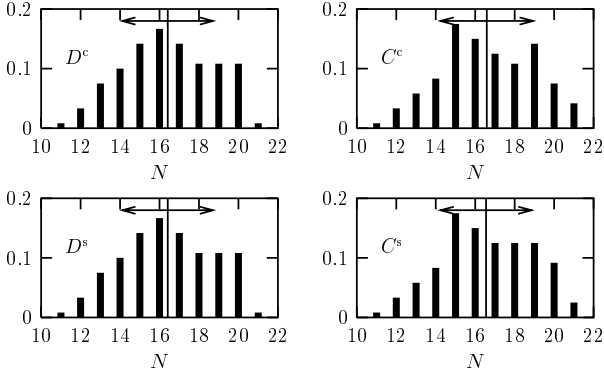}
\caption{Normalized repartition of $D$ and $C$, charge and spin couplings in Fig.~\ref{fig:flot_debut_c} and Fig.~\ref{fig:flot_debut_s} at time $s=100$. Each bar represents the normalized number of couplings taking the value they would have on the SU$(N)$ (+,-) fixed direction (see Table \ref{tab:dirfixeN}). The fine vertical line represent the mean $N$, and the arrows represent the variance.}
\label{fig:histo}
\end{figure}

\begin{figure}[h]
\includegraphics[width=8cm]{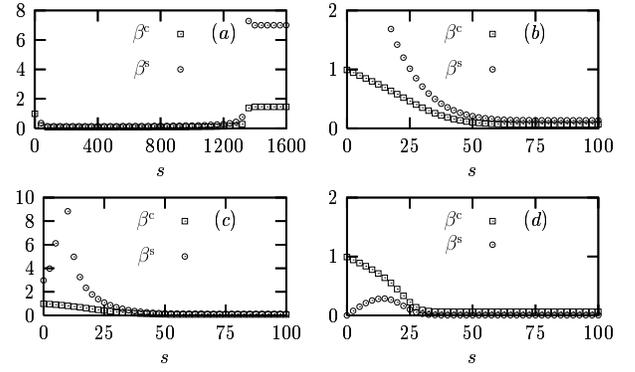}
\caption{Evolution of $\beta^{\rm c}$ and $\beta^{\rm s}$ (see text) with time, for different initial conditions, and for $N=16$. $(a)$ and $(b)$~: $G^{\rm c}=-G^{\rm s}=0.1$ $(b)$ is a zoom on small times of $(a)$. At $s=0$, $\beta^{\rm s}=\infty$. This is peculiar to the initial condition for which $G^{\rm c}=-G^{\rm s}$. In $(c)$ we show what happens for $G^{\rm c}=-G^{\rm s}/2=0.1$ at $s=0$. Finally in $(d)$ we show the flow for initial condition $G^{\rm c}=G^{\rm s}=0.1$}
\label{fig:coop_sur_pei}
\end{figure}

To understand better the pattern of the $A(I)$'s, $C(I,J)$'s and $D(I,J)$'s couplings, we notice that for most values of $I$ and $J$, the sign of these couplings, in both charge and spin sector are well defined, and that the magnitudes do not vary strongly in a given sector, as shown on Figs.~\ref{fig:flot_debut_c} and \ref{fig:flot_debut_s}. So in the sequel we shall consider fixed directions with $A$, $C$ and $D$ couplings chosen not to depend on $I$ and $J$. As will be shown in Sec.~\ref{sec:fixeddir} below, this leads us to situations where $A^{{\rm c},{\rm s}}=C^{{\rm c},{\rm s}}+D^{{\rm c},{\rm s}}$. If such a relation holds, the effective Hamiltonian exhibits a large SU$(N)$ symmetry in the space of ``patches''. This phenomenon of enlarged symmetries is rather common in one-dimensional quantum field theories. A striking example is the appearance of a SO$(8)$ symmetry in a two channel model which may be relevant to the description of low-energy properties of metallic carbon nanotubes.\cite{Lin98} Most of the sequel will be dedicated to these special points for which a large amount of information may be obtained by analytical calculations, at least in the large $N$ limit. In Fig.~\ref{fig:histo}, we compare the actual distribution of dominant couplings of $C$ and $D$ type (both for charge and spin sectors) in the intermediate regime (for $s=100$) with the expected values under the assumption of SU$(N)$ symmetry.
Clearly the actual flow doesn't reach exactly an SU$(N)$ symmetrical fixed direction, but the histograms of $C$ and $D$ couplings are well centered around the predicted value for SU$(N)$ with $N=16$. As we shall see, the Luttinger type couplings $C^{{\rm c},{\rm s}}$ are of order $1/N$ whereas $A^{{\rm c},{\rm s}}$ and $D^{{\rm c},{\rm s}}$ are of order unity if $N$ is large. This remark suggests another way to analyze the intermediate fixed direction. If the $C(I,J)$ are assumed to vanish, the $D(I,J)$ couplings are renormalized in the Peierls channel only. So an RPA type approximation should give a rather good picture for the vicinity of these fixed directions. To test this idea, we present in Fig.~\ref{fig:coop_sur_pei} a comparison between the average contributions of the Cooper and the Peierls channels to the flow equations, as a function of the new ``time'' variable $s$. 
More precisely, we define for each channel (charge c or spin s) a ratio $\beta$ measuring the relative importance between Cooper and Peierls channels as follows. We first rewrite the flow equations in the charge (spin) channel, distinguishing between the Peierls and Cooper contributions, in the form~: $\partial_t g_i^{\rm c(s)} = A_{ijk}^{\rm C, c(s)}g_j g_k - A_{ijk}^{\rm P, c(s)}g_j g_k$. With obvious notations, this is cast in a vectorial form as~: $\partial\boldsymbol{g}^{\rm c(s)}=\boldsymbol{A}^{\rm C, c(s)}(\boldsymbol{g})-\boldsymbol{A}^{\rm P, c(s)}(\boldsymbol{g})$. Finally in each channel we define $\beta^{\rm c(s)}=\|\boldsymbol{A}^{\rm C, c(s)}\| / \|\boldsymbol{A}^{\rm P, c(s)}\|$, where $\|.\|$ is the Euclidean norm. Clearly, the Peierls channel dominates the flow as soon as $s\gtrsim 50$ and for $s\lesssim 1200$ (see Fig.~\ref{fig:coop_sur_pei}$(b)$). One can numerically check that in the intermediate regime, the values of the $\beta$ ratios scale as $1/N$ when one varies $N$. In this situation, neglecting the Cooper terms yields a smooth manifold of possible fixed directions in the space of $D(I,J)$ couplings, as will be shown below in Sec.~\ref{sec:RPA}. This manifold contains the special SU$(N)$ symmetry points discussed above as special isolated points. So, the RPA picture shows that the SU$(N)$ symmetry doesn't really emerge (see again Fig.~\ref{fig:histo}). But these special points are nevertheless very interesting, since the actual flows get close to them in the intermediate regime, and their underlying symmetry is a source of tremendous analytical simplification.

\begin{figure}[h]
\includegraphics[width=8cm]{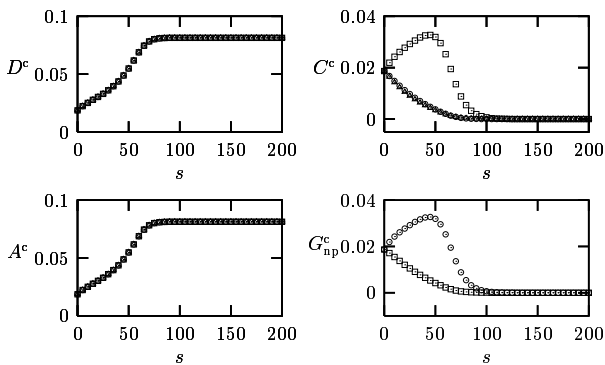}
\caption{Same as Fig.~\ref{fig:flot_debut_c} in the RPA. We have not represented the flow up to $s=1600$ as in Fig.~\ref{fig:flot_glob_c} because the couplings do not evolve any more. It is remarkable that {\em all} $D$ and $A$ couplings remain equal.}
\label{fig:flot_debut_RPA_c}
\end{figure}

\begin{figure}[h]
\includegraphics[width=8cm]{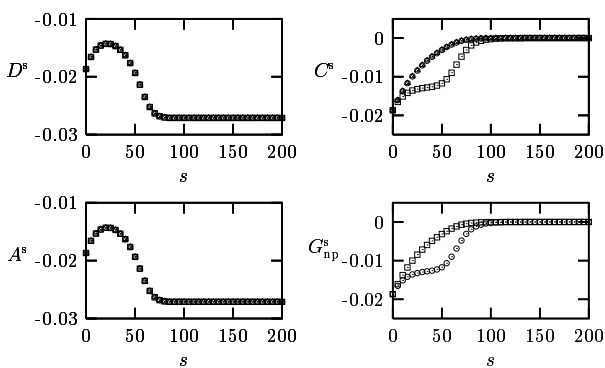}
\caption{Same as Fig.~\ref{fig:flot_debut_s} in the RPA.}
\label{fig:flot_debut_RPA_s}
\end{figure}

Before going further, we would like to stress the following important point, actually already made by Zheleznyak et al.\cite{Zheleznyak97} Although the intermediate regime is very well described by a RPA picture, it is quantitatively wrong to neglect the Cooper channel in the early stages of the flow, as indicated by Fig.~\ref{fig:coop_sur_pei}. This would give a higher value of $\Lambda_{\rm c}$ than obtained from the complete ``parquet'' solution of the one-loop equations. But also, as seen when comparing Figs.~\ref{fig:flot_debut_RPA_c} and \ref{fig:flot_debut_RPA_s} with Figs.~\ref{fig:flot_debut_c} and \ref{fig:flot_debut_s}, the initial transient has to be accurately treated in order to determine the actual fixed direction chosen by the system in its RPA like intermediate regime.
The interference between the Cooper and Peierls channels is indeed responsible for the observed splitting among the values of the dominant charge and spin couplings in this regime. A simple RPA (i.e. no Cooper terms at all) starting from $k$-independent couplings preserves this constant character of $A$ and $D$ couplings along the flow, by contrast to what happens in the ``real'' one-loop flow.

To summarize, we shall follow two complementary approaches. In Sec.~\ref{sec:RPA} we shall first set up a perturbation analysis around the RPA in the Peierls channel, the Cooper channel being considered as a small perturbation. In this picture, the RPA exhibits a continuous manifold of fixed directions, which may be destabilized by the weak Cooper terms, but this final stage involves very long RG time scales. Alternatively, we may say that the system in the intermediate regime reaches a neighborhood of a fixed direction (for the full RG flow this time) with SU$(N)$ symmetry. This direction admits a set of $(N-1)$ relevant perturbations which become marginal in the infinite $N$ limit. This will be the object of Secs.~\ref{sec:high_sym_FD} and \ref{sec:stability}. We shall first of all set up our notations in the 1D case and study in detail its set of fixed points in the usual sense.

\section{ONE-LOOP RG FOR $N$ COUPLED CHAINS}
\label{sec:1loopRGforNcoupch}
\subsection{GENERAL SETTING AND RG EQUATIONS}
\label{sec:gen_setting}
We have seen that pairs of nested flat regions on the Fermi surface produce a complex dependence of the electron interaction as a function of the typical energy scale. These phenomena arise from the mutual couplings between the particle-particle and particle-hole channels. These main features are already present in one-dimensional systems, for which RG equations are a little simpler. Furthermore, the previous discussion of our numerical results suggests that many couplings become negligible in comparison to a much smaller set of dominant couplings, as the system flows towards lower energies. We have seen that often these dominant couplings correspond at least approximately to a very large SU($N$) symmetry of the effective interaction, at some intermediate stage along the flow. Keeping only these couplings brings dramatic simplification, especially in the one-dimensional case of $N$ coupled chains, where analytic approaches become available. 

\begin{figure}[h]
\includegraphics[width=8cm]{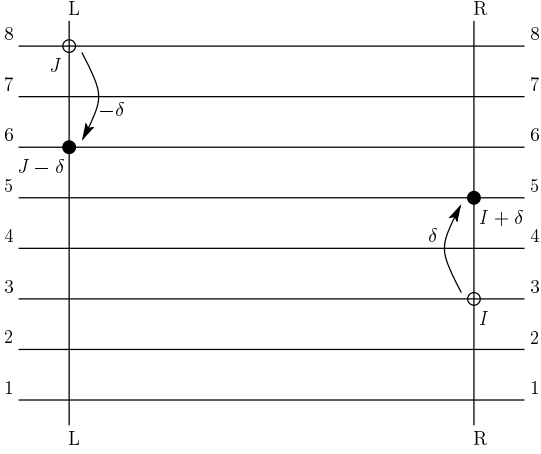}
\caption{Labelling of the patches in the 1D case for $N=8$, according to the interpretation in terms of chains. Arrows illustrate the $G_\delta(I,J)=G_2(3,8)$ coupling (see text).}
\label{fig:param_1D}
\end{figure}

To be specific, we shall concentrate on systems of $N$ chains without transverse hopping (to generate a flat Fermi surface) and coupled by two-particle interactions between right and left movers, with a small transferred momentum along the chain directions. Each chain may be seen as a device to describe the electronic states in the vicinity of a point chosen on a flat section of the Fermi surface. In what follows, the chain direction is understood to be perpendicular to the Fermi surface. The kinetic energy is taken to be $H_0$, with~:
\begin{eqnarray}
\label{eq:H_0}
H_0=v_{\rm F} \sum_{k,I,\sigma} \Bigl\lbrace  (k-k_{\rm F}) c^\dagger_{{\rm R},I,\sigma} (k) c_{{\rm R},I,\sigma} (k) \nonumber\\
- (k+k_{\rm F}) c^\dagger_{{\rm L},I,\sigma} (k) c_{{\rm L},I,\sigma} (k) \Bigl\rbrace \;.
\end{eqnarray}
Here $v_{\rm F}$ and $k_{\rm F}$ stand for the Fermi velocity and wave-vector. The operators $c^\dagger_{{\rm R}({\rm L}),I,\sigma} (k)$ and $c_{{\rm R}({\rm L}),I,\sigma} (k)$ are usual creation and destruction operators for right (left)  moving spin $1/2$ fermions. $k$ is the momentum along the chains, $I$ is a ``color'' index which refers to a given patch on the Fermi surface (according to the previous section) and $\sigma$ is the $z$-component of the spin. The sum over $k$ runs for $k=2\pi m/L$, $m$ being any integer, and $L$ denoting the length of the chains. We interpret $I$ as the momentum in the transverse direction of the chains. It is chosen in the set $\lbrace 1,\ldots,N \rbrace$ (see Fig.~\ref{fig:param_1D}). The most general spin rotation invariant two-particle interaction Hamiltonian, involving scattering events between right and left movers may be written as~:
\begin{eqnarray}
\label{eq:H_int}
H_{\rm int}=\frac{2\pi v_{\rm F}}{L N} \sum_{I,J,\delta,q\neq 0} \Bigl\lbrace  {\tilde{G}}_\delta^{\rm c}(I,J) \rho_{{\rm R},I,\delta}(q) \rho_{{\rm L},J,-\delta}(-q)\nonumber\\
+ {\tilde{G}}_\delta^{\rm s}(I,J) \boldsymbol{S}_{{\rm R},I,\delta}(q) \boldsymbol{S}_{{\rm L},J,-\delta}(-q)\Bigl\rbrace \;.\hspace{0.4cm}
\end{eqnarray}
In the summation, both $I$ and $J$ belong to $\lbrace 1,\ldots,N \rbrace $ and $\delta$ is chosen in such a way that both $I+\delta$ and $J-\delta$ belong to this set of patches (see the illustration on Fig.~\ref{fig:param_1D}). Note that this way of writing the two-particle interaction enforces momentum conservation, in both perpendicular ($q$ variable) and parallel ($\delta$ variable) directions with respect to the Fermi surface. The restrictions on $I$, $J$, $I+\delta$ and $J-\delta$ imply that we are ruling out Umklapp processes along the Fermi surface. Therefore this model is not relevant for quasi one-dimensional conductors which also exhibit disconnected Fermi surfaces with nearly flat curves running all across the first Brillouin zone. Again, our chains are introduced to describe processes in the vicinity of the flat regions of a connected Fermi surface. This is the reason we introduce this kind of open boundary conditions at the end points of the flat regions, which correspond to chains labelled by 1 or $N$. The quantities $\rho_{{\rm R},I,\delta}(q)$ and $\boldsymbol{S}_{{\rm R},I,\delta}(q)$ are generalized charge and spin right currents (and similarly for the L side) defined by~:
\begin{eqnarray}
\label{eq:currents1}
\rho_{{\rm R},I,\delta}(q)&=&\sum_{k,\tau} c^\dagger_{{\rm R},I+\delta,\tau} (k+q) c_{{\rm R},I,\tau} (k)\;,\\
\label{eq:currents2}
S^a_{{\rm R},I,\delta}(q)&=&\sum_{k,\tau,\tau'} c^\dagger_{{\rm R},I+\delta,\tau} (k+q) \sigma^a_{\tau \tau'} c_{{\rm R},I,\tau'} (k)\;,
\end{eqnarray}
where $a=x,y,z$ and $\sigma^a$ are the usual Pauli matrices. The functions ${\tilde{G}}_\delta^{\rm c}(I,J)$ and ${\tilde{G}}_\delta^{\rm s}(I,J)$ will be referred to as charge and spin couplings. With our normalization of $H_{\rm int}$, these couplings are dimensionless. In this Hamiltonian, we have neglected processes which involve all incoming and outgoing particles on the same branch since they don't appear in the flow of the couplings at the one-loop level. They commute with $H_0$, so their main effect is to lift some degeneracies between various excited states of $H_0$ which often amounts to changing Fermi velocities. The renormalization of these velocities cannot be addressed within a one-loop approximation. 
\begin{figure}[h]
\includegraphics[width=8cm]{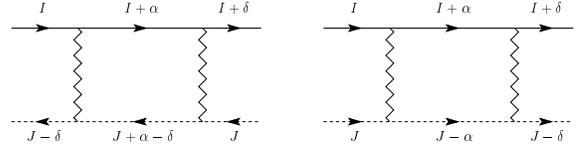}
\caption{The particle-hole and particle-particle graphs that contribute to the 1D RG equations. We have not indicated any spin indices.}
\label{fig:feynman}
\end{figure}
So for simplicity, we assume all our Fermi velocities to be equal and constant along the RG flow. We are also discarding Umklapp processes along the chains, which play a dramatic role only for a discrete set of band fillings.

The total Hamiltonian $H=H_0+H_{\rm int}$ describes a renormalizable field theory. The one-loop flow equations are obtained by simply computing the two graphs shown on Fig.~\ref{fig:feynman}, and we get~:
\begin{widetext}
\begin{equation}
\label{eq:eqRGchaines}
\left\lbrace
\begin{array}{r}
\partial_t G_\delta^{\rm c}(I,J)=\frac{1}{N}\sum_\alpha 
\Bigl\lbrace G_\alpha^{\rm c}(I,J+\alpha-\delta) G_{\delta-\alpha}^{\rm c}(I+\alpha,J)+3 G_\alpha^{\rm s}(I,J+\alpha-\delta) G_{\delta-\alpha}^{\rm s}(I+\alpha,J)\\
- G_\alpha^{\rm c}(I,J) G_{\delta-\alpha}^{\rm c}(I+\alpha,J-\alpha)-3 G_\alpha^{\rm s}(I,J) G_{\delta-\alpha}^{\rm s}(I+\alpha,J-\alpha)\Bigl\rbrace\vspace{0.3cm}\\
\partial_t G_\delta^{\rm s}(I,J)=\frac{1}{N}\sum_\alpha \Bigl\lbrace 2 G_\alpha^{\rm s}(I,J+\alpha-\delta) G_{\delta-\alpha}^{\rm s}(I+\alpha,J) + G_\alpha^{\rm s}(I,J+\alpha-\delta) G_{\delta-\alpha}^{\rm c}(I+\alpha,J)\\
+ G_\alpha^{\rm c}(I,J+\alpha-\delta) G_{\delta-\alpha}^{\rm s}(I+\alpha,J) +2 G_\alpha^{\rm s}(I,J) G_{\delta-\alpha}^{\rm s}(I+\alpha,J-\alpha)\\ 
- G_\alpha^{\rm s}(I,J) G_{\delta-\alpha}^{\rm c}(I+\alpha,J-\alpha) - G_\alpha^{\rm c}(I,J) G_{\delta-\alpha}^{\rm s}(I+\alpha,J-\alpha)\Bigl\rbrace
\end{array}\right..
\end{equation}
\end{widetext}
We have set $G=2\pi v_{\rm F} \tilde{G}$ in order to get rid of factors $2\pi v_{\rm F}$. As before the $\partial_t$ symbol means the derivative with respect to $t=\ln (\Lambda_0/\nu)$, where $\Lambda_0$ is a ``bare'' (i.e. microscopic) cut-off, and $\nu \leq \Lambda_0$ is the typical energy scale of the process of interaction. In the sequel, we shall also analyze the simpler case of spinless fermions, which is simply obtained by taking $G_\delta^{\rm s}(I,J)=0$ in Eq.~(\ref{eq:eqRGchaines}). In this case, $G_\delta^{\rm c}(I,J)$ will simply be written as $G_\delta(I,J)$, for which the one-loop flow reads~:
\begin{eqnarray}
\label{eq:eqRGchaines_spinless}
\partial_t G_\delta(I,J)=\frac{1}{N}\sum_\alpha 
\Bigl\lbrace G_\alpha(I,J+\alpha-\delta) G_{\delta-\alpha}(I+\alpha,J) \nonumber\\
- G_\alpha(I,J) G_{\delta-\alpha}(I+\alpha,J-\alpha) \Bigl\rbrace\;.\hspace{0.6cm}
\end{eqnarray}
Note that the dominant couplings in the intermediate regime introduced in Sec.~\ref{subsec:numres2D} correspond in this language to : $A(I)=G_0(I,J)$, $C(I,J)=G_0(I,J)$ and $D(I,J)=G_{J-I}(I,J)$. It is easy to check they form a closed set under Eq.~(\ref{eq:eqRGchaines}).

\subsection{FIXED POINT SOLUTIONS AND THEIR STABILITY}
\subsubsection{``$k$-space'' Luttinger liquids}

As usual, it is convenient to begin the study of these flows by searching for fixed points (i.e. true fixed points, not fixed directions). This discussion is logically independent of the main line of this work, but we include it here, since it is a rather simple and natural question to consider. Furthermore it illustrates, in a different manner, some difficulties which arise in taking the large $N$ limit (see below). In general, we have found a continuous family of fixed points which may be seen as generalized Luttinger liquids. They are characterized by the property that all couplings for which $\delta\neq 0$ vanish. We remind that $\delta$ denotes the transferred momentum {\em along} the Fermi surface. Furthermore, the spin couplings ${G_0^{\rm s}}^*(I,J)$ for $\delta=0$ all vanish (the star characterizes the fixed points). So these fixed points are parametrized by a two-parameter family of couplings ${G_0^{\rm c}}^*(I,J)$ (still for $\delta=0$). Physically, they are Luttinger liquids in ``$k$-space'', in close resemblance with P.W. Anderson's idea of what he called a ``tomographic'' Luttinger liquid.\cite{Anderson91} The corresponding fixed point Hamiltonian may be cast as~:
\begin{eqnarray}
\label{eq:hamLLG}
H=\frac{2 \pi v_{\rm F}}{L}\sum_{I,\sigma,q>0} \Bigl\lbrace \rho_{{\rm R},I}(q)\rho_{{\rm R},I}(-q)+\rho_{{\rm L},I}(-q)\rho_{{\rm L},I}(q) \Bigl\rbrace\nonumber\\
+ \frac{1}{NL}\sum_{I,J,q\neq 0} {G_0^{\rm c}}^*(I,J) \rho_{{\rm R},I}(q)\rho_{{\rm L},J}(-q)\;.\hspace{0.8cm}
\end{eqnarray}
Here, we have used the notation $\rho_{{\rm R},I}(q)$ for $\rho_{{\rm R},I,\delta=0}(q)$ and similarly for left movers. Since it is quadratic in density modes, it is easily diagonalized, yielding a decoupled set of $N$ branches of non-interacting sound-like bosonic modes. Usual bosonization techniques may be used to obtain exact low-energy asymptotic expressions for the corresponding fermion propagator and various correlation functions. But we shall not develop this here.

Let us now investigate the linear stability of these generalized Luttinger liquid fixed points. To do this, we parametrize the set of couplings as~:
\begin{equation}
\left\lbrace
\begin{array}{l}
G_0^{\rm c}(I,J)={G_0^{\rm c}}^*(I,J)+g_0^{\rm c}(I,J)\\
G_\delta^{\rm c}(I,J)=g_\delta^{\rm c}(I,J), \mbox{ for } \delta\neq 0\\
G_\delta^{\rm s}(I,J)=g_\delta^{\rm s}(I,J), \mbox{ for any }\delta
\end{array}\right..
\end{equation}
Here $g_\delta^{{\rm c},{\rm s}}(I,J)$ are assumed to be a small perturbation. Taking the flow equations and keeping only the linear terms in $g$ variables gives~:
\begin{equation}
\label{eq:flotlinLLG}
\left\lbrace
\begin{array}{l}
\partial_t g_0^{\rm c}(I,J)= O(g^2)\\
\partial_t g_0^{\rm s}(I,J)= O(g^2)\\
\partial_t g_\delta^{\rm c}(I,J)=-\frac{1}{N} B^*(I,J,\delta) g_\delta^{\rm c}(I,J)\\
\partial_t g_\delta^{\rm s}(I,J)=-\frac{1}{N} B^*(I,J,\delta) g_\delta^{\rm s}(I,J)
\end{array}\right.,
\end{equation}
where~:
\begin{eqnarray}
B^*(I,J,\delta)={G_0^{\rm c}}^*(I,J)-{G_0^{\rm c}}^*(I,J-\delta)\hspace{1.4cm}\nonumber\\
-{G_0^{\rm c}}^*(I+\delta,J)+{G_0^{\rm c}}^*(I+\delta,J-\delta)\;.
\end{eqnarray}
So in order to be stable with respect to small perturbations, these fixed points are required to satisfy the following inequalities~: 
\begin{equation}
\label{eq:stabLLG}
B^*(I,J,\delta) >0 \mbox{ for any } \delta\;.
\end{equation}
It is easy to show that this condition is satisfied for any $\delta$ provided it is satisfied for $\delta=1$. Note that the perturbations at zero momentum transfer $g_0^{{\rm c},{\rm s}}(I,J)$ remain marginal around these Luttinger fixed points. If we examine their flow in more detail, we find on the one hand that $\partial_t g_0^{\rm c}(I,J)$ involves only irrelevant couplings $g_\alpha^{\rm c}(P,Q)$ with $\alpha\neq 0$. Therefore $\partial_t g_0^{\rm c}(I,J)$ is expected to be small, provided the bare couplings satisfy the stability condition stated above. On the other hand, we obtain~: $\partial_t g_0^{\rm s}(I,J)=(4/N) {g_0^{\rm s}(I,J)}^2$ + irrelevant terms. Assuming the irrelevant terms vanish as the energy cut-off goes to zero, we see that the flow remains in the attraction domain of the generalized Luttinger liquid provided all the spin couplings remain negative. Otherwise, some of these couplings are expected to become marginally relevant, with the opening of a spin gap as a likely consequence.

The main result of this discussion is the possibility (for any finite $N$) to stabilize generalized Luttinger liquids in ``$k$-space'', characterized by processes involving momentum transfers exclusively perpendicular to the Fermi surface. These fixed points are parametrized by the interaction function ${G_0^{\rm c}}^*(I,J)$ which plays the role of the Landau parameters for the Fermi liquid fixed point. Stability is ensured by the not too stringent Eq.~(\ref{eq:stabLLG}), restricted to the case $\delta=1$. The main difficulty is that our results are valid for a finite number $N$ of Fermi points. For a realistic system, we would like to take the $N\to\infty$ limit. But this is tricky, as indicated by the $1/N$ factor in the RHS of the linearized RG equations (\ref{eq:flotlinLLG}). In this limit, the $\delta\neq 0$ couplings become marginal instead of being irrelevant as in the finite $N$ case. In the infinite $N$ limit, ${G_\delta}^*(I,J)$ is replaced by a function ${G_d}^*(u,v)$ of three continuous variables, where $u$ and $v$ denote the transverse momenta of the incoming particles, and $d$ the transverse component (i.e. parallel to the flat Fermi surface) of the momentum transfer. The RG equations take the form~:
\begin{eqnarray}
\partial_t G_d^{\rm c}(u,v)=\int{\rm d}\alpha \Bigl\lbrace G_\alpha^{\rm c}(u,v+\alpha-d)G_{d-\alpha}^{\rm c}(u+\alpha,v)\nonumber\\
-G_\alpha^{\rm c}(u,v)G_{d-\alpha}^{\rm c}(u+\alpha,v-\alpha)\Bigl\rbrace\;.\hspace{0.4cm}
\end{eqnarray}
To simplify this discussion we assume the spin couplings are vanishing. The tomographic Luttinger liquids discussed above correspond to ${G_d^{\rm c}}^*(u,v)=\delta(d){G}^*(u,v)$, and the stability condition becomes~:
\begin{equation}
\frac{{\partial}^2 G^*}{\partial u \partial v}(u,v)<0\;.
\end{equation}
But clearly these are singular couplings. The main question which has to be addressed is the following~: suppose the bare (microscopic) couplings are smooth functions of the three variables $d$, $u$ and $v$~; is it possible to build up a $\delta(d)$ singularity in a finite RG time ? Although we can't make any rigourous statement, our experience with numerical solutions of RG flows, attempting to increase $N$ as far as $N=20$, indicates that these singularities do not appear, unless they are already present in the bare couplings. So we are tempted to believe that tomographic Luttinger liquids do not appear as fixed points for systems with smooth microscopic interactions in $k$-space, that is with short range interactions in real space. This example shows clearly that a lot of caution is required in extrapolating numerical results obtained for a finite $N$ to the continuous Fermi surface relevant to real life situations.

\subsubsection{Comparison with ``real space'' Luttinger liquids}

We would like for completeness to make contact with the physics of strongly anisotropic conductors. In the previous discussion, chains are seen as a mathematical device to discretize a continuous Fermi surface of a two-dimensional conductor along its flat regions. Let us now consider a one-dimensional system made of a series of $N$ parallel identical chains, and let us assume periodic boundary conditions in the perpendicular direction with respect to the chains. To simplify, we assume there is no single-particle hopping from one chain to the next. As a result, we obtain two flat Fermi surface segments along the $\hat{y}$-axis (if $\hat{x}$ is the chain direction) running along the whole Brillouin zone, from $k_y=-\pi$ to $k_y=\pi$ (we choose the unit length to be the separation between two neighboring chains). We consider the following interaction~:
\begin{eqnarray}
H_{\rm int} = \int_0^L {\rm d}x \sum_{m,m',n,n'} F(m',m;n',n) \nonumber\\
\times :\psi^\dagger_{{\rm R},m'}(x)\psi_{{\rm R},m}(x)\psi^\dagger_{{\rm L},n'}(x)\psi_{{\rm L},n}(x):\;,
\end{eqnarray}
where $:{\mathcal O}:$ denotes normal ordering of opertator ${\mathcal O}$. $\psi_{{\rm R},m}(x)$ is a fermion destruction operator at point $x$, on chain labelled by the integer $m$. To simplify this discussion, we shall omit the spin degree of freedom. The non-interacting part we choose to consider is as usual~:
\begin{eqnarray}
H_0 = v_{\rm F} \int_0^L {\rm d}x \sum_m \Bigl\lbrace : \psi^\dagger_{{\rm R},m}(x) \frac{1}{i}\frac{\partial}{\partial x} \psi_{{\rm R},m}(x) : \nonumber\\
- : \psi^\dagger_{{\rm L},m}(x) \frac{1}{i}\frac{\partial}{\partial x} \psi_{{\rm L},m}(x) :\Bigl\rbrace\;.
\end{eqnarray}
Translation invariance along the $\hat{y}$ direction requires $F(m'+s,m+s;n'+s,n+s)=F(m',m;n',n)$. It is easy to write down RG equations for the $F$ function to one-loop order, and we get~:
\begin{widetext}
\begin{equation}
\partial_t F(m',m;n',n)=\sum_{m'',n''} \Bigl\lbrace F(m',m'';n'',n) F(m'',m;n',n'') - F(m',m'';n',n'') F(m'',m;n'',n)\Bigl\rbrace\;.
\end{equation}
\end{widetext}
A generalized Luttinger fixed point in real space is obtained for $F(m',m;n',n)=\delta_{m',m}\delta_{n',n} F(m-n)$. This implies that the total particle number in each chain is conserved. In the vicinity of such a generalized Luttinger fixed point, we may write~: $F(m',m;n',n)=\delta_{m',m}\delta_{n',n} F(m-n)+f(m',m;n',n)$. Here $f(m',m;n',n)$ is assumed to be small in comparison to $F(m-n)$. Linearizing the flow equations gives~:
\begin{equation}
\label{eq:flot_lin_pour_f}
\partial_t f(m',m;n',n)=-B(m',m;n',n) f(m',m;n',n)\;,
\end{equation}
with $B(m',m;n',n)=F(m'-n') + F(m-n)- F(m'-n) - F(m-n')$.
The Luttinger fixed point characterized by the function $F$ is locally stable if for any four-uple $(m',m,n',n)$ we have 
\begin{equation}
\label{eq:stabLLG_RS}
B(m',m;n',n)>0\;.
\end{equation}
Taking $m'=m+1$ and $n'=n+1$ yields~: $2 F(m-n)-F(m-n+1)-F(m-n-1)>0$. But if $m'=m+1$ and $n'=n-1$, the same condition gives~: $2 F(m-n+1)-F(m-n+2)-F(m-n)<0$. So it is clearly impossible to find a function $F$ such that all the inequalities of Eq.~(\ref{eq:stabLLG_RS}) would be satisfied. In more physical terms, real space Luttinger liquids are always destabilized at low energies by interaction processes which don't conserve the total particle number in each chain. At this point, we should emphasize that our analysis is strictly speaking valid only in the vicinity of the free electron fixed point, i.e. when $F(m-n)$ is always small. Otherwise, the generalized Luttinger liquid would have to be treated non-perturbatively (using bosonization), which modifies Eq.~(\ref{eq:flot_lin_pour_f}) for the anomalous dimension of perturbing operators. Within this non-perturbative analysis, some recent work\cite{Vish01} by Vishwanath and Carpentier suggests the possibility of stabilizing a real space Luttinger liquid. But this requires very special properties on the $F$ function. The main difference between real space and $k$-space Luttinger liquids comes from the number of free parameters which are required to describe the fixed point interaction. In the first case, the $F$ function depends on one variable, whereas in the second case, $G^*$ involves two variables, which leaves more possibilities to achieve local stability of the ``tomographic'' fixed points (at finite $N$).

If we have $N$ coupled chains with periodic boundary conditions in the transverse direction, we can go from the real space to the $k$-space description by Fourier transform, according to~: 
\begin{equation}
c^\dagger_{{\rm R},J}(k)=\frac{1}{\sqrt{L N}}\sum_{m=0}^{N-1} \int_0^L {\rm d}x\, e^{i\left(kx + \frac{2\pi}{N}Jm\right)} \psi^\dagger_{{\rm R},m}(x)\;.
\end{equation}
The real space Luttinger fixed point with the $F$ function yields the following interaction Hamiltonian, written in $k$-space~:
\begin{equation}
\label{eq:HintLLG}
H_{\rm int}= \frac{1}{L N} \sum_{I,J,\delta,q\neq 0} \tilde{F}(\delta) \rho_{{\rm R},I,\delta}(q) \rho_{{\rm L},J,-\delta}(-q)\;.
\end{equation}
So, according to our general notation, we get $G_\delta(I,J)=\tilde{F}(\delta)=\sum_{m=0}^{N-1} \exp(-i 2\pi\delta m/N) F(m)$, which depends only on the transferred momentum $\delta$ along the flat Fermi surface. In Eq.~(\ref{eq:HintLLG}), we drop the restriction that $I+\delta$ and $J-\delta$ should belong to the set $\lbrace1,\ldots,N\rbrace$, since we have the natural periodicity in $k$-space~: $c^\dagger_{{\rm R},J}(k)=c^\dagger_{{\rm R},J+N}(k)$.

To conclude both discussions, we would like to stress that each form of Luttinger liquid is very unlikely to appear as the true low-energy fixed point of a real two-dimensional system. For an anisotropic metal with flat regions on the Fermi surface, the main difficulty arises in considering the limit where the number of patches becomes infinite, in the case of non-singular microscopic interactions. For quasi one-dimensional conductors, we have shown that there are always some unstable directions around the corresponding Luttinger liquid Hamiltonian, even if we do not consider inter-chain hopping. In the remaining part of this article, we shall therefore consider only strong coupling regimes, but still within the perturbative RG approach.

\section{STRONG COUPLING REGIME~: THE RPA}
\label{sec:RPA}

In this section we shall focus on the last two stages of a typical RG flow as shown on Figs.~\ref{fig:flot_glob_c} and \ref{fig:flot_glob_s}. The first task is to understand the plateau observed for intermediate RG times. The comparison between relative contributions to the flow of Peierls and Cooper channels (see Fig.~\ref{fig:coop_sur_pei}) clearly shows that the former is dominant in this regime, with subleading corrections of order $1/N$ (see also Figs.~\ref{fig:flot_debut_RPA_c},\ref{fig:flot_debut_RPA_s} and Table \ref{tab:dirfixeN}). 

\subsection{RPA FIXED DIRECTIONS AND THEIR STABILITY}
\begin{figure}[h]
\includegraphics[width=8cm]{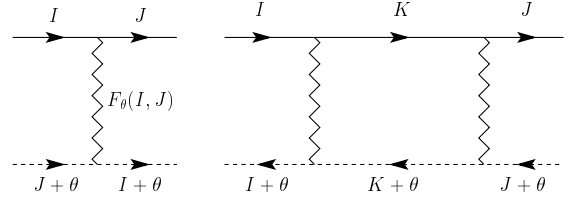}
\caption{On the left, we show the interaction process with a notation adapted to the RPA. On the right, we give the second order contribution in the Peierls channel which involves $G_{K-I}(I,K+\theta) G_{J-K}(K,J+\theta)$, and which is rewritten as $F_\theta(I,K) F_\theta(K,J)$}
\label{fig:couplageF}
\end{figure}

It is first useful to present a complete description of the set of fixed directions within the RPA. In a second step, a local stability analysis around these directions will be sketched. The RPA considered here consists simply in dropping the contributions from the Cooper channel in the RHS of the RG flow Eq.~(\ref{eq:eqRGchaines}). As illustrated on Fig.~\ref{fig:couplageF}, the notation is greatly simplified if we replace expressions like $G_{J-I}(I,J+\theta)$ by quantities of the form $F_\theta(I,J)$, since $\theta$ is the component of the transferred momentum along the flat Fermi surface, and is therefore conserved in the elementary particle-hole bubble. 
In the same spirit, we decompose the possible particle-hole states into singlet and triplet subspaces, which corresponding couplings are renormalized independently. More specifically, we set~:
\begin{eqnarray}
\left\lbrace
\begin{array}{l}
F_\theta^{\rm sing} (I,J)=G_{J-I}^{\rm c} (I,J+\theta) + 3 G_{J-I}^{\rm s}(I,J+\theta)\\ 
F_\theta^{\rm trip} (I,J)=G_{J-I}^{\rm c} (I,J+\theta) - G_{J-I}^{\rm s}(I,J+\theta)
\end{array}\right..\hspace{0.6cm}
\end{eqnarray}
With these definitions the RPA approximation reads~:
\begin{equation}
\partial_t F_\theta ^a (I,J) = \sum_{
\shortstack{ \scriptsize $1\leq K \leq N$\\ 
\scriptsize $1\leq K+\theta \leq N$}} 
F_\theta ^a (I,K)F_\theta ^a (K,J)\;.
\end{equation}
Here $a$ denotes either the singlet or triplet channel. This has the generic form $\partial_t M=M^2$, where $M$ is a symmetrical matrix of size $N_\theta=N-\theta$, with real coefficients. It is straighforward to solve these simplified flow equations for any initial condition $M(t=0)=M_0$. There exists an orthogonal matrix $P$ and a diagonal matrix $\Delta_0$ such that $M_0=P \Delta_0 P^{-1}$. For any positive time, $M(t)=P \Delta(t) P^{-1}$, with $[\Delta(t)]_{i i}= [\Delta_0]_{i i}/(1-t[\Delta_0]_{i i})$. In this situation, if some initial couplings $[\Delta_0]_{i i}$ are positive, the corresponding diagonal element diverges at time $t_i=[\Delta_0]_{i i}^{-1}$. Let us consider the set ${\mathcal I}$ of $i$'s such that $[\Delta_0]_{i i}$ reaches its maximal positive value. For any $i_1,i_2\in {\mathcal I}$,  $[\Delta(t)]_{i_1 i_1}= [\Delta(t)]_{i_2 i_2}$. And it is clear that the ratio $[\Delta(t)]_{j j}/[\Delta(t)]_{i i}$ vanishes in the limit where $t$ approaches $t_i$, provided $j\notin {\mathcal I}$ and $i\in {\mathcal I}$. This shows that the flow in $M$-space is appoaching a fixed direction along $M^*$ given by $M^*=P \mathbb{I}_{\mathcal I} P^{-1}$. The matrix $\mathbb{I}_{\mathcal I}$ is the diagonal matrix such that $[\mathbb{I}_{\mathcal I}]_{i i}=1$ if $i\in {\mathcal I}$, and $[\mathbb{I}_{\mathcal I}]_{i i}=0$ if $i\notin {\mathcal I}$. Let us denote by $m$ the number of elements of ${\mathcal I}$ (we are assuming here that ${\mathcal I}$ is not empty, so $m\geq 1$~; of course $m\leq N_\theta)$. The fixed direction along $M^*$ can be interpreted as an orthogonal projector on a $m$ dimensional subspace of the $N_\theta$ dimensional space on which $M$ matrices operate. In this way, for a given $m$ satisfying $1\leq m\leq N_\theta$, we obtain a smooth manifold of fixed directions, which may be identified with the smooth manifold of $m$ dimensional subspaces imbedded in a $N_\theta$ dimensional vector space. This object is usually called a Grassmann manifold ($G_m(\mathbb{R}^{N_\theta})$ being a standard notation) and its dimension is $m(N_\theta-m)$. For a given fixed direction $M^*$ such that its largest positive eigenvalue is $m$ times degenerate, it is simple to characterize its attraction basin along the RPA coupling flow. Using a permutation of the columns of $P$, it is always possible to assume ${\mathcal I}=\lbrace 1,\ldots,m \rbrace$. In the basis of $\mathbb{R}^{N_\theta}$ defined by the columns of $P$, the set of matrices which flow at large enough times towards the direction defined by $M^*$ is given by~:
\begin{equation}
M=\left(
\begin{array}{cc}
\mathbb{I}_m & 0\\
0 & \tilde{M}\\
\end{array}
\right)\;,
\end{equation}
where $\mathbb{I}_m$ is the $m\times m$ unit matrix, and $\tilde{M}$ is any symmetrical real matrix of size $N_\theta-m$ whose eigenvalues are all strictly smaller than one. In this basis, $M^*$ corresponds to $\tilde{M}=0$. So this attraction basin is an open set of matrices which is also a $(N_\theta-m)(N_\theta-m+1)/2$ dimensional manifold. The dimension of this attraction manifold is the largest when $m=1$. This analysis shows we expect the $N_\theta(N_\theta+1)/2$ variables involved in the RPA flow equations in the vicinity of a $m$-type fixed direction $M^*$ to fall in three classes (note that the notion of relevance or irrelevance refers to the behavior of the RPA flow in the space of directions and not to the global magnitude of the couplings)~:\\
- Irrelevant variables. We have $(N_\theta-m)(N_\theta-m+1)/2$ of them, and they correspond to changing the initial condition of the flow within the attraction basin of $M^*$.\\
- Marginal variables. They are associated to small changes of $M_0$ within the Grassmann manifold $G_m(\mathbb{R}^{N_\theta})$, therefore providing $m(N_\theta-m)$ such variables.\\
- Relevant variables. By a simple minded counting argument, we expect their number to be $m(m+1)/2$.

This picture is confirmed and refined by the following stability analysis. Let us consider a fixed direction $M^*=P \mathbb{I}_{\mathcal I} P^{-1}$ as above. For the sake of simplicity, we shall assume ${\mathcal I}=\lbrace 1,\ldots,m \rbrace$ (and thus $\mathbb{I}_{\mathcal I}=\mathbb{I}_m$). This fixed direction is associated to the one parameter family of solutions $M(t)=M^*/(t_{\rm c}-t)$ of the equation $\partial_t M=M^2$. Now let us consider weakly perturbed solutions of the form $M(t)=P \lbrace \mathbb{I}_m/(t_{\rm c}-t)+N(t) \rbrace P^{-1}$. Linearizing the differential equation with respect to $N(t)$, we obtain~:
\begin{equation}
\partial_t N(t)=\frac{\mathbb{I}_m N(t) + N(t) \mathbb{I}_m}{t_{\rm c}-t}\;.
\end{equation}
Solving this linearized equation gives~:
\begin{equation}
\label{eq:RPA_lin_general}
\left\lbrace
\begin{array}{ll}
{[N(t)]}_{i j}=[N(0)]_{i j} \left(\frac{t_{\rm c}}{t_{\rm c}-t}\right)^2 & \mbox{if } 1\leq i,j\leq m,\\ \\
{[N(t)]}_{i j}=[N(0)]_{i j} \left(\frac{t_{\rm c}}{t_{\rm c}-t}\right) & \mbox{if } i\leq m \mbox{ and } j\geq m+1\\
& \mbox{ or } i\geq m+1 \mbox{ and } j\leq m,\\ \\
{[N(t)]}_{i j}=[N(0)]_{i j} & \mbox{if } m+1\leq i,j\leq N_\theta.
\end{array}\right.
\end{equation}
The three cases correspond respectively to the relevant, marginal and irrelevant variables introduced before, justifying the counting already given there. 

In the sequel, we shall often perform a local stability analysis around fixed directions of the form $g_i(t)=u_i/(t_{\rm c}-t)$ for flow equations of the form $\partial_t g_i=A_{i j k} g_j g_k$. This can be done by setting $g_i(t)=u_i/(t_{\rm c}-t)+\delta g_i(t)$, and by linearizing the evolution equation with respect to $\delta g_i$. We obtain~:
\begin{equation}
\label{eq:eqRGlin_et_matriceB}
\partial_t \delta g_i=f(t) B_{i k} \delta g_k, \mbox{ with } \left\lbrace
\begin{array}{l}
f(t)=(t_{\rm c}-t)^{-1}\\
B_{i k} = (A_{i j k}+A_{i k j})u_j
\end{array}\right.\;.
\end{equation}
The linear stability is governed by the eigenvalues of the matrix $B$, which explicitly depends on the fixed direction. Let us assume $B$ can be put in diagonal form, with eigenvectors ${\boldsymbol{v}}^{(\alpha)}$, such that $B {\boldsymbol{v}}^{(\alpha)}=\lambda_\alpha {\boldsymbol{v}}^{(\alpha)}$. If $\boldsymbol{\delta g}(t)$ is along ${\boldsymbol{v}}^{(\alpha)}$, that is $\boldsymbol{\delta g}(t)={\mathcal G}(t){\boldsymbol{v}}^{(\alpha)}$, the linearized equation implies that~: ${\mathcal G}(t)={\mathcal G}(0)[t_{\rm c}/(t_{\rm c}-t)]^{\lambda_\alpha}$. Setting $\varepsilon={\mathcal G}(0)$, we have~:
\begin{equation}
g_i(t)=f(t) \left\lbrace u_i+\varepsilon t_{\rm c} {\left( \frac{t_{\rm c}}{t_{\rm c}-t}\right)}^{\lambda_\alpha-1} v^{(\alpha)}_i + O(\varepsilon^2) \right\rbrace\;.
\end{equation}
The fixed direction along $u_i$ will be stable if the term proportional to $u_i$ grows faster than the one proportional to $v^{(\alpha)}_i$, that is if all the eigenvalues $\lambda_\alpha$ of $B$ are smaller than 1. In the language already introduced, eigendirections with $\lambda_\alpha>1$, resp. $\lambda_\alpha=1$, resp. $\lambda_\alpha<1$ will be referred to as relevant, resp. marginal, resp. irrelevant. Note that $\boldsymbol{u}$ is always an eigenvector of the $B$ matrix with $\lambda=2$. This can be interpreted simply by a shift in the value of $t_{\rm c}$~:
\begin{equation}
\frac{u_i}{t_{\rm c}+\delta t_{\rm c}-t}=\frac{u_i}{t_{\rm c}-t} \left\lbrace 1-\frac{\delta t_{\rm c}}{t_{\rm c}-t}+O(\delta t_{\rm c}^2)\right\rbrace\;.
\end{equation}
Applying this terminology, we find the $\lambda$ spectrum for these RPA fixed directions is very degenerate, with only three eigenvalues~: $\lambda=2$ for relevant variables, $\lambda=1$ for marginal ones, and $\lambda=0$ for irrelevant directions (remember Eq.~(\ref{eq:RPA_lin_general})). Note that for a fixed direction with $m=1$, we have only the trivial eigenvalue at $\lambda=2$, whereas the marginal $\lambda=1$ is $N_\theta-1$ times degenerate. We should mention that similar stability analyses have been given by various authors\cite{Lin00,Konik00} in the context of coupled chains of interacting fermions.

\subsection{INFLUENCE OF THE COOPER TERMS ON THE RPA FLOW}
\label{sec:influence_cooper}
After this detailed description of the flow pattern in the RPA, let us describe the influence, on this flow, of a small perturbation in the Cooper channel. Intuitively, the Grassmann manifold $G_m(\mathbb{R}^{N_\theta})$ of fixed directions will evolve into an attracting manifold along which the flow in direction space is slow, by comparison to the variation of both irrelevant and relevant variables. In the remaining part of this section, we shall briefly discuss how we may construct this slow manifold using perturbation theory with respect to the Cooper channel. Then the dynamics along this manifold will be explicited. We shall consider the flow in direction space, of the form~:
\begin{equation}
\label{eq:eqRG_dir_normeN}
\partial_s h_i = A_{ijk} h_j h_k-\frac{1}{N^2}(A_{jkl} h_j h_k h_l) h_i\;.
\end{equation}
For convenience we have set $\| \boldsymbol{h} \|=N$ instead of unity. This is because fixed directions in the RPA obtained from $u_i=A_{ijk}^{\rm RPA} u_j u_k$ satisfy $\| \boldsymbol{u} \|=N$ as will be seen below. Now let us consider the flow with $A_{ijk}=A_{ijk}^{(0)}+\varepsilon A_{ijk}^{(1)}$, where $A^{(0)}$ refers to the Peierls channel and $A^{(1)}$ to the Cooper channel. We are interested in the situation where $\varepsilon=1$, but it will be convenient to treat $\varepsilon$ as a free small parameter. Let us choose a fixed direction $u_i=D(I,J)$ for the RPA flow, of the $m=1$ type according to the above-given classification. The matrix $D$ is then a projector on a one-dimensional space, so there is a $N$ dimensional vector $\phi(I)$ such that $D(I,J)=\phi(I)\phi(J)$ and $\sum_{I=1}^N {\phi(I)}^2=N$. Here, for simplicity, we consider the case $\theta=0$ so $N_\theta=N$. We then have $\sum_{I,J} {D(I,J)}^2=N^2$, so the norm of this vector $D$ of couplings is equal to $N$. 

To construct the point of the slow manifold which evolves smoothly from this fixed direction as $\varepsilon$ is switched on, we have to take a small perturbation $\varepsilon v_i=\varepsilon(\tilde{D}(I,J),\tilde{C}(I,J))$ which belongs to the irrelevant subspace (i.e. the eigenspace of $B^{(0)}$ with $\lambda=0$, according to the previous paragraph), incorporate $u_i+\varepsilon v_i$ in the RHS of Eq.~(\ref{eq:eqRG_dir_normeN}), and enforce that the projection of the result on the $\lambda=0$ subspace vanishes, neglecting terms of order $O(\varepsilon^2)$. We are following here the approach developped by N. Fenichel\cite{Fenichel79} (see also the work by Van Kampen\cite{VanKampen85}). We therefore obtain~:
\begin{widetext}
\begin{equation}
\left\lbrace
\begin{array}{l}
\tilde{D}(I,I)=-\frac{1}{N}\left[ {\phi(I)}^4 - \frac{2}{N}{\phi(I)}^6 + \frac{1}{N^2} {\phi(I)}^2 \sum_K {\phi(K)}^6 \right]\\ \\
\tilde{D}(I,J)=\frac{1}{N^2}\phi(I)\phi(J)\left[ {\phi(I)}^4+{\phi(J)}^4-\frac{1}{N} \sum_K {\phi(K)}^6\right]\\ \\
\tilde{C}(I,J)=-\frac{1}{N}{\phi(I)}^2 {\phi(J)}^2
\end{array}
\right..
\end{equation}
\end{widetext}
We may easily check that $\sum_J \tilde{D}(I,J) \phi(J)=0$, so $\tilde{D}$ belongs to the $\lambda=0$ subspace. These expressions show that $C(I,J)$ terms of relative strength $1/N$ are generated, as well as a small difference between diagonal and off-diagonal terms of the $D(I,J)$ matrix. These general trends will be confirmed by the explicit examples of fixed directions (including the Cooper terms to all orders in $\varepsilon$) given in the following section. 

Now according to the intuitive picture already given, we expect that Cooper terms will induce a slow drift motion of the $\phi(I)$ vector which doesn't depend on time if $\varepsilon=0$. To compute this slow motion, we simply have to consider the Cooper term for the unperturbed fixed direction $u_i$, namely to compute $A_{ijk}^{(1)} u_j u_k-\frac{1}{N^2}(A_{jkl}^{(1)} u_j u_k u_l) u_i$, and project this on the marginal (i.e. $\lambda=1$) subspace. This procedure gives~:
\begin{eqnarray}
\partial_s D(I,J)=-\frac{1}{N^2}\phi(I)\phi(J)\Bigg[ {\phi(I)}^4+{\phi(J)}^4\hspace{1cm}\\
-\frac{2}{N}\sum_L {\phi(L)}^6 \Bigg]\;.\nonumber
\end{eqnarray}
It is easy to interpret this as a small evolution of $\phi(I)$, so that
\begin{eqnarray}
\partial_s D(I,J)=\partial_s \phi(I) \phi(J)+\phi(I)\partial_s \phi(J), \mbox{ with}\\ 
\label{eq:derive}
\partial_s \phi(I)=-\frac{1}{N^2}\phi(I)\left[ {\phi(I)}^4-\frac{1}{N}\sum_L {\phi(L)}^6\right]\;.
\end{eqnarray}
Note that the norm of $\phi(I)$ is unchanged under this slow motion since we have $\sum_I \phi(I) \partial_s \phi(I)=0$. 

\begin{figure}[h]
\includegraphics[width=8cm]{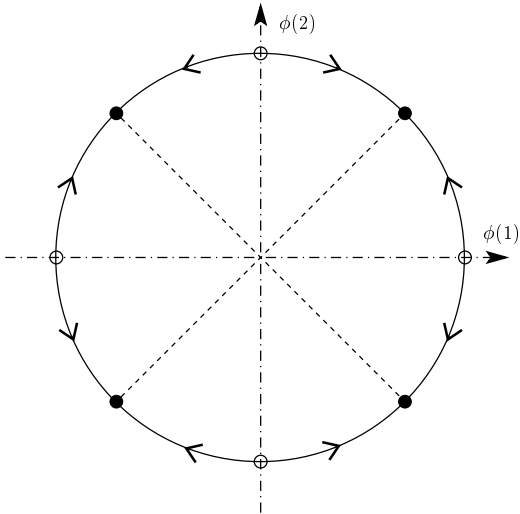}
\caption{Flow pattern in the $(\phi(1),\phi(2))$ space, in the $N=2$ case.}
\label{fig:phi}
\end{figure}

Eq.~(\ref{eq:derive}) has two interesting consequences. First, it is clear that fixed directions no longer form a continuous space, as was the case with flow equations involving only a single channel. According to Eq.~(\ref{eq:derive}), fixed directions are obtained only if the $p$ non-vanishing components of $\phi(I)$ all have the same absolute value, equal to $\sqrt{N/p}$. In the vicinity of such a fixed direction, it is easy to check that the tangent space to the slow manifold splits into $p-1$ eigenvalues $\lambda=1-4\varepsilon/p^2+O(\varepsilon^2)$ and $N-p$ eigenvalues $\lambda=1+\varepsilon/p^2+O(\varepsilon^2)$. The special case where $p=N$ and $\phi(I)=1$ independently of $I$ is specially interesting, since it corresponds to an effective interaction with SU$(N)$ symmetry. A detailed analysis will be given in the next section. We have represented what happens when $N=2$ on Fig.~\ref{fig:phi}, in the $(\phi(1),\phi(2))$ plane.
The circle is of radius $\sqrt{2}$, and is the continuous set of fixed point in the RPA. The small filled (resp. non-filled) circles are the stable (resp. unstable) fixed points, when the Cooper channel is turned on. The arrows represent the direction of the flow. 

As a second important consequence of Eq.~(\ref{eq:derive}), it is now possible to estimate the duration of the intermediate regime of the complete RG flow (see Figs.~\ref{fig:flot_glob_c} to \ref{fig:flot_debut_s}). As already discussed earlier, the dominant couplings in this regime are of the $D(I,J)$ type. A very natural way to interpret the plateau regime seen on Figs.~\ref{fig:flot_glob_c} to \ref{fig:flot_debut_s} involves the slow motion just derived here. At the beginning of the plateau, $D(I,J)$ couplings are of the same order of magnitude, independently of $I$ and $J$. But as the fictitious time $s$ increases, there is a trend to eliminate couplings for which $I$ or $J$ is not one of the end points of the flat Fermi surface segments. In the language developped in this section, the direction of the $\phi(I)$ vector evolves slowly in this regime under the residual influence of the Cooper channel. During this phase of the flow, the magnitude of the coupling vector increases tremendously. Let us estimate this semi-quantitatively. From Eq.~(\ref{eq:derive}), we see that we have to wait for a $s$ interval of order $N^2$ to get a change in a given component $\phi(I)$ of order unity. With the normalization of this section, we have $h_i(t)=N g_i(t)/\|\boldsymbol{g}(t)\|$ and $ds/dt=\|\boldsymbol{g}(t)\|/N$. So the evolution of $\|\boldsymbol{g}(t)\|$ is given by~:
\begin{equation}
\partial_s \ln \|\boldsymbol{g}\|=\frac{1}{N^2} A_{ijk}h_i h_j h_k\;.
\end{equation}
To get an estimate, we approximate the RHS by its value obtained if $\varepsilon=0$, namely keeping only the Peierls channel, and assuming $D(I,J)=\phi(I)\phi(J)$, $C(I,J)=0$ and $\sum {\phi(I)}^2=N$. With these approximations, $\partial_s \ln \|\boldsymbol{g}\| \simeq 1$, or $\|\boldsymbol{g}(s)\|=\|\boldsymbol{g}(0)\| \exp(s)$. Since $s$ increases by an amount of order $N^2$, we see that the magnitude of the coupling vector is multiplied by a factor of the form $\exp(\gamma N^2)$, where $\gamma$ is a positive number which is expected not to depend on $N$. Unfortunately, this restricts tremendously the possibility to observe the final stage of the flow in physical systems. Indeed, our whole approach relies on a one-loop approximation, and therefore requires $\|\boldsymbol{g}\|/N$ to remain small. This criterion breaks down as $s$ goes to infinity. But even to observe the complete plateau regime in a physical system would require extremely weak initial couplings, of the order of $\exp(-\gamma N^2)$, which doesn't make much physical sense, since $N$ is in principle infinite. This shows that for all practical purposes, the system is in many respects ``frozen'' at the entrance of the plateau regime, therefore justifying the description already given by Zheleznyak et al. Note however that we cannot rule out at this stage the possibility of a cross-over from a spin density wave state to a d-wave state superconductor at lower energies which would be induced by higher order terms in the RG flow, beyond one-loop. But we don't see any strong argument to support this scenario. Any attempt to investigate higher order corrections is severely limited by the divergency of the couplings in the lowest order approximation.

In the following sections, we shall study in great detail some fixed directions where $\phi(I)$ is non-vanishing for $p$ values of $I$, focusing on extreme cases $p=N$ and $p=2$. $p=2$ corresponds to the final regime of the one-loop flow, whereas $p=N$ is a particularly interesting fixed direction, since the flow passes in a rather close neighborhood of this point at the beginning of the intermediate regime. Its investigation is greatly simplified by the fact it exhibits a global SU$(N)$ symmetry for the effective couplings.\\ \\

\section{STUDY OF FIXED DIRECTIONS THAT ARE HIGHLY SYMMETRICAL}
\label{sec:high_sym_FD}
\subsection{DETERMINATION OF THE SYMMETRIC FIXED DIRECTIONS}
\label{sec:fixeddir}

We now aim at studying in greater detail the intermediate regime, for which the couplings show an approximate SU$(N)$ symmetry. As usual, the use of a symmetry greatly simplifies the calculation, so that we will restrict ourselves to the case where the couplings are exactly SU$(N)$ symmetric.

Fixed directions are found when solving the equation $u_i=A_{ijk} u_j u_k$, where we have used the general notation. In the particular case we are interested in, we simply have to consider Eq.~(\ref{eq:eqRGchaines}), and replace $\partial_t G$ in the LHS with $G$. Considering only the principal couplings of the type $A$, $C$ and $D$, as discussed in Sec.~\ref{subsec:numres2D}, and setting all others to zero, we get the following system~:
\begin{widetext}
\begin{equation}
\left\lbrace
\begin{array}{l}
NA^{\rm c}=(N-1)({D^{\rm c}}^2+3{D^{\rm s}}^2)\\
NA^{\rm s}=4{A^{\rm s}}^2 +2(N-1)({D^{\rm s}}^2+D^{\rm s}D^{\rm c})\\
NC^{\rm c}=-({D^{\rm c}}^2+3{D^{\rm s}}^2)\\
NC^{\rm s}=4{C^{\rm s}}^2 +2({D^{\rm s}}^2-D^{\rm s}D^{\rm c})\\
ND^{\rm c}=(N-2)({D^{\rm c}}^2+3{D^{\rm s}}^2)+2(A^{\rm c}-C^{\rm c})D^{\rm c}+6(A^{\rm s}-C^{\rm s})D^{\rm s}\\
ND^{\rm s}=2(N-2)({D^{\rm s}}^2+D^{\rm s}D^{\rm c})+2(A^{\rm s}-C^{\rm s})D^{\rm c}+2(A^{\rm c}-C^{\rm c})D^{\rm s}+4(A^{\rm s}+C^{\rm s})D^{\rm s}
\end{array}
\right..
\end{equation}
\end{widetext}
The SU$(N)$ symmetry is imposed by the relations~: $A^{\rm c,s}=C^{\rm c,s}+D^{\rm c,s}$. Indeed, the interaction Hamiltonian in Eq.~(\ref{eq:H_int}), in the charge and spin sector, is then of the following schematic form~:
\begin{widetext}
\begin{eqnarray}
\label{eq:Hint_schematic}
H_{\rm int} \sim C\sum_{q\neq 0} \Big[ \sum_{I,k} c^\dagger_{{\rm R},I} (k+q)c_{{\rm R},I} (k) \Big] \Big[ \sum_{J,k'} c^\dagger_{{\rm L},J} (k'-q)c_{{\rm L},J} (k') \Big] \hspace{4cm}\nonumber\\
- D \sum_{q\neq 0 ,k,k'} \Big[ \sum_{J} c^\dagger_{{\rm R},J} (k+q)c_{{\rm L},J} (k') \Big] \Big[ \sum_{I} c^\dagger_{{\rm L},I} (k'-q)c_{{\rm R},I} (k) \Big]\;. 
\end{eqnarray}
\end{widetext}

 The spinless case is obtained when setting all spin couplings to zero. A little bit of algebra leads to the results listed in Table \ref{tab:dirfixeN}.
\begin{table}
\caption{Values of the couplings for the SU$(N)$ fixed directions.}
\label{tab:dirfixeN}
\begin{tabular}{|c|c|c|c|}
\hline
& Spinless & (+,+) & (+,-) \\ \hline
$A^{\rm c}$ & $1-1/N$ & $(1-1/N)/4$ & $3/4[1-1/N-2/N^2$\hspace{2cm}\\
&&&\hspace{2cm}$+O(1/N^3)]$ \\ \hline
$C^{\rm c}$ & $-1/N$ & $-1/(4N)$ & $-3/(4N)+O(1/N^3)$ \\ \hline
$D^{\rm c}$ & $1$ & $1/4$ & $3/4[1-2/N^2+O(1/N^4)]$ \\ \hline
$A^{\rm s}$ & $0$ & $1/4$ & $-1/4[1-2/N+2/N^2$\hspace{2cm}\\
&&&\hspace{2cm}$+O(1/N^3)]$ \\ \hline
$C^{\rm s}$ & $0$ & $0$ & $1/(2N)+O(1/N^3)$ \\ \hline
$D^{\rm s}$ & $0$ & $1/4$ & $-1/4[1+2/N^2+O(1/N^4)]$ \\ \hline
\end{tabular}
\end{table}
In the case of spin 1/2 electrons, we find two fixed directions, named $(+,+)$ and $(+,-)$, where the signs are those of $(D^{\rm c},D^{\rm s})$. In the spinless and $(+,+)$ cases, the results are exact, whereas in the $(+,-)$ case, we gave a $1/N$ expansion of the results valid up to order 3. In fact the $D$'s are solution of a polynomial equation of order 4 and thus have complicated exact expressions. The $(+,+)$ fixed direction corresponds to a charge density wave, whereas the $(+,-)$ fixed direction corresponds to a spin density wave. This will be shown below, when we compute the response function in these cases.

Another simple and important case is the fixed direction that involves only the 2 patches at the border of the Fermi surface, that we have called ``hot spots'' in Sec.~\ref{subsec:numres2D}. We have indeed seen that this fixed direction was reached in the final part of the flow, after the flow has spent quite a long time near the $(+,-)$ fixed direction. We have looked for such ``hot spots'' fixed directions, and we have found three of them, respecting a SU$(2)$ symmetry. The resulting couplings (divided by $N$) are shown in Table \ref{tab:dirfixe2}.
\begin{table}
\caption{Values of the couplings for the SU$(2)$ fixed directions (``hot spots'' case) .}
\label{tab:dirfixe2}
\begin{tabular}{|c|c|c|c|}
\hline
& Spinless & (+,+) & (+,-) \\ \hline
$A^{\rm c}/N$ & $1/4$ & $1/16$ & $1/16$ \\ \hline
$C^{\rm c}/N$ & $-1/4$ & $-1/16$ & $-1/16$ \\ \hline
$D^{\rm c}/N$ & $1/2$ & $1/8$ & $1/8$ \\ \hline
$A^{\rm s}/N$ & $0$ & $1/8$ & $0$ \\ \hline
$C^{\rm s}/N$ & $0$ & $0$ & $1/8$ \\ \hline
$D^{\rm s}/N$ & $0$ & $1/8$ & $-1/8$ \\ \hline
\end{tabular}
\end{table}
Notice that most of these results can simply be obtained by setting $N=2$ in Table \ref{tab:dirfixeN}. The only exception is in the $(+,-)$ case, where the whole $1/N$ series would have to be summed.

\subsection{RESPONSE FUNCTIONS FOR THESE FIXED DIRECTIONS}
\label{sec:reponse}
\subsubsection{Notations and general setting}
\begin{figure}[h]
\includegraphics[width=8cm]{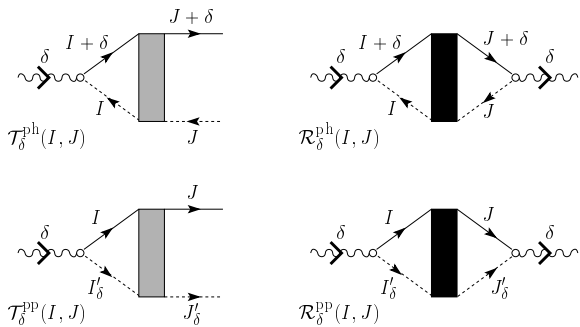}
\caption{Representation of the triangular vertices and susceptibilities in the particle-hole and particle-particle channels. We have used the notation $I_\delta'=N+1-I+\delta$ and similarly for $J_\delta'$. $I_{\delta=0}'$ would be equal to $-I$ if the patch numbers ran from $-N/2$ to $N/2$ instead of $1$ to $N$. For the sake of simplicity, we have not mentioned whether the singlet or triplet channel is involved, and we have only indicated the momenta parallel to the Fermi surface (i.e. the patch numbers), but not the perpendicular ones, nor the energies.}
\label{fig:reponse}
\end{figure}

The formalism of the renormalization of the response functions is well known (see for example Zheleznyak et al\cite{Zheleznyak97}), but let us briefly summarize it and fix our notations. We shall here use the RG to follow the evolution of the response functions with the decreasing of a typical energy scale $\nu$. The high energy cut-off $\Lambda_0$ is fixed, and here we set $t=\ln (\Lambda_0/\nu)$, so that $t$ increases when $\nu$ is lowered. The RG equations for the couplings are just the same as before (see Eq.~(\ref{eq:eqRGchaines})). In this case, $\nu$ is the energy scale at which the interaction process takes place. We now consider the response functions. We introduce the triangular vertices and the susceptibilities (see Fig.~\ref{fig:reponse}). From the point of view of a renormalizable field theory, these are the only composite operators that need being renormalized (see for instance\cite{Itzykson}). More physically, the triangular vertices describe the interaction between the fermions and an external field, whereas the susceptibilities give the linear response of the system when a small external field is applied. When a susceptibility diverges, the system gets spontaneously ordered, even in a vanishing external field. The common wisdom, when several susceptibilities diverge, is to assume that the most divergent susceptibility determines what phase transition is observed.

As is usual, we introduce different response functions, corresponding to different physical situations. These can be either of the particle-particle (Cooper) or of the particle-hole (Peierls) type, i.e. they concern either the superconducting instability or the density wave instability. In our notation, we will associate $^{\rm pp}$ and $^{\rm ph}$ superscripts. This is shown on Fig.~\ref{fig:reponse}. Notice on this figure that we have only indicated the momenta parallel to the Fermi surface, that is the patch numbers. None of these momenta is summed over, contrarily to the momenta perpendicular to the Fermi surface, and the energies that appear in the loops. For the external legs, we choose the configuration that gives the biggest divergence. We thus choose all external fermion momenta on the Fermi surface, and the momentum entering the vertices are chosen equal to $2 k_{\rm F}$ (resp. 0) in the particle-hole (resp. particle-particle) channel. The energy entering the vertices is chosen to be $\nu$ as discussed above. 

We insist it is not sufficient to tell what comes in and out in the triangular vertices (the patches involving $J$, $J+\delta$ or $J_\delta'$). It is necessary to keep track of what happens at the vertex pictured by a circle, i.e. of the non-diagonal character of the triangular vertices. If we consider the patch number as a tag for different fermion species, this phenomenon is known as operator mixing in the field theory literature.\cite{Itzykson} Notice also that since our patch numbers run from $1$ to $N$, it is $I_\delta'=N+1-I+\delta$ that appears, instead of $-I+\delta$ as would be the case if the patch were numbered from $-N/2$ to $N/2$ (in the even $N$ case). 

We furthermore distinguish between the singlet and the triplet channels, respectively giving rise to singlet and triplet superconductivity in the particle-particle channel, and to charge and spin density wave in the particle-hole channel. We will add a $_{\rm S}$ or $_{\rm T}$ subscript in our notation for the response functions to indicate what channel is under consideration. For example, the spin density wave susceptibility is denoted by ${\mathcal R}_{{\rm T},\delta}^{\rm ph}(I,J)$. From the spin rotation invariance of the interaction, it is clear that singlet and triplet channels do not mix. It is thus interesting to separate the interaction process in singlet and in triplet channels, instead of the charge and spin couplings. This is done differently in the particle-hole and the particle-particle cases, because the rotation group doesn't act in the same way in both cases (the representations are different). The resulting couplings in the four distinct channels are given in Table \ref{tab:singtrip}.
\begin{table}
\caption{Dictionary between the charge-spin and the singlet-triplet ways of separating the couplings.}
\label{tab:singtrip}
\begin{tabular}{|c|c|c|}
\hline
& particle-hole & particle-particle\\ \hline
singlet & $G^{\rm c}+3 G^{\rm s}$ & $G^{\rm c}-3 G^{\rm s}$ \\ \hline
triplet & $G^{\rm c}- G^{\rm s}$ & $G^{\rm c}+ G^{\rm s}$ \\ \hline
\end{tabular}
\end{table}

\begin{figure}[h]
\includegraphics[width=8cm]{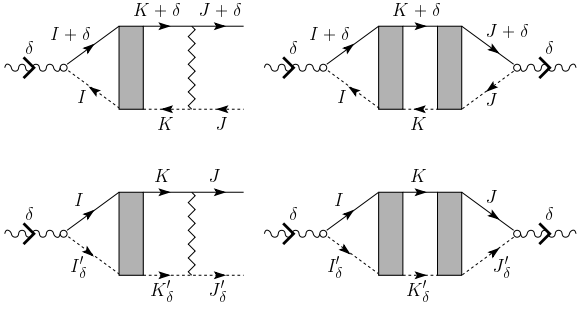}
\caption{Feynman diagrams that contribute to the renormalization of the triangular vertices (left) and the susceptibilities (right) in the particle-hole channel (top) and particle-particle channel (bottom). The $K$ variable is summed over.}
\label{fig:renormreponse}
\end{figure}

We now come to the RG equations for the response functions. The different Feynman diagrams that contribute to the renormalization of the different response functions are shown on Fig.~\ref{fig:renormreponse}.
It can be shown that the most divergent responses in each channel always appear for $\delta=0$ so that we will assume $\delta=0$ in all the following, and we will not write the $\delta$ subscript anymore. The RG equations in the singlet and triplet channels are the same and read (we drop the $_{\rm S}$ and $_{\rm T}$ subscripts here)~:
\begin{equation}
\label{eq:RGeqreponses}
\left\lbrace
\begin{array}{l}
\partial_t {\mathcal T}^{\rm ph}(I,J)=\frac{1}{N}\sum_K {\mathcal T}^{\rm ph}(I,K) G_{J-K}(K,J)\\ \\
\partial_t {\mathcal T}^{\rm pp}(I,J)=-\frac{1}{N}\sum_K {\mathcal T}^{\rm pp}(I,K) G_{J-K}(K,K')\\ \\
\partial_t {\mathcal R}^{\rm ph}(I,J)=\sum_K {\mathcal T}^{\rm ph}(I,K) {\mathcal T}^{\rm ph}(J,K) \\ \\
\partial_t {\mathcal R}^{\rm pp}(I,J)=\sum_K {\mathcal T}^{\rm pp}(I,K) {\mathcal T}^{\rm pp}(J,K)
\end{array}
\right.,
\end{equation}
where $K'=N+1-K$. It should be noticed that in fact, there should be two different triangular vertices in each channel. Let us consider the particle-particle (and for example singlet) channel. The two external fermions can go towards the vertex or go away from the vertex. It is however quite easy to see that those two vertices are related through complex conjugation. But here our vertices are all real. Indeed they are real at $t=0$ (i.e. $\nu=\Lambda_0$), their value being their non-interacting one, that is 1 for the diagonal elements and 0 for the others. Furthermore, from the renormalization equations that do not involve any complex numbers, they have to remain real along the flow. Let us also remark that the initial values of the ${\mathcal R}$'s are all zero, which is their non-interacting value, when evaluated at $\nu=\Lambda_0$. In Eq.~(\ref{eq:RGeqreponses}), the $G$'s are of course channel dependent, and are found thanks to Table \ref{tab:singtrip}.
\subsubsection{Analytical results}
The previous equations are directly usable in a numerical simulation, once the $t$ variable is changed for the $s$ variable, which is simply done by dividing these equations by the norm ${\mathcal N}$ of the coupling vector. Before showing numerical results, it is interesting to give analytical results that are available for the fixed directions we found in Sec.~\ref{sec:fixeddir}. Let us briefly explain how this works. Eq.~(\ref{eq:RGeqreponses}) can easily be cast in a matrix form~:
\begin{equation}
\label{eq:RGeqreponses_matrix}
\left\lbrace
\begin{array}{l}
\partial_t {\mathcal T}^{\rm ph}={\mathcal T}^{\rm ph} {\mathcal M}^{\rm ph}\\
\partial_t {\mathcal T}^{\rm pp}={\mathcal T}^{\rm pp} {\mathcal M}^{\rm pp}\\
\partial_t {\mathcal R}^{\rm ph}={\mathcal T}^{\rm ph} \;^{\rm t} {\mathcal T}^{\rm ph}\\
\partial_t {\mathcal R}^{\rm pp}={\mathcal T}^{\rm pp} \;^{\rm t}{\mathcal T}^{\rm pp}
\end{array}
\right.,
\end{equation}
where $^{\rm t}{\mathcal T}$ is the transpose of ${\mathcal T}$. The definitions of the ${\mathcal M}$'s are clear, and they are symmetric thanks to the particle-hole symmetry of the couplings (coming from the hermiticity of the Hamiltonian)~: $G_\delta(I,J)=G_{-\delta}(J,I)$. For a fixed direction, we know that ${\mathcal M}={\mathcal U}/(t_{\rm c}-t)$, where ${\mathcal U}$ is a $t$ independent matrix that defines the fixed direction (it is formed with the $u_i$'s). We diagonalize this matrix, setting ${\mathcal U}={\mathcal P}{\mathcal D}{\mathcal P}^{-1}$, where ${\mathcal D}$ is diagonal, and ${\mathcal P}$ is an orthogonal matrix. Denoting $\tilde{{\mathcal T}}={\mathcal P}^{-1}{\mathcal T}{\mathcal P}$, and similarly for the ${\mathcal R}$'s, we get~:
\begin{equation}
\label{eq:RGeqreponses_diag}
\left\lbrace
\begin{array}{l}
\partial_t \tilde{{\mathcal T}}^{\rm ph}=\frac{1}{t_{\rm c}-t}\tilde{{\mathcal T}}^{\rm ph} {\mathcal D}^{\rm ph}\\
\partial_t \tilde{{\mathcal T}}^{\rm pp}=\frac{1}{t_{\rm c}-t}\tilde{{\mathcal T}}^{\rm pp} {\mathcal D}^{\rm pp}\\
\partial_t \tilde{{\mathcal R}}^{\rm ph}=\tilde{{\mathcal T}}^{\rm ph} \;^{\rm t} \tilde{{\mathcal T}}^{\rm ph}\\
\partial_t \tilde{{\mathcal R}}^{\rm pp}=\tilde{{\mathcal T}}^{\rm pp} \;^{\rm t} \tilde{{\mathcal T}}^{\rm pp}
\end{array}
\right..
\end{equation}
From these equations and the initial conditions ${\mathcal T}(t=0)={\mathbb I}$ and ${\mathcal R}(t=0)=0$, it is evident that the $\tilde{{\mathcal T}}$'s and the $\tilde{{\mathcal R}}$'s are diagonal at any time. One of the consequences is that ${\mathcal T}=\;^{\rm t}{\mathcal T}$. Let us suppose that $[{\mathcal D}]_{ii}=\delta_i\neq 1/2$. It is easy to check that one gets~:
\begin{eqnarray}
{[\tilde{{\mathcal T}}]}_{ii}&=&{\left( \frac{t_{\rm c}}{t_{\rm c}-t}\right)}^{\delta_i}, \mbox{ and }\\
{[\tilde{{\mathcal R}}]}_{ii}&=&\frac{t_{\rm c}}{2\delta_i-1}\left[ {\left(\frac{t_{\rm c}}{t_{\rm c}-t}\right)}^{2\delta_i-1} -1\right]\;.
\end{eqnarray}
If $\delta_i=1/2$, we get a logarithm instead of power functions for the susceptibilities (which is the case in the purely 1D case). In order to get the critical exponents $2\delta_i-1$, we thus only have to diagonalize a matrix. We have computed these exponents, and in Table \ref{tab:exposants} we give the biggest exponent of the four channels, for the four fixed directions we have previously found. All the other exponents can be shown to be equal to -1, up to some $1/N$ corrections. The results in the spinless case are given in Table \ref{tab:exposants_spinless}. We also find the other exponents are nearly equal to -1 in the SU$(N)$ spinless case. 
\begin{table}
\caption{Biggest exponents of the susceptibilities, for the four fixed directions involving spinning electrons.}
\label{tab:exposants}
\begin{tabular}{|c|c|c|c|c|}
\hline
& SU$(N)$ & SU$(N)$ & SU$(2)$ & SU$(2)$  \\
& (+,+) & (+,-) & (+,+) & (+,-)\\ \hline
ph, S & $1-1/(2N^2)$ & & 7/8 &\\ \hline
ph, T & & $1-9/(2N^2)+O(1/N^4)$ & &\\ \hline
pp, S & & & & 7/8\\ \hline
pp, T & & & &\\ \hline
\end{tabular}
\end{table}
\begin{table}
\caption{Biggest exponents of the susceptibilities, for the two fixed directions involving spinless electrons.}
\label{tab:exposants_spinless}
\begin{tabular}{|c|c|c|}
\hline
& SU$(N)$ & SU$(2)$ \\ \hline
ph & $1-2/N^2$ & 1/2\\ \hline
pp & & 1/2 \\ \hline
\end{tabular}
\end{table}
We thus see that in the SU$(N)$ case, the (+,+) fixed direction is of the charge density wave type, whereas the (+,-) fixed direction is of the spin density wave type. No superconducting instability arises in this case. In the SU$(2)$ case, i.e. the ``hot spots'' regime, the (+,+) fixed direction is still of the charge density wave type, but the (+,-) fixed direction is of the singlet superconductivity type. Let us emphasize that even if the exponent 7/8 in the SU$(2)$, particle-hole, singlet, (+,+) case, can be found in taking $N=2$ in $1-1/(2N^2)$, the $N=2$ case is different from the $N\neq 2$ case, as can be seen from the (+,-) fixed direction where the dominant instability is not of the same type. In the spinless case, the SU$(N)$ fixed direction gives rise to a charge density wave, and the SU$(2)$ fixed direction exhibits a perfect competition between charge density wave and superconductivity.

In all three SU$(N)$ fixed directions, the most divergent susceptibility is obtained by summing all the susceptibilities~: ${\mathcal R}^*=\sum_{I,J} {\mathcal R}(I,J)$. This corresponds to a uniform sign of the particle-hole or particle-particle wave function. This is also true in the SU$(2)$ case, except for the (+,-) fixed direction, where the most divergent susceptibility is found to be~: ${\mathcal R}^*= {\mathcal R}(1,1)-{\mathcal R}(1,N)-{\mathcal R}(N,1)+{\mathcal R}(N,N)$. This can be checked to correspond to a ``d-wave'' type particle-particle wave function.

\begin{figure}[h]
\includegraphics[width=8cm]{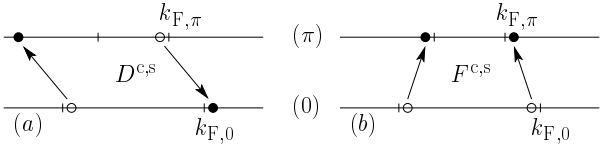}
\caption{$(a)$ The $D^{{\rm c,s}}$ processes are no longer available at low energy on a 2-leg ladder since $k_{{\rm F},0}\neq k_{{\rm F},\pi}$. $(b)$ But transverse umklapps are now possible ; they are denoted by $F^{{\rm c,s}}$.}
\label{fig:lienDF}
\end{figure}

The d-wave superconducting correlations found for the SU$(2)$ (+,-) fixed direction looks similar to the C1S0 phase described by various authors for doped 2-leg Hubbard ladders. Since ladder systems have been extensively studied,\cite{Khveshchenko94,Balents96,Schulz96,Schulz98} it seems useful to explain the connection between our systems and 2-leg ladders. In a 2-leg ladder, the single-particle spectrum is composed of two energy bands with the following dispersion relation~: $\varepsilon_\alpha(k)=-2t\cos(k)-t_\perp\cos(\alpha)$. Here $k$ is the momentum along the ladder, and $\alpha\in\{0,\pi\}$ is the transverse momentum. If $t_\perp$ is not too large and the filling factor not too large nor too small, both bands cross the Fermi energy, at wave-vectors $k_{{\rm F},\alpha}$. If $t_\perp$ is positive, we have $k_{{\rm F},\pi}<k_{{\rm F},0}$. From the fact that $k_{{\rm F},\pi}\neq k_{{\rm F},0}$, the $D$ processes become irrelevant at low energy, as illustrated on Fig.~\ref{fig:lienDF}$(a)$. On the other hand, this system allows for transverse Umklapp processes, which are depicted on Fig.~\ref{fig:lienDF}$(b)$, and these will be denoted here by $F^{{\rm c,s}}$.

Of course, $A^{{\rm c,s}}(0)$, $A^{{\rm c,s}}(\pi)$, $C^{{\rm c,s}}(0,\pi)=C^{{\rm c,s}}(\pi,0)=C^{{\rm c,s}}$ processes remain available. To simplify the comparison with our model, we shall assume the 2-leg ladder is not half-filled so that longitudinal Umklapp processes are irrelevant. We shall also neglect the difference between the two Fermi velocities, and assume that $A^{{\rm c,s}}(0)=A^{{\rm c,s}}(\pi)=A^{{\rm c,s}}$. Clearly, our model of coupled chains in $k$-space in the hot spot regime, where only chains labelled by 1 and $N$ are involved in the dominant scattering processes is not directly equivalent to an effective model for a doped 2-leg ladder. The $D^{{\rm c,s}}$ couplings in the former case are replaced by the $F^{{\rm c,s}}$ ones in the latter. However, it is possible to connect both situations by a unitary transformation $U$. We shall define $U$ as follows~:
\begin{eqnarray}
&&U c_{{\rm R},I,\mu}(k) U^{-1}=c_{{\rm R},I,\mu}(k), \mbox{ }(I\in\{1,N\})\\
&&U c_{{\rm L},I,-\mu}(-k) U^{-1}=\mu c^\dagger_{{\rm L},I,\mu}(k-2k_{\rm F}).
\end{eqnarray}
Here, we have assumed that the spin index $\mu$ takes two possible values $\pm 1$. Using the definitions for the charge and spin currents (Eqs.~\ref{eq:currents1} and \ref{eq:currents2}), it is clear they are unchanged by $U$ on the right-moving side, but they are transformed on the left-moving one according to~:
\begin{eqnarray}
&&U \rho_{{\rm L},I,\delta}(q) U^{-1}=-\rho_{{\rm L},I+\delta,-\delta}(q),\\
&&U S^a_{{\rm L},I,\delta}(q) U^{-1}=S^a_{{\rm L},I+\delta,-\delta}(q).
\end{eqnarray}
Furthermore, because of particle-hole symmetry, $U$ commutes with the kinetic energy. Starting from a model with couplings $A^{{\rm c,s}}$, $C^{{\rm c,s}}$, and $D^{{\rm c,s}}$, and after applying the unitary transformation $U$, we get a model with couplings $\tilde{A}^{{\rm c,s}}$, $\tilde{C}^{{\rm c,s}}$, and $\tilde{F}^{{\rm c,s}}$, given by~:
\begin{eqnarray}
&&\tilde{A}^{\rm c}=-A^{\rm c};\, \tilde{C}^{\rm c}=-C^{\rm c};\ \tilde{F}^{\rm c}=-D^{\rm c} \mbox{ and } \nonumber\\
&&\tilde{A}^{\rm s}=A^{\rm s};\ \tilde{C}^{\rm s}=C^{\rm s};\ \tilde{F}^{\rm s}=D^{\rm s}
\end{eqnarray}
Our (+,-) SU$(2)$ fixed direction can be reached directly from the particular choices of initial couplings such that $A^{{\rm c,s}}(1)=A^{{\rm c,s}}(N)=G^{{\rm c,s}}$, $C^{{\rm c,s}}(1,N)=C^{{\rm c,s}}(N,1)=G^{{\rm c,s}}$, and $D^{{\rm c,s}}(1,N)=D^{{\rm c,s}}(N,1)=G^{{\rm c,s}}$, all other bare couplings being set to zero. After the $U$ transformation, we get a model equivalent to a doped 2-leg ladder. In the g-ology notation used by Schulz,\cite{Schulz96}, the bare couplings are $g_1=-2 G^{\rm s}$ and  $g_2=-G^{\rm c}-G^{\rm s}$. The (+,-) SU$(2)$ fixed direction is the stable low-energy attractor when $G^{\rm c}>0$ and $G^{\rm s}<0$, which translates into $g_1>\max (0,2g_2)$ in the 2-leg ladder terminology. According to Schulz,\cite{Schulz96} the 2-leg ladder is in the staggered flux phase regime (see Fig.~1 in Schulz's paper\cite{Schulz98}). This is in perfect agreement with our d-wave superconductivity dominant correlation for the (+,-) SU$(2)$ fixed direction. Indeed, the d-wave order parameter reads~:
\begin{eqnarray}
\sum_k c^\dagger_{{\rm R},1,\uparrow}(k_{\rm F}+k) c^\dagger_{{\rm L},N,\downarrow}(-k_{\rm F}-k)\nonumber\\
 - c^\dagger_{{\rm R},1,\downarrow}(k_{\rm F}+k) c^\dagger_{{\rm L},N,\uparrow}(-k_{\rm F}-k)\nonumber\\
 - c^\dagger_{{\rm R},N,\uparrow}(k_{\rm F}+k) c^\dagger_{{\rm L},1,\downarrow}(-k_{\rm F}-k)\\
 + c^\dagger_{{\rm R},N,\downarrow}(k_{\rm F}+k) c^\dagger_{{\rm L},1,\uparrow}(-k_{\rm F}-k).
\end{eqnarray}
The unitary transformation $U$ turns this into~:
\begin{eqnarray}
\sum_{k,\sigma} c^\dagger_{{\rm R},1,\sigma}(k_{{\rm F}}+k) c_{{\rm L},N,\sigma}(-k_{{\rm F}}+k)\nonumber\\
 - c^\dagger_{{\rm R},N,\sigma}(k_{{\rm F}}+k) c_{{\rm L},1,\sigma}(-k_{{\rm F}}+k).
\end{eqnarray}
To translate this in the language of a 2-leg ladder we assume that chain 1 corresponds to $\alpha=0$ and chain N to $\alpha=\pi$. Shifting a little bit the $k_{{\rm F}}$ values, which is harmless since only the marginal couplings at zero energy have been considered, we get for the order parameter :
\begin{eqnarray}
\sum_{k,\sigma} c^\dagger_{{\rm R},0,\sigma}(k_{{\rm F},0}+k) c_{{\rm L},\pi,\sigma}(-k_{{\rm F},\pi}+k)\nonumber\\
 - c^\dagger_{{\rm R},\pi,\sigma}(k_{{\rm F},\pi}+k) c_{{\rm L},0,\sigma}(-k_{{\rm F},0}+k).
\end{eqnarray}
In real space, this reads~:
\begin{equation}
\sum_{x,\sigma} e^{i (k_{{\rm F},0}+k_{{\rm F},\pi})x} [\psi^\dagger_1(x)\psi_2(x)-\psi^\dagger_2(x)\psi_1(x)],
\end{equation}
where $\psi^\dagger_{i,\sigma}(x)$ creates an electron with spin $\sigma$ along the rung $x$ of the ladder, and on chain number $i$. Clearly, this corresponds to a staggered current pattern.

This phase has many interesting properties. It has only one gapless mode, which is the even charge mode. All other modes are gapped. As shown by Schulz,\cite{Schulz98} this system exhibits a large S0$(6)$ symmetry, which contains our SU$(2)\times$SU$(2)$. But a detailed analysis of the energy spectrum of this model is beyond the scope of this article.

\subsubsection{Numerical results}
\begin{figure}[h]
\includegraphics[width=8cm]{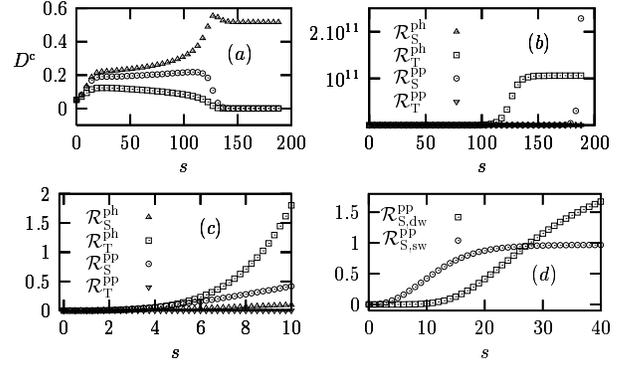}
\caption{Flows of response functions for $N=8$ and initial condition $G^{\rm c}=-G^{\rm s}=0.1$. $(a)$~: flow of couplings $D^c$ as in Fig.~\ref{fig:flot_glob_c}. $(b)$~: flow of the biggest response function in each channel. $(c)$~: zoom of $(b)$ for small time $s$. $(d)$~: competition at small time $s$, in the particle-particle, singlet channel, of the ``d-wave'' and ``s-wave'' types of response functions.}
\label{fig:flot_rep}
\end{figure}

In this section we will show some numerical flows of the various response functions ${\mathcal R}$, for the spinful case, and for initial conditions such that the couplings get close to the SU$(N)$ (+,-) fixed direction, in the intermediate regime. We consider a number of patches $N=8$ and not $N=16$ as in Figs.~\ref{fig:flot_glob_c} to \ref{fig:flot_debut_s}. The reason is only to shorten the computation time. Indeed, the final growth of the ``d-wave'' superconducting response function does not occur immediately after the transition from SU$(N)$ to SU$(2)$, but only after a delay time that is growing with $N$. The flows are shown on Fig.~\ref{fig:flot_rep}. 
In Fig.~\ref{fig:flot_rep}$(a)$ we show the evolution of the $D^{\rm c}$ couplings, exactly as in Fig.~\ref{fig:flot_glob_c},  so that one can read the times at which the intermediate approximate SU$(N)$ regime is reached, and at which the transition to the SU(2) fixed direction takes place. The flows of the biggest response function in each of the four channels (particle-particle/particle-hole and singlet/triplet) are represented in Fig.~\ref{fig:flot_rep}$(b)$. We clearly see that during the intermediate regime, the biggest response fuction is in the particle-hole, triplet channel, in other word a spin density wave. We have checked that it corresponds to an approximately uniform ``s-wave'' type response function. It would be exactly uniform only if the SU$(N)$ symmetry was exactly realized. After the transition to SU(2), this response function saturates while the particle-particle, singlet response function grows and becomes bigger than the former. We have also checked that it corresponds to a ``d-wave'' type response function. We have shown in Fig.~\ref{fig:flot_rep}$(c)$ a zoom of Fig.~\ref{fig:flot_rep}$(b)$ so that one can see the initial competition between the four channels. Finally, Fig.~\ref{fig:flot_rep}$(d)$ shows the competition between two superconducting response functions. The first one (represented with circles) is of the ``s-wave'' type and is the biggest in the beginning of the flow, while the second one (represented with squares) is the ``d-wave'' type response function we have previously already mentioned, that becomes the biggest after a while. Even if those two response functions are much smaller than the spin density wave one, it is interesting to see how in their competition, the ``hot spots'' already manifest themselves in the intermediate regime.

\subsection{PHYSICAL PICTURE FOR THE FIXED DIRECTIONS WITH SU$(N)$ SYMMETRY}
\label{sec:phys_pict}
In this section, we shall attempt to discuss some properties of these fixed directions which are not easily grasped within the perturbative RG approach followed in most of this work. For instance, it has been known for several decades that many electronic systems exhibit spin and charge decoupling in one dimension. In the present case, the transverse momentum is an additional degree of freedom which we shall denote by ``color'' in this discussion. A natural question is therefore~: do we have decoupled charge, spin and color excitations in these systems ? It is useful in this respect to introduce charge, spin and color currents, which may be defined as~:
\begin{equation}
\left\lbrace \begin{array}{l}
\rho_{\rm R}(q)=\sum_{k,I,\lambda} c^\dagger_{{\rm R},I,\lambda}(k+q) c_{{\rm R},I,\lambda}(k) \mbox{ for charge}\\
S^a_{\rm R}(q)=\sum_{k,I,\lambda,\mu} c^\dagger_{{\rm R},I,\lambda}(k+q) \sigma^a_{\lambda \mu}c_{{\rm R},I,\mu}(k) \mbox{ for spin}\\
J^{IJ}_{\rm R}(q)=\sum_{k,\lambda} c^\dagger_{{\rm R},I,\lambda}(k+q) c_{{\rm R},J,\lambda}(k) \mbox{ for color}
\end{array}\right..
\end{equation}
As it is well known, it is possible to express the kinetic energy part $H_0$ of the system Hamiltonian as a sum of three quadratic terms in these currents~:\cite{Tsvelik} $H_0=H_{0,{\rm R}}+H_{0,{\rm L}}$, with~:
\begin{widetext}
\begin{eqnarray}
H_{0,{\rm R}}=\frac{\pi v_{\rm F}}{MN}\sum_{q\in \frac{2\pi}{L}\mathbb{Z}}~:\rho_{\rm R}(q)\rho_{\rm R}(-q): + \frac{\pi v_{\rm F}}{2(M+N)}\sum_{q,a} :S^a_{\rm R}(q) S^a_{\rm R}(-q): \nonumber\\
+\frac{\pi v_{\rm F}}{M+N}\sum_{q,I,J} \left\lbrace :J^{IJ}_{\rm R}(q)J^{JI}_{\rm R}(-q):-\frac{1}{N}:J^{II}_{\rm R}(q)J^{JJ}_{\rm R}(-q):\right\rbrace\;.
\end{eqnarray}
\end{widetext}
For the right moving branch, normal-ordered products are defined according to $:A(q)B(q'):=A(q)B(q')$ if $q\geq q'$ and $B(q')A(q)$ if $q<q'$. Note that similar expressions may be written for the left moving branch. We have introduced $M=2$ for SU$(2)$ spins, to illustrate what would happen for SU$(M)$ spins. It is important to note that each term in $H_{0,{\rm R}}$ commutes with the other two terms. This property is often stated as charge-spin-color decoupling. Now, the most general interaction which preserves SU$(2)\times$SU$(N)$ symmetry and this decoupling, may be written as~:
\begin{widetext}
\begin{equation}
H_{\rm int} = g_{\rm charge} \sum_q \rho_{\rm L}(-q)\rho_{\rm R}(q)+g_{\rm spin} \sum_{q,a} S^a_{\rm L}(-q) S^a_{\rm R}(q) + g_{\rm color} \sum_{q,I,J} \left\lbrace J^{IJ}_{\rm L}(-q)J^{JI}_{\rm R}(q)-\frac{1}{N} J^{II}_{\rm L}(-q)J^{JJ}_{\rm R}(q)\right\rbrace\;.
\end{equation}
\end{widetext}
These interaction terms may be cast into the parametrization used in this section, and the precise mapping reads~:
\begin{equation}
\left\lbrace \begin{array}{l}
C^{\rm c}-C^{\rm s}=g_{\rm charge}-\frac{1}{N} g_{\rm color}-g_{\rm spin}\\
D^{\rm c}-D^{\rm s}=g_{\rm color}\\
C^{\rm s}=g_{\rm spin}\\
D^{\rm s}=0
\end{array}\right..
\end{equation}
So the fixed directions with SU$(2)\times$SU$(N)$ symmetry discussed here cannot be written in this charge-spin-color decoupled form, since $D^{\rm s}\neq 0$ for both (+,+) and (+,-) fixed directions, as seen on Table \ref{tab:dirfixeN}. We do have a weaker form of decoupling, where the charge bosons $\rho_{{\rm RL}}(q)$ decouple and remain in a gapless Luttinger liquid state. But spin and color sectors are strongly mixed. Actually the (+,+) direction exhibits a larger SU$(2N)$ symmetry than the expected SU$(2)\times$SU$(N)$. To show this, let us define generalized SU$(2N)$ currents by~: 
\begin{equation}
E_{\rm R}^{I,\lambda;J,\mu}(q)=\sum_k c^\dagger_{{\rm R},I,\lambda}(k+q) c_{{\rm R},J,\mu}(k)\;,
\end{equation}
and let us consider the U$(1)\times$SU$(2N)$ invariant interaction~:
\begin{widetext}
\begin{equation}
H_{\rm int}=g_{\rm charge} \sum_q \rho_{\rm L}(-q)\rho_{\rm R}(q)+\tilde{g} \sum_{q,I,J,\lambda,\mu}\left\lbrace E_{\rm L}^{I,\lambda;J,\mu}(-q)E_{\rm R}^{J,\mu;I,\lambda}(q)-\frac{1}{2N}E_{\rm L}^{I,\lambda;I,\lambda}(-q)E_{\rm R}^{J,\mu;J,\mu}(q)\right\rbrace\;.
\end{equation}
\end{widetext}
It is easy to show that such an interaction Hamiltonian corresponds to the following values of the charge and spin couplings~:
\begin{equation}
\left\lbrace \begin{array}{l}
C^{\rm c}-C^{\rm s}=g_{\rm charge}-\frac{1}{2N} \tilde{g}\\
D^{\rm c}-D^{\rm s}=0\\
C^{\rm s}=0\\
2D^{\rm s}=\tilde{g}
\end{array}\right..
\end{equation}
From Table \ref{tab:dirfixeN}, it is clear that the (+,+) direction can be written this way, with $g_{\rm charge}=0$ and $\tilde{g}=1/2$. This model is simply the SU$(2N)$ Chiral-invariant Gross-Neveu model, which was diagonalized by Andrei and Lowenstein\cite{Andrei79} using the Bethe Ansatz for many-body wavefunctions. From their results, a rather simple picture emerges, involving a massive SU$(2N)$ ``spin'' sector, completely decoupled from a single massless bosonic mode (the charge mode already discussed). In the $N\to\infty$ limit, this physical picture evolves smoothly towards a mean-field description, characterized by the appearance of a non-vanishing expectation value for the charge density wave operator $\sum_k c^\dagger_{{\rm R},I,\lambda}(k+k_{\rm F}) c_{{\rm L},I,\lambda}(k-k_{\rm F})$. This mean-field shows itself in the behavior of the particle-hole response function in the singlet sector, which diverges near the instability with an exponent equal to $1-1/(2N)^2$ (see Table \ref{tab:exposants}). As $N$ goes to infinity, the mean-field (or RPA) value 1 is indeed recovered. For large but finite $N$, long wavelength fluctuations of the order parameter prevent a true spontaneous symmetry breaking. However, as discussed long ago by several authors,\cite{Wiegmann77,Witten78} some qualitative conclusions of the mean-field analysis are correct, as for example the prediction of a non-vanishing energy gap in the single-electron spectral function. Note that the original electrons are no longer the fundamental excitations, and they split into a massless charge part and a massive neutral spinful part, so their spectral function does not exhibit a sharp quasi-particle pole.

Let us now focus on the (+,-) fixed directions. As hinted by Eq.~\ref{eq:Hint_schematic}, the corresponding interaction Hamiltonian is of the form~: 
\begin{eqnarray}
H_{\rm int}=g\int_0^L {\rm d}x :\psi^\dagger_{{\rm R},I,\lambda}(x)\sigma^a_{\lambda \lambda'} \psi_{{\rm L},I,\lambda'}(x) \hspace{1cm}\nonumber\\
\times \psi^\dagger_{{\rm L},J,\mu}(x)\sigma^a_{\mu \mu'} \psi_{{\rm R},J,\mu'}(x):\;.
\end{eqnarray}
In a mean-field picture, the corresponding order parameter is $\xi^a(x)=<\psi^\dagger_{{\rm R},I,\lambda}(x)\sigma^a_{\lambda \lambda'} \psi_{{\rm L},I,\lambda'}(x)>$. $\boldsymbol{\xi}(x)$ is a vector with three complex coordinates. It transforms under the SO$(3)$ rotation group corresponding to global rotations in spin space. Furthermore, under a chiral transformation which changes $\psi^\dagger_{{\rm R},I,\lambda}$ into $\exp(i\theta /2)\psi^\dagger_{{\rm R},I,\lambda}$ and $\psi^\dagger_{{\rm L},I,\lambda}$ into $\exp(-i\theta /2)\psi^\dagger_{{\rm L},I,\lambda}$, $\boldsymbol{\xi}(x)$ is replaced by $\exp(i\theta)\boldsymbol{\xi}(x)$. Note that a global chiral transformation corresponds roughly to a uniform translation of the spin density wave condensate. So the mean-field approximation breaks two independent continuous symmetries of the underlying Hamiltonian. It is therefore natural to derive an effective action for the long wavelength fluctuations of the order parameter $\boldsymbol{\xi}(x)$. This problem has been addressed by several authors in the context of dynamical properties of spin density waves. For a recent source, with several references to earlier work, see for instance the article by Sengupta and Dupuis.\cite{Sengupta00} In order to keep the presentation simple, we shall simply quote the results obtained in the long wavelength approximation, keeping only massless modes. It turns out that these modes correspond to order parameter configurations of the form $\boldsymbol{\xi}(x)=\rho \exp(i\theta(x))\boldsymbol{n}(x)$, where $\rho$ is a positive number found by solving mean-field equations, $\theta(x)$ is the ``phason'' field, and $\boldsymbol{n}(x)$ is a real unit vector, giving the direction of the staggered magnetization. Integrating out fermions provides the following effective action~:
\begin{eqnarray}
S_{\rm eff}(\theta,\boldsymbol{n})=\frac{N}{4\pi}\int {\rm d}x{\rm d}t \left(\partial_\mu \theta \partial^\mu \theta +\partial_\mu \boldsymbol{n} \partial^\mu \boldsymbol{n}\right) \\
+ \mbox{ less relevant terms.}\nonumber
\end{eqnarray}
As expected, the phason field remains gapless, and its fluctuations preserve the chiral symmetry of the system at low energy. Now, the magnetic local order parameter will also remain in a symmetric phase with respect to spin rotations. However, by contrast to the phason, it will develop a spectral gap $M\sim m\exp(-N)$, where $m$ is the fermion mass gap (see for instance Polyakov's book,\cite{Polyakov} chap. 2). $M$ is therefore a spin gap. Its appearance cannot be detected in the simple perturbative approach in terms of the original fermions. Indeed, the correlation functions computed in Sec.~\ref{sec:reponse} above were diverging at the typical energy scale $m$ at which particle-hole bound states occur. Clearly, our one-loop fermionic RG misses long wavelength fluctuations of these composite bosons completely. Coming to real systems, we see that this spin gap is expected to vanish quickly in the $N\to\infty$ limit, which again has to be taken in order to describe two-dimensional systems with a continuous Fermi surface (this exponential variation of the spin gap with $N$ is very similar to the behavior found for an even number of weakly coupled antiferromagnetic spin chains; see for instance the work by Chakravarty\cite{Chakravarty96}). To conclude this analysis, we emphasize that most physical properties of the (+,-) fixed directions are well captured, as $N$ becomes infinite, by a standard mean-field treatment of the Peierls instability.

\section{STABILITY ANALYSIS OF THE SYMMETRICAL FIXED DIRECTIONS}
\label{sec:stability}
\subsection{LINEAR STABILITY ANALYSIS}
\label{sec:stab_lin}
In this section, we shall study the stability of the SU$(N)$ and SU$(2)$ fixed directions studied in the previous section. We have already studied stability questions within the general setting of the RPA in Sec.~\ref{sec:RPA}. This stability analysis was performed around the fixed directions obtained with a small contribution of the Cooper channel (remember we introduced a small $\varepsilon$ as a way to tune the influence of the Cooper channel with respect to the Peierls one). Here we directly study the stability of the SU$(N)$ and SU$(2)$ fixed directions, for which no assumption about the smallness of the Cooper channel is made. The very possibility to compute the eigenvalues in a linear stability analysis in these cases comes from the gigantic simplifications gained from the symmetries. We recall that with the general notations (see Sec.~\ref{sec:RPA}), the eigenvalues are those of the matrix $B$ (Eq.~\ref{eq:eqRGlin_et_matriceB}).

We shall briefly explain how to compute the eigenvalues and their degeneracies, in the SU$(N)$ spinless case only. The other cases are either much simpler (SU$(2)$), or the algebra is much heavier but the ideas just as simple (SU$(N)$ with spin). The fixed direction only involves the principal couplings $A$, $C$ and $D$. It can be shown that a perturbation in the non-principal couplings is irrelevant, so that we will restrict ourselves to the principal couplings. We set $D(I,J)=D/(t_{\rm c}-t)+\delta D(I,J)$, where $D=1$ is the value of the $D$-couplings on the fixed direction (see Table \ref{tab:dirfixeN}). Doing the same with the $A$'s and $C$'s, linearizing Eq.~(\ref{eq:eqRGchaines_spinless}) (it would be Eq.~(\ref{eq:eqRGchaines}) in the case with spin), and writing down the eigenvalue problem yields~:
\begin{equation}
\left\lbrace
\begin{array}{l}
\lambda \delta A(I)=\frac{2D}{N} S(I)\\
\lambda \delta C(I,J)=-\frac{2D}{N} \delta D(I,J)\\
\lambda \delta D(I,J)= \frac{D}{N}[S(I)+S(J)\\
\hspace{3cm}+\delta A(I)+\delta A(J)-2 \delta C(I,J)]
\end{array}
\right.,
\end{equation}
with the definition $S(I)=\sum_{K\neq I} \delta D(I,K)$. It is first obvious that $\lambda=0$ is a $N$ times degenerate solution (we can arbitrarily fix the $\delta A(I)$'s). If $\lambda\neq 0$, the first two equations define $\delta A(I)$ and $\delta C(I,J)$ as functions of $\delta D(I,J)$, so that the third equation can be rewritten solely with the $D$'s~:
\begin{eqnarray}
(\lambda-\frac{2D}{N})(\lambda+\frac{2D}{N}) \delta D(I,J)=\hspace{3cm}\nonumber\\
\frac{D}{N}(\lambda+\frac{2D}{N}) (S(I)+S(J))\;.\hspace{0.5cm}
\end{eqnarray}
It is now clear that either $\lambda=-\frac{2D}{N}$, which is satisfied for any $\delta D(I,J)$ so that it is $N(N-1)/2$ times degenerate, or $\lambda\neq -\frac{2D}{N}$, in which case one can further simplify the problem in~: $(\lambda-\frac{2D}{N}) \delta D(I,J)=D(S(I)+S(J))/N$. $\lambda=\frac{2D}{N}$ is a solution, provided $S(I)=0$ for any $I$ (this is true because $N\geq 3$). The restrictions imposed on the $\delta D(I,J)$ imply the degeneracy is $N(N-3)/2$. Let us now suppose $\lambda$ is not equal to any of the eigenvalues we already found. We have to solve an equation of the type $f(\lambda)\delta D(I,J)=g(\lambda) (S(I)+S(J))$, so that the $S(I)$'s completely determine the $\delta D(I,J)$. Here $f$ and $g$ are very simple functions of $\lambda$ ($f(\lambda)=\lambda-2D/N$, and $g=D/N$ does not even depend on $\lambda$), but in the general case with spin, one is also confronted to such a problem, that is why we keep quite general arguments. Summing this equation first over $J$, then over $I$, and noting $S=\sum_I S(I)$, we get two compatibility conditions~:
\begin{eqnarray}
{[f(\lambda)-(N-2)g(\lambda)]}S(I)&=&g(\lambda)S, \mbox{ and}\\
{[f(\lambda)-2(N-1)g(\lambda)]}S&=&0.
\end{eqnarray}
These equations, in the spinless case read $(\lambda-D)S(I)=DS/N$ and $(\lambda-2D)S=0$. Thus, either $\lambda=D$ with the condition $S=0$, that is the degeneracy is $(N-1)$, corresponding to the free choice of $(N-1)$ variables $S(I)$, or $\lambda\neq D$ which automatically implies $\lambda=2D$ (otherwise we get a zero eigenvector), and the degeneracy is 1 because once $S$ is given, everything is known. As $D=1$, we thus have found $(N-1)$ eigenvalues equal to 1, and 1 eigenvalue equal to 2. This is in agreement with the results previously obtained in Sec.~\ref{sec:RPA}. Notice that we find exactly the eigenvalue 1, whereas in Sec.~\ref{sec:RPA}, we found $(N-1)$ eigenvalues $1-4\varepsilon/N^2$. This does not invalidate those previous results, because they were obtained in a perturbation expansion in $\varepsilon$, which is in fact an expansion in $\varepsilon^{n_1}/N^{n_2}$ with $n_2\geq n_1$. Thus our exact result simply shows that the $(n_1=1,n_2=2)$ term we calculated will be compensated by the $(n_1=2,n_2=2)$ term, once $\varepsilon$ is set equal to 1. Let us also notice that all the other eigenvalues we calculated exactly are either 0, or tend to 0 in the $N\to\infty$ limit, as we expect from the RPA results, where only eigenvalues equal to 0, 1 and 2 appeared. We furthermore recover the degeneracies predicted in the RPA calculation, once the infinite $N$ limit has been taken.

We list all the eigenvalues and their degeneracies we found for all six fixed directions, in Table \ref{tab:eigenvalues}.
\begin{table*}
\caption{List of the eigenvalues for the six fixed directions. We give them in the form (eigenvalue, degeneracy). Notice that all the results are exact, except for the SU$(N)$, (+,-) case, where we give a $1/N$ expansion of the results.}
\label{tab:eigenvalues}
\begin{tabular}{|c|c|c|c|c|c|}
\hline
SU$(N)$ & SU$(N)$ & SU$(N)$ & SU$(2)$ & SU$(2)$ & SU$(2)$ \\ 
spinless & (+,+) & (+,-) & spinless & (+,+) & (+,-) \\ \hline
$(-2/N,N(N-1)/2)$ & $(-1/N,N(N-1))$ & $(-3/N,N(N-1)/2)$ & $(-1,1)$ & $(-1/2,2)$ & $(-1/2,2)$ \\ \hline
$(0,N)$ & $(0,N)$ & $(-\sqrt{3}/N,N)$ & $(0,2)$ & $(0,2)$ & $(0,3)$ \\ \hline
$(2/N,N(N-3)/2)$ & $(1/N,N(N-1))$ & $(-1/N,N(N-1)/2)$ & & $(1/2,2)$ & $(1/2,2)$ \\ \hline
& & $(0,N)$ & & & \\ \hline
& & $(1/N,N(N-1)/2)$ & & & \\ \hline
& & $(\sqrt{3}/N,N)$ & & & \\ \hline
& & $(3/N,N(N-1)/2)$ & & & \\ \hline
$(1,N-1)$ & $(1,N-1)$ & $(1+2/N^2,N-1)$ & & $(1,1)$ & \\ \hline
$(2,1)$ & $(2,1)$ & $(2,1)$ & $(2,1)$ & $(2,1)$ & $(2,1)$ \\ \hline
\end{tabular}
\end{table*}
In agreement with the discussion in Sec.~\ref{sec:phys_pict}, the SU$(N)$ (+,+) direction is very similar to the spinless case with SU$(2N)$ symmetry. It is clear that both situations lead to the same eigenvalues. The degeneracies are not identical since in the spinful case, we always enforce the SU$(2)$ spin symmetry of the pair interaction, which has no counterpart for the SU$(2N)$ spinless problem. It should be noticed that in the SU$(N)$, (+,-) case, we do not give exact results, but a $1/N$ expansion of the eigenvalues. We only give the lowest order term in this expansion, except for the eigenvalue approaching 1 in the infinite $N$ limit, because this eigenvalue $\lambda\simeq 1+2/N^2$ is greater than 1 for any finite $N$. This explains the instability of the spin density wave fixed direction, that one observes in the numerical simulation. Furthermore, the absence of eigenvalue greater or equal to 1 in the SU$(2)$, (+,-) case, shows that this fixed direction is stable, which is confirmed by the numerical results, since after leaving the spin density wave regime, the flow ends in the ``hot spots'' regime. Let us finally notice that the $\lambda=1+2/N^2+O(1/N^3)$ eigenvalue is also compatible with the RPA result $\lambda^{\rm RPA}\simeq 1-4\varepsilon/N^2$, in the sense the $1/N$ term is zero in both results. The difference in the $1/N^2$ terms comes, as in the spinless case, from the $(\varepsilon/N)^2$ term that we have not computed.

\subsection{SIMPLE INTERPRETATION OF THE $\lambda=1$ EIGENVALUES}
We have already interpreted the occurence of $\lambda=1$ (or $\lambda=1+O(1/N)$) thanks to the results of Sec.~\ref{sec:RPA}. We will now show that these eigenvalues simply correspond to a lowering of symmetry, from SU$(N)$ to SU$(N-1)$. In other words, it simply corresponds to the possibility of ``turning off'' one chain among $N$. 

To see this, it is possible to compute the eigenvectors associated to these eigenvalues, and see their influence on the fixed direction. This can be checked to confirm our statement. We here give a much simpler proof, restricting ourselves to the spinless case. Let us suppose we isolate the chain whose number is $I_0$. We assume a permutation symmetry among the $N-1$ other chains. We still denote by $A=A(I)$, $C=C(I,J)$ and $D=D(I,J)$ the couplings involving only the $N-1$ equivalent chains ($I,J\neq I_0$), and we now denote $a=A(I_0)$, $c=C(I_0,J)$ and $d=D(I_0,J)$. Eq.~(\ref{eq:eqRGchaines_spinless}) then takes the form~:
\begin{equation}
\label{eq:eqRGchaineI_0}
\left\lbrace
\begin{array}{ll}
N\partial_t A&=(N-2)D^2+d^2 \\
N\partial_t a&=(N-1)d^2 \\
N\partial_t C&=-D^2 \\
N\partial_t c&=-d^2 \\
N\partial_t D&=(N-3)D^2+d^2+2(A-C)D \\
N\partial_t d&=(N-2)dD+d(a+A-2c)
\end{array}
\right..
\end{equation}
These equations admit a few fixed directions. Among them, we of course have the SU$(N)$ fixed direction, with $a=A=1-1/N$, and so on. We have the SU$(N-1)$ fixed direction with $a=c=d=0$ and $A=1-1/(N-1)$, etc, for which the chain $I_0$ is switched off. We in fact have one more SU$(N-1)$ fixed direction, with $A$, $C$, $D$ taking their SU$(N)$ values $1-1/N$,$-1/N$ and $1$, and $a=A$, $c=C$ and $d=-D$. This comes from the symmetry of Eq.~(\ref{eq:eqRGchaineI_0}) under $d\to -d$. It is also reminiscent of the RPA discussion of Sec.~\ref{sec:RPA}, where the components $\phi(I)$ were only fixed in absolute value, their signs being free.

If we study the linear stability of the SU$(N)$ fixed direction, we get the following six eigenvalues~: $-2/N$ twice, $0$ twice, $1$ once and $2$ once. These eigenvalues are in perfect agreement with what we found in the previous section (see Table \ref{tab:eigenvalues}). In the $(A,a,C,c,D,d)$ basis, the $\lambda=1$ eigenvector reads~: 
\begin{eqnarray}
\boldsymbol{v}&=&(v_A,v_a,v_C,v_c,v_D,v_d)\\
&=&\left(-\frac{2}{N},\frac{2(N-1)}{N},\frac{4}{N(N-2)},-\frac{2}{N},-\frac{2}{N-2},1\right)\nonumber\;. 
\end{eqnarray}
It is straightforward to check that $v_A=v_C+v_D$ and $v_a\neq v_c+v_d$. Thus, if we add the vector defining the SU$(N)$ fixed direction and $\varepsilon \boldsymbol{v}$, with arbitrary $\varepsilon$, the SU$(N-1)$ symmetry will be maintained while the SU$(N)$ symmetry will be broken. Of course we cannot obtain the vector defining the SU$(N-1)$ fixed direction simply by adjusting $\varepsilon$, because the passage from SU$(N)$ to SU$(N-1)$ symmetry is non-perturbative. We can however tell $\boldsymbol{v}$ does not connect the SU$(N)$ fixed direction to the second SU$(N-1)$ one (with $a=A$, $c=C$, $d=-D$), because from the coefficients of $\boldsymbol{v}$, we see that a change in $d$ implies a change in $a$ of the same order of magnitude. Therefore, we are led to conclude that a perturbation along a $\lambda=1$ eigenvector should lower the symmetry, in a case where this eigenvector is marginally relevant. In order to confirm this picture, we have to go beyond linear perturbation theory which only tells us the $\lambda=1$ eigenvectors are marginal.

\subsection{NON-LINEAR STABILITY ANALYSIS}
\label{sec:nl_stab_an}
\begin{figure}[h]
\includegraphics[width=8cm]{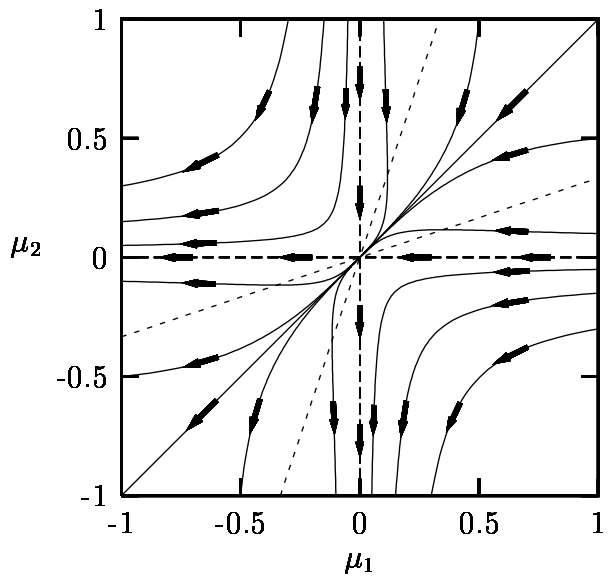}
\caption{Structure of the flow for a perturbation along two marginal eigenvectors. The arrows indicate the direction of the flow along the trajectories. The dashed lines (with no arrows) are the lines of equation $\mu_2=p \mu_1$ and $\mu_2=\mu_1$/p.}
\label{fig:NL}
\end{figure}

It is not possible to simply use Eq.~(\ref{eq:eqRGchaineI_0}) in order to perform a non-linear stability analysis. Indeed, as our problem is non-linear, the effect of a perturbation along two $\lambda=1$ eigenvectors (that lower the symmetry) cannot be determined by linear superposition. But neither are we forced to use the full RG Eq.~(\ref{eq:eqRGchaines_spinless}) and perturb the fixed direction by all $(N-1)$ marginal eigenvectors. From the symmetry between the chains, it suffices to restrict the perturbation along two eigenvectors. Let us demonstrate this with the general notations introduced in Sec.~\ref{sec:RPA}. We will assume that we only perturb the fixed direction in the marginal subspace ${\mathcal E}_1$. From the flow equation, the coupling vector will also aquire components along the other eigenvectors, but these will be supposed to be negligible, which is reasonable since they will go to zero on a time scale that is short compared to the slow evolution in ${\mathcal E}_1$. We thus suppose $g_i(t)=f(t)(u_i+V_i(t))$, with $\boldsymbol{V}\in {\mathcal E}_1$. We now plug this in the flow equation $\partial_t g_i=A_{ijk}g_j g_k$ and get~:
\begin{equation}
\partial_s V_i={\mathcal P}_1\left[ A_{ijk} V_j(s) V_k(s)\right]\;,
\end{equation}
where $s(t)=\ln [t_{\rm c}/(t_{\rm c}-t)]$ is the time scale adapted to the slow motion in ${\mathcal E}_1$, and where ${\mathcal P}_1$ is the projector on ${\mathcal E}_1$. It should be noticed that if the RHS vanishes, our approach is not valid anymore, and one has to take the influence of the terms we neglected into account. If we now decompose $V_i(s)=\sum_\alpha \mu_\alpha(s)v_i^{(\alpha)}$, where $\alpha$ runs from 1 to $N-1$, and where the $\boldsymbol{v}$'s are marginal eigenvectors that lower the symmetry from SU$(N)$ to SU$(N-1)$, we get~:
\begin{equation}
\label{eq:eq_flot_mu}
\partial_s \mu_\alpha=\sum_{\beta} A_\alpha^{\beta} \mu_\beta^2(s)+\sum_{\gamma>\beta} B_\alpha^{\beta\gamma} \mu_\beta (s) \mu_\gamma (s)\;.  
\end{equation}
In this equation, the coefficients $A$ and $B$ are defined by ${\mathcal P}_1[A_{ijk} v_j^{(\beta)} v_k^{(\beta)}]=\sum_\alpha A_\alpha^{\beta} v_i^{(\alpha)}$, and ${\mathcal P}_1[(A_{ijk} +A_{ikj})v_j^{(\beta)} v_k^{(\gamma)}]=\sum_\alpha B_\alpha^{\beta\gamma} v_i^{(\alpha)}$. We know from the Appendix that RG flows cannot break a symmetry. This imposes severe conditions on $A$ and $B$. If one perturbs the fixed direction along only one eigenvector ($\boldsymbol{v}^{(\beta)}$), that is one singles out only one chain, the remaining SU$(N-1)$ symmetry must be conserved, thus $A_\alpha^{\beta}=a \delta_{\beta,\alpha}$. $a$ does not depend on $\alpha$ because of the permutation symmetry between the chains. If one now perturbs the fixed direction along two eigenvectors ($\boldsymbol{v}^{(\beta)}$ and $\boldsymbol{v}^{(\gamma)}$), and with the same perturbation strength along each vector, the remaining SU$(N-2)$ symmetry as well as the permutation symmetry among the two vectors are conserved, so that $B_\alpha^{\beta\gamma}=b(\delta_{\beta,\alpha}+\delta_{\gamma,\alpha})$. Again, $b$ does not depend on $\alpha$ because of the permutation symmetry between the chains. Eq.~(\ref{eq:eq_flot_mu}) now takes the much simpler form~: 
\begin{equation}
\label{eq:eq_flot_mu2}
\partial_s \mu_\alpha=a \mu_\alpha^2 + b \sum_{\beta\neq\alpha} \mu_\alpha\mu_\beta.
\end{equation}
We have thus demonstrated that the non-linear evolution near the SU$(N)$ fixed direction is captured by only two numbers, $a$ and $b$, which can most easily be determined by restricting oneself to two marginal eigenvectors, that is generalizing Eq.~(\ref{eq:eqRGchaineI_0}), with two singularized chains $I_1$ and $I_2$. This tremendously reduces the dimension of the space we have to work with, since it is 11 in the spinless case, and 22 in the spin 1/2 case. The results are given in Table \ref{tab:a_&_b}.
\begin{table}
\caption{$a$ and $b$ coefficients in the spinless and spin 1/2, (+,+) cases.}
\label{tab:a_&_b}
\begin{tabular}{|c|c|c|}
\hline
& SU$(N)$ & SU$(N)$ \\
& spinless & (+,+) \\ \hline
$a$ & $-\frac{4}{(N-2)(N+2)}$ & $-\frac{4}{(N-1)(N+1)}$ \\ \hline
$b$ & $\frac{8}{(N-2)^2(N+2)}$ & $\frac{8}{(N-2)(N-1)(N+1)}$ \\ \hline
\end{tabular}
\end{table}
We see that there is no big difference between the spinless and spinning case. What is important is that in both cases, $a$ is negative and behaves as $1/N^2$ when $N$ is large, and $b$ is positive and behaves as $1/N^3$. Let us see what happens for a perturbation along two eigenvectors. Eq.~(\ref{eq:eq_flot_mu2}) gives~: $\partial_s \mu_1=b \mu_1 (\mu_2-p \mu_1)$ and $\partial_s \mu_2=b \mu_2 (\mu_1-p \mu_2)$, with $p=-a/b$ ($p>1$ for $N>4$). We have performed a numerical simulation of this flow (for the spinless values of $a$ and $b$, and for $N=8$), and the result is shown on Fig.~\ref{fig:NL}. 
The dashed lines (with no arrows) are the lines of equation $\mu_2=p \mu_1$ and $\mu_2=\mu_1$/p. The first of these lines determines the set of points where the RHS of the flow equation for $\mu_1$ vanishes. It thus delimitates two areas, one of growing $\mu_1$ (under this line), and one of decreasing $\mu_1$ (above this line). The same analysis applies for the second line (symmetric of the first one, with respect to the first bisector), with $\mu_1$ replaced by $\mu_2$. The arrows give the direction of the flow on the trajectories. It is important to notice that the point $(\mu_1=0,\mu_2=0)$, corresponding to the SU$(N)$ symmetry, is a fixed point of the flow. The main consequence of our analysis is to show that this fixed point is unstable, except for the case where both $\mu_1$ and $\mu_2$ are positive. If one had taken a perturbation along all marginal eigenvectors, the stability condition would have been that all $N-1$ $\mu$'s should be positive. The relative size of the stable subset thus goes to zero as $1/2^N$, so that in the large $N$ limit, nearly all initial conditions lead to an instability of the SU$(N)$ fixed direction. 

At this point, the reader would be right to point the fact that we said in Sec.~\ref{subsec:numres2D}, that the SU$(N)$ (+,+) fixed direction was numerically stable, and that our analytical results seem to refute the numerics. In fact, this fixed direction {\em is found to be unstable, both analytically and numerically}. At the beginning of this article, we didn't want to go into too much detail, so we did say the fixed direction was numerically stable. Anyway, this was not so much of a ``lie''. Indeed, before analytical results were available, we believed it was stable, because the numerical simulations didn't show any sign of instability, for RG times of the same order than the ones needed to see the (+,-) instability. But once the analytical study pointed to instability, we tried to work with much bigger RG times, and we indeed observed an instability. We should also mention that this instability could only be seen for quite small values of $N$, because otherwise the precision of the machine was not sufficient to detect it. Finally, let us mention the following point. In the case of the (+,-) fixed direction, the linear instability results in a direct transition between the SU$(N)$ and SU$(2)$ fixed directions. In the two other cases, the non-linear instability leads to a gradual elimination of pairs of chains, beginning by the centre (resp. the edges) for the spinless (resp. (+,+)) case. We shall not go further in the analysis of these transitions, and turn to our last section.

\section{SYMMETRY RESTORATION AND THE LARGE $N$ LIMIT}
\label{sec:symmetry_restoration}
The subject of symmetry restoration through RG flows is an interesting question, which several authors have already addressed.\cite{Lin98,Azaria98,Konik00,Arrigoni99,Schulz98} We shall here take advantage of what we have said before, and show some implications on symmetry restoration, in the large $N$ limit. By symmetry restoration, we mean that the vector of couplings flows towards a symmetric fixed direction. We will focus on the case of the anisotropic Gross-Neveu model, studied by Azaria et al\cite{Azaria98} and which is related to the physics of coupled chains at half filling. By contrast to the statements made by these authors, we believe that a careful analysis of one-loop RG flows is sufficient to indicate whether symmetry restoration is to be expected in the large $N$ limit.

Let us first set up our notations for this peculiar model (the general notations are the same as in the previous sections). We consider a 1D Gross-Neveu model, with $N$ types of fermions, numbered from $0$ to $N-1$. The 0-fermions are singled out in order to get an anisotropic Gross-Neveu model. We thus get three possible density-density interaction terms. Denoting the size of the system by $L$, we get the following Hamiltonian~:
\begin{eqnarray}
\label{eq:HGN}
H=\int_0^L {\rm d}x \Biggl\lbrace {\bar\psi}_0 \gamma^\mu i \partial_\mu \psi_0+ {\bar\psi}_a \gamma^\mu i \partial_\mu \psi_a \nonumber\\
+ \frac{\tilde{g}_0}{N} {({\bar\psi}_0 \psi_0)}^2 + \frac{\tilde{g}_c}{N} ({\bar\psi}_0 \psi_0)({\bar\psi}_a \psi_a) +  \frac{\tilde{g}}{N} {({\bar\psi}_a \psi_a)}^2 \Biggl\rbrace \;,
\end{eqnarray}
with $a=1,\ldots,N-1$ ($N\geq 2$), and where the sum over $a$ is to be understood. The $1/N$ factors have been included to yield a good thermodynamical limit. Setting $g=\tilde{g}/(2\pi)$ for the three coupling constants, we get the following one-loop RG equations~:\cite{Azaria98}
\begin{equation}
\label{eq:RG_GN}
\left\{
\begin{array}{l}
\partial_t g=-4\left( 1-\frac{2}{N}\right) g^2-\frac{1}{N}{g_c}^2\vspace{0.1cm}\\
\partial_t g_c=-2\left( 2-\frac{3}{N}\right)g g_c-\frac{2}{N}g_0 g_c\vspace{0.1cm}\\
\partial_t g_0=-(1-\frac{1}{N}){g_c}^2
\end{array}
\right.,
\end{equation}
where $t=\ln(\Lambda_0/\nu)$, as in Sec.~\ref{sec:gen_setting}. These equations admit a line of fixed points, given by $g=g_c=0$. We also find three fixed directions given by 
\begin{eqnarray}
\label{eq:dir_fix_GN_1}
&&(u,u_c,u_0)=\frac{N}{4(N-2)}(-1,0,0)\;,\\
\label{eq:dir_fix_GN_2}
&&(u,u_c,u_0)=\frac{N}{4(N-1)}(-1,\pm 2,-1)\;.
\end{eqnarray}
We shall only be interested in the last one, with $u_c=2u=2u_0$, corresponding to symmetric coupling constants involving all fermions. In the first case, $\psi_0$ fermions do not interact. We have performed a linear stability analysis around the symmetric fixed direction, exactly as is Sec.~\ref{sec:stab_lin}, and found the following eigenvalues for the matrix $B$~:
\begin{equation}
\label{eq:eigenvalues_GN}
2, \mbox{ } -\frac{1}{N-1}\sim -\frac{1}{N}, \mbox{ and } \frac{1-2/N}{1-1/N}\sim 1-\frac{1}{N}\;.
\end{equation}
Notice that for $N=4$, one gets the results already found by Lin.\cite{Lin00} We thus see that the symmetric fixed direction is locally stable, but it becomes marginal in the infinite $N$ limit, as the $\lambda\sim 1-1/N$ eigenvalue tends to 1 in this limit. This picture is the same as the one we developed in the previous sections, except it provides an example where an eigenvalue is strictly {\em smaller} than 1 when $N$ is finite, and goes to 1 when $N\to\infty$.

\begin{figure}[h]
\includegraphics[width=8cm]{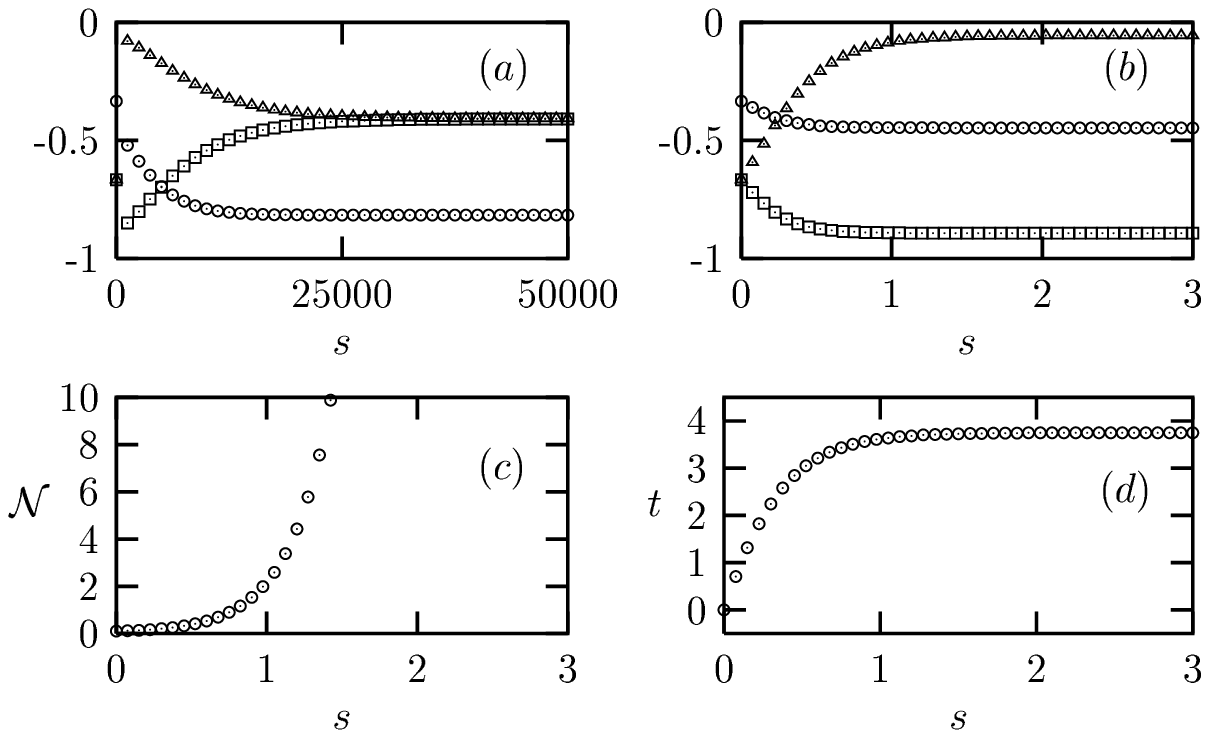}
\caption{RG flow of different quantities. $(a)$~: flow of the three couplings $h$, $h_c$, and $h_0$ represented respectively with squares, circles and triangles. $(b)$ is a zoom on short time $s$ of $(a)$. $(c)$ shows the evolution of the norm, and $(d)$ gives the relation between the two RG times $s$ and $t$.}
\label{fig:GN_fixed_direction}
\end{figure}

We shall consider the RG flows for two initial conditions. Let us first consider an initial condition, for which all couplings are negative and $g=2g_c=g_0$ (we have not set $h_c(0)=h(0)$ because in this case, $h_c \simeq h$ at the beginning of the flow, and this makes it difficult to read the figures). Notice that the sign of $g_c$ will only distinguish between the two fixed directions of Eq.~(\ref{eq:dir_fix_GN_2}), so that we can choose $g_c<0$. From the signs of the couplings and the RG flow equations, it is clear that the norm will quickly grow. The numerical flow in direction space has been computed with a large value of $N$ ($10^4$), and the results are shown on Fig.~\ref{fig:GN_fixed_direction}. Notice it is possible to use such large values of $N$ because here $N$ is only a parameter, that does not determine the dimension of the coupling space, which is always equal to 3.
One thus sees that in the very end of the flow (Fig.~\ref{fig:GN_fixed_direction}$(a)$), the symmetry is restored, since $g=g_0$ and $g_c=2g$. However, the physics will be given by the beginning of the flow. It is indeed clear from Fig.~\ref{fig:GN_fixed_direction}$(c)$ that the norm will have exploded before the symmetry is restored. In Fig.~\ref{fig:GN_fixed_direction}$(b)$, the couplings seem to be on a fixed direction. But we have seen there are only three fixed directions (Eqs.~(\ref{eq:dir_fix_GN_1}) and (\ref{eq:dir_fix_GN_2})). In fact, what we see here is a RPA like flow. Indeed, if one takes the infinite $N$ limit in Eq.~(\ref{eq:RG_GN}), one easily sees that we get an infinity of fixed directions, which will all be marginal. This is also confirmed by the presence of the $\lambda\simeq 1-1/N\to 1$ eigenvalue. This first example is very close, in spirit, to what we have dealt with in the previous sections of this article. 

\begin{figure}[h]
\includegraphics[width=8cm]{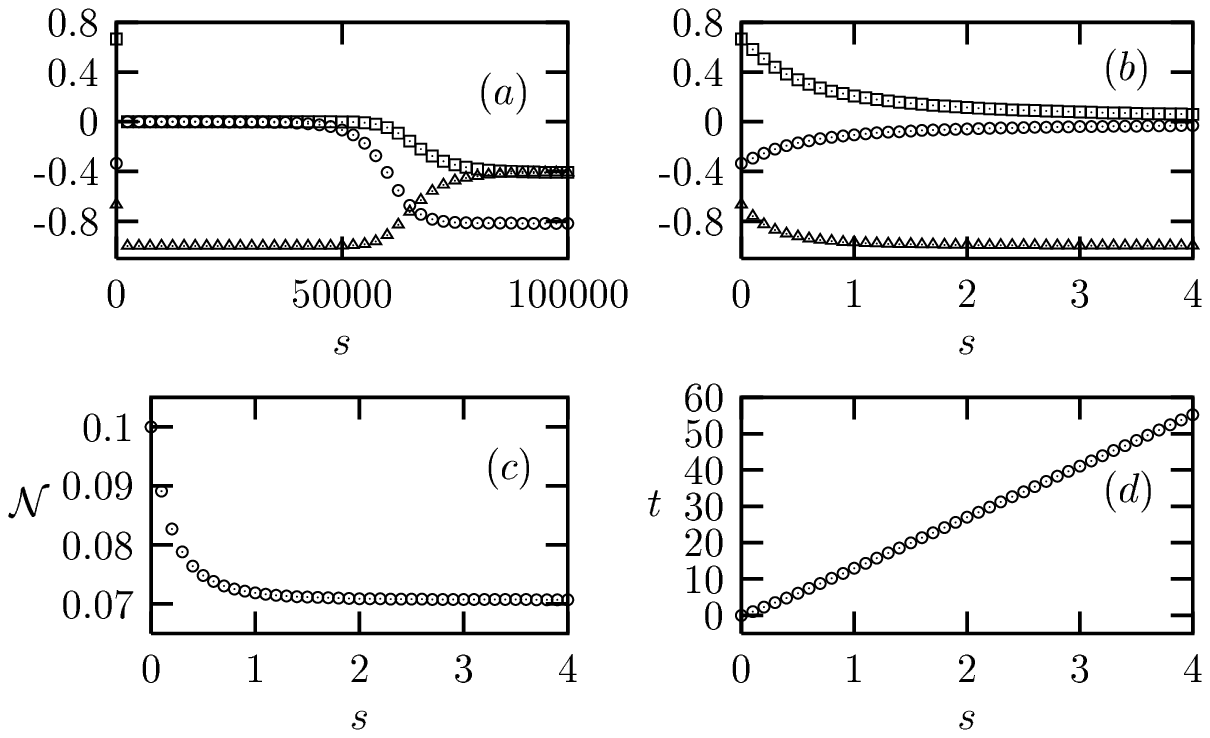}
\caption{RG flow of different quantities. $(a)$~: flow of the three couplings $h$, $h_c$, and $h_0$ represented respectively with squares, circles and triangles. $(b)$ is a zoom on short time $s$ of $(a)$. $(c)$ shows the evolution of the norm, and $(d)$ gives the relation between the two RG times $s$ and $t$.}
\label{fig:GN_fixed_point}
\end{figure}

Let us now consider another initial condition for which the flow passes in a close neighborhood of the line of fixed points. This is what Azaria et al have considered. We thus choose $g>0$ and $g_0<0$. The condition $g>0$ implies that the absolute values of both $g$ and $g_c$ will decrease in the early stages of the flow, and $g_0<0$ implies the instability of the intermediate fixed point. Again, the sign of $g_c$ is chosen to be negative. Precisely, in the space of normalized couplings, we set~: $h(0)=-2 h_c(0)=-h_0(0)$ (again we have not set $h_c(0)=h(0)$ to make the figures clear). The corresponding flows are shown on Fig.~\ref{fig:GN_fixed_point}.
One thus sees that in the very end of the flow (Fig.~\ref{fig:GN_fixed_point}$(a)$), the symmetry is again restored. However, one has to look carefully at what happens at the beginning of the flow. It is clear from Fig.~\ref{fig:GN_fixed_point}$(b)$ and $(c)$ that the flow gets close to the line of fixed points. It spends a very long time there ($s \lesssim 50000$). In the same time, $t$ grows linearly with $s$ (Fig.~\ref{fig:GN_fixed_point}$(d)$), with a proportionality coefficient given by the inverse of the norm of the couplings (remember from Sec.~\ref{sec:oneloopRGeq} that ${\rm d}s=\mathcal{N}(t) {\rm d}t$). This shows that the symmetry restoration will only take place at a very small energy scale (which goes to zero as $\exp (-N)$). Furthermore, we have seen that the symmetric fixed direction is locally stable, but that it becomes marginal in the $N\to\infty$ limit. Thus, the time needed to get close to this fixed direction, with a given accuracy, grows with $N$. But when the couplings are near this fixed direction, the norm grows nearly as fast as if the couplings were on the fixed direction, that is it explodes very quickly. We thus conclude that to observe a symmetry restoration (with a given accuracy) in the domain of validity of the one-loop RG, the couplings must be sent to zero as $N$ goes to infinity. In other words, if the initial couplings are finite, and $N$ is taken to infinity, the validity of the RG equations will break down before the symmetry is restored. We thus have found two indicators, {\em given by the RG itself}, showing that the symmetry restoration will not physically occur.

For both initial conditions we have chosen, we have seen that the RG restores the symmetry. But this happens in a regime where the pertubation theory is not valid anymore (and in the very far infra red for the second initial condition). What is important is that the RG itself tells us not to believe in the symmetry restoration.

\section{CONCLUSION}
In this article, we have revisited the old problem of interacting electrons with a short range repulsion in the weak coupling limit, but with a perfectly nested Fermi surface in two dimensions. Whereas a mean-field analysis predicts a spin density wave instability at low enough temperature, we have focused our study on the competition between this instability and the onset of superconductivity, which is known as the ``parquet'' problem. In purely one-dimensional sytems, this competition  is a key ingredient in stabilizing the famous Luttinger liquid phase. In two dimensions, a careful analysis\cite{Zheleznyak97} has shown that after a high-energy transient regime, the dominant effective couplings indeed correspond to a spin density wave state. According to this important  work, the main effect of the competition between various instabilities is merely to lower the value of the critical temperature, in comparison to the mean-field estimate. In the present paper, extending our previous numerical study of RG flow equations,\cite{VdA01} we have examined in greater detail the evolution of the effective couplins in the vicinity of the critical energy scale associated to the onset of long-range order. By contrast to the results of Zheleznyak et al,\cite{Zheleznyak97} we find a third regime, in which the spin density wave instability slowly gives way to a d-wave superconductor. This final stage of the flow can be described as a ``hot-spot'' regime since it involves large couplings only when initial and final particles are located near the end points of the flat regions on the Fermi surface.

We have analyzed the residual perturbation from particle-particle scattering on the mean-field like spin density wave regime, and we have found that its main effect is indeed to gradually reduce the width of the Fermi surface regions with large $2k_{\rm F}$ couplings, as the typical energy scale is reduced. However the same considerations show this effect is strongly dependent on the number $N$ of Fermi surface patches which are always introduced in the numerical solution of the RG flows. We have shown that the flow (in the space of directions for the coupling vector) slows down tremendously in the spin density wave regime, as $N$ goes to infinity. Therefore, the magnitude of the coupling vector increases by a factor of order $\exp(\gamma N^2)$ (where $\gamma>0$) during this phase before the system reaches the third regime. So we conclude that initial couplings have to be extremely small (of the order of $\exp(-\gamma N^2)$) if the cross-over from spin density wave to d-wave superconductivity is to be predicted in a reliable way from the one-loop approximation. For real 2D systems with flat Fermi surface, $N$ is in fact infinite, so the more traditional picture is recovered.\cite{Dzyaloshinskii72,Gorkov75,Zheleznyak97}

As a final note, we would like to make a brief remark on the various numerical computations of RG flow for the 2D Hubbard model which have appeared over the past few years.\cite{Zanchi96,Zanchi00,Halboth00,Furukawa98,Honerkamp01} Since they introduce by numerical necessity a finite set of patches, do they suffer from the same limitation which has been discussed here~? We are inclined to say no in fact, since in most of these works, the bare couplings are not chosen to be very weak. The instability occurs then relatively quickly, so that all the subtle effects connected with the finiteness of $N$ do not have time to influence the flow in a dramatic way.

\section*{ACKNOWLEDGEMENTS}
We have benefitted from discussions with several colleagues including D. Baeriswyl, B. Binz, B. Delamotte, D. Mouhanna, M. Tissier, J. Vidal and D. Zanchi whom we would like to thank here.

\section*{APPENDIX~: ONE-LOOP RENORMALIZATION GROUP AND CONTINUOUS SYMMETRIES}
%
%
As noticed before by Lin, Balents and Fisher,\cite{Lin98} the non-interacting Hamiltonian $H_0$ for a system of coupled chains possesses a very large symmetry. For a system of $N$ species of spin 1/2 fermions with the same Fermi velocity, it is easy to exhibit a Lie algebra ${\mathcal L}$  of generators commuting with $H_0$ and isomorphic to ${{\rm so}(4N)}_{\rm R} \oplus {{\rm so}(4N)}_{\rm L}$. This very large symmetry is of course dramatically broken by the interaction part of the Hamiltonian. But many people have noticed a general trend towards enhanced symmetries of the effective couplings, as the RG flow runs towards low energies.\cite{Lin98,Azaria98,Konik00,Arrigoni99,Schulz98} This phenomenon is certainly related to an important property of RG flows~: ``The action of the generators of ${\mathcal L}$ commutes with the RG flow''. If we denote the bare coupling vector at scale $\Lambda_0$ by $g_0$, and the corresponding effective coupling at scale $\Lambda$ by $\bar{g}(g_0,\Lambda_0,\Lambda)$, then for any transformation $a$ belonging to ${\mathcal L}$, we have~: $a(\bar{g}(g_0,\Lambda_0,\Lambda))=\bar{g}(a(g_0),\Lambda_0,\Lambda)$. A proof of this statement is given below (for the one-loop approximation). We have an important simple consequence of this which is~: ``if the effective couplings are invariant under the action of a generator of ${\mathcal L}$ at some scale, they will remain invariant at any other scale''. This combined property of RG flows and symmetry operations has been used in Sec.~\ref{sec:nl_stab_an}. For any subalgebra ${\mathcal A}$ of ${\mathcal L}$, we consider the subspace ${\mathcal A}^{\rm c}$ of all the two-particle interaction terms which commute with any element of ${\mathcal A}$. We know that the RG flow leaves ${\mathcal A}^{\rm c}$ globally invariant in the sense that it takes any element of ${\mathcal A}^{\rm c}$ at scale $\Lambda_0$ into another element of ${\mathcal A}^{\rm c}$ at scale $\Lambda$. Conversely, for any subspace $S$ in the space of all two-particle interaction terms, we may define the subalgebra $S^{\rm c}$ of all generators of ${\mathcal L}$ which commute with any element of $S$. Clearly $S$ is included in ${(S^{\rm c})}^{\rm c}$. We don't expect $S={(S^{\rm c})}^{\rm c}$ to hold for any subspace $S$, but if there is a subalgebra ${\mathcal A}$ of ${\mathcal L}$ such that $S={\mathcal A}^{\rm c}$, then $S={(S^{\rm c})}^{\rm c}$. Qualitatively, the RG flow has a tendency to gradually eliminate all the non-relevant couplings with respect to the low-energy fixed point. Here a fixed point may be defined only in the projective coupling space, since the magnitude of the couplings is found to diverge at a finite energy scale, within the one-loop approximation. So the flow is confined to the vicinity of smaller and smaller subspaces $S_n$ as more and more irrelevant directions disappear. This decreasing sequence of subspaces corresponds to an increasing sequence of subalgebras ${\mathcal A}^{\rm c}_n$. So the mechanism of symmetry generation at low energies seems to fit this simple-minded picture.

Let us now turn to a more precise discussion of these points. Let us assume that our fermions can be in $N'=2N$ internal states (combining chain and spin degrees of freedom), which corresponding creation operators will be denoted by $c^\dagger_{{\rm R},\alpha}(k)$ and $c^\dagger_{{\rm L},\alpha}(k)$. Let us consider the right moving branch, with the kinetic Hamiltonian $H_0=v_{\rm F} \sum_{k,\alpha} (k-k_{\rm F}):c^\dagger_{{\rm R},\alpha}(k) c_{{\rm R},\alpha}(k):$\;. From this expression, it is clear that the following operators~:
\begin{equation}
\left\lbrace 
\begin{array}{l}
P_{{\rm R},\alpha,\beta}=\sum_k c^\dagger_{{\rm R},\alpha}(k_{\rm F}+k) c^\dagger_{{\rm R},\beta}(k_{\rm F}-k) \\
D_{{\rm R},\alpha,\beta}=\sum_k :c^\dagger_{{\rm R},\alpha}(k) c_{{\rm R},\beta}(k):
\end{array}\right.
\end{equation}
commute with $H_0$. It is also well established\cite{Lin98} that they form a closed Lie algebra isomorphic to $\mathbb{C} \sum_\alpha D_{{\rm R},\alpha,\alpha} \oplus {\rm so}(2N')$. A given Hermitian generator $G$ in this Lie algebra operates on the fermion operators according to~:
\begin{equation}
\left\lbrace 
\begin{array}{l}
{[G,c^\dagger_{{\rm R},\beta}(k)]}=A^{\rm R}_{\alpha,\beta}c^\dagger_{{\rm R},\alpha}(k)+B^{\rm R}_{\alpha,\beta}c^\dagger_{{\rm R},\alpha}(2k_{\rm F}-k)\\
{[G,c_{{\rm R},\beta}(k)]}=C^{\rm R}_{\alpha,\beta}c^\dagger_{{\rm R},\alpha}(2k_{\rm F}-k)+D^{\rm R}_{\alpha,\beta}c^\dagger_{{\rm R},\alpha}(k)
\end{array}\right..\hspace{2cm}
\end{equation}
From $G=G^\dagger$, we get $D^{\rm R}_{\alpha,\beta}=-\bar{A}^{\rm R}_{\alpha,\beta}$ and $C^{\rm R}_{\alpha,\beta}=-\bar{B}^{\rm R}_{\alpha,\beta}$. Furthermore, the canonical anticommutation relations for the electron fields are preserved under the action of an infinitesimal transformation, which implies~:
\begin{eqnarray}
\left\lbrace 
\begin{array}{l}
\label{eq:contraintes1}
A^{\rm R}_{\alpha,\beta}+D^{\rm R}_{\beta,\alpha}=0\\
B^{\rm R}_{\alpha,\beta}+B^{\rm R}_{\beta,\alpha}=0\\
C^{\rm R}_{\alpha,\beta}+C^{\rm R}_{\beta,\alpha}=0
\end{array}\right..
\end{eqnarray}
Of course, we can construct similar symmetry generators for the left moving branch. In the presence of interactions, the huge $\mathcal{L}={\rm so}(2N')_{\rm R} \oplus {\rm so}(2N')_{\rm L}$ symmetry algebra of $H_0$ is no longer preserved in general, but for some special choices of the pair interaction, some non-trivial subalgebras of $\mathcal{L}$ may still commute with the full Hamiltonian. It is therefore useful to study the action of the generators of $\mathcal{L}$ on the pair interaction. In general, we may write the full bare Hamiltonian as~: $H=H_0+\frac{1}{L} \sum_{k,k',q} \sum_{\alpha,\beta,\gamma,\delta} F(\gamma,\delta | \alpha,\beta) c^\dagger_{{\rm R},\gamma}(k+q) c^\dagger_{{\rm L},\delta}(k'-q) c_{{\rm L},\beta}(k') c_{{\rm R},\alpha}(k)$, where the coefficients $F(\gamma,\delta | \alpha,\beta)$ parametrize the interaction. Taking $G$ in $\mathcal{L}$ as before, and computing $[G,H]$ yields the transformation law $F\to F+\delta F$ of the interaction function under the operations of $\mathcal{L}$. A simple computation gives~:
\begin{widetext}
\begin{equation}
\label{eq:deltaF}
\delta F(\gamma,\delta | \alpha,\beta)=A^{\rm R}_{\gamma,\lambda} F(\lambda,\delta | \alpha,\beta) + A^{\rm L}_{\delta,\lambda} F(\gamma,\lambda | \alpha,\beta) + D^{\rm R}_{\alpha,\lambda} F(\gamma,\delta | \lambda,\beta) + D^{\rm L}_{\beta,\lambda} F(\gamma,\delta | \alpha,\lambda)\;.
\end{equation}
\end{widetext}
Note that in general some terms which do not conserve the total particle number such as $c^\dagger c^\dagger c^\dagger c$ or $c^\dagger c c c $ are also generated. We shall restrict ourselves to situations where those terms vanish. This corresponds to the following constraints~:
\begin{eqnarray}
\left\lbrace 
\begin{array}{l}
\label{eq:contraintes2}
B^{\rm R}_{\gamma,\lambda} F(\lambda,\delta | \alpha,\beta) - B^{\rm R}_{\alpha,\lambda} F(\lambda,\delta | \gamma,\beta)=0\\
B^{\rm L}_{\delta,\lambda} F(\gamma,\lambda | \alpha,\beta) - B^{\rm L}_{\beta,\lambda} F(\gamma,\lambda | \alpha,\delta)=0\\
C^{\rm L}_{\beta,\lambda} F(\gamma,\delta | \alpha,\lambda) - C^{\rm L}_{\delta,\lambda} F(\gamma,\beta | \alpha,\lambda)=0\\
C^{\rm R}_{\alpha,\lambda} F(\gamma,\delta | \lambda,\beta) - C^{\rm R}_{\gamma,\lambda} F(\alpha,\delta | \lambda,\beta)=0
\end{array}\right..
\end{eqnarray}
Now, the one loop flow Eqs.~(\ref{eq:eqRGchaines}) have the following form~:
\begin{eqnarray}
\partial_t F(\gamma,\delta | \alpha,\beta) &=& F*F \nonumber\\
&=& F(\gamma,\mu | \lambda,\beta) F(\lambda,\delta | \alpha,\mu)\hspace{1cm}\\
&&\hspace{0.5cm}- F(\gamma,\delta | \lambda,\mu) F(\lambda,\mu | \alpha,\beta)\nonumber\;.
\end{eqnarray}

We are now ready to make the two following statements~:

\textit{i)} Suppose the action of a generator $G$ of $\mathcal{L}$ on an interaction function $F$ satisfies the conditions of Eqs.~(\ref{eq:contraintes2}). Then $F*F$ also satisfies these conditions.

\textit{ii)} Choosing $G$ and $F$ as in \textit{i)}, the action of $G$ commutes with an infinitesimal RG transformation, namely~: $(F+\delta F)*(F+\delta F) = F*F +\delta(F*F) +O({\delta F}^2)$. Or equivalently~: $\delta(F*F) = F*\delta F + \delta F *F$ (we use the notation $\delta F$ as in Eq.~(\ref{eq:deltaF}) above). It is a simple exercise to check that \textit{i)} and \textit{ii)} hold. We just need the last two relations of Eq.~(\ref{eq:contraintes1}) for \textit{i)} and the first one for \textit{ii)}.

The consequences of \textit{i)} and \textit{ii)} are immediately extended to finite (instead of infinitesimal) RG transformations. So the property that a generator $G$ in $\mathcal{L}$  preserves the total particle number conservation is valid at any stage along a RG flow trajectory, provided it holds for the initial couplings. Similarly, invariance of an interaction function along a transformation $G$ propagates well along any RG flow trajectory.\\

\end{document}